\documentclass[aps,preprint,prd,floatfix,amsmath,amssymb,superscriptaddress]{revtex4-1}
\usepackage{pstricks}
\usepackage{color}
\usepackage{float}
\usepackage[ruled]{algorithm2e}
\usepackage{algorithmic}
\usepackage{graphicx}
\usepackage{dcolumn}
\usepackage{multirow}
\usepackage{bm}
\usepackage{kpfonts}
\usepackage{subfigure}
\usepackage{titlesec}

\titleformat{\paragraph}
{\normalfont\normalsize\bfseries}{\theparagraph}{1em}{}
\titlespacing*{\paragraph}
{0pt}{3.25ex plus 1ex minus .2ex}{1.5ex plus .2ex}

\newcommand{\m}{\mathcal}
\newcommand{\ord}{\alpha\alpha_s^2}
\newcommand{\Oa}{\m{O}(\ord)}
\newcommand{\ra}{\rightarrow}
\newcommand{\lra}{\leftrightarrow}

\def\mathswitchr#1{\relax\ifmmode{\mathrm{#1}}\else$\mathrm{#1}$\fi}
\def\mathswitch#1{\relax\ifmmode#1\else$#1$\fi}
\newcommand{\PW}{\mathswitchr W}
\newcommand{\mm}{\m{M}}

\newcommand{\ppar}{\partial}

\newcommand{\ew}{\mathrm{ew}}
\newcommand{\bew}{b^{\ew}}

\newcommand{\NCf}{\mathswitch {N_{\mathrm{C}}^f}}
\newcommand{\Nc}{\mathswitch {N_{\mathrm{c}}}}
\newcommand{\scrs}{\scriptscriptstyle}
\newcommand{\sw}{\mathswitch {s_{\scrs\PW}}}
\newcommand{\cw}{\mathswitch {c_{\scrs\PW}}}

\newcommand{\ttb}{\mathswitchr {t\bar{t}}}
\newcommand{\dijet}{\mathrm{dijet}}
\newcommand{\qqb}{\mathswitchr {q\bar{q}}}

\newcommand{\rR}{\mathrm{R}}
\newcommand{\rL}{\mathrm{L}}

\newcommand{\GeV}{\mathrm{GeV}}

\def\si{\sigma}

\def\la{\lambda}

\def\Li#1{\mathrm{Li}_{#1}}

\def\dl#1{\log^2(#1)}

\def\qi{q_1^{\tau}}
\def\qf{q_2^{\lambda}}

\def\wk{\mathswitchr {wk}}
\def\Sudakov{\mathswitchr {Sudakov}}
\def\add{\mathswitchr {add}}
\def\prod{\mathswitchr {prod}}
\def\cwk{C^{\wk}}
\def\brc#1{\left(#1\right)}
\def\Lm{\log\brc{\frac{M_Z^2}{M_W^2}}}

\def\Mtt{M(t\bar t)}

\def\PTt{p_{T}(t)}

\def\bI{{\bf I}}
\def\bT{{\bf T}}
\def\bV{{\bf V}}

\def\EW{\mathrm{EW}}

\def\tree{\mathrm{tree}}
\def\pb{\mathrm{pb}}
\def\brc#1{\left(#1\right)}

\floatstyle{ruled}
\newfloat{Feynman Diagram}{thp}{lop}
\floatname{Feynman Diagram}{Feynman Diagram}
\newfloat{Crossing}{thp}{lop}
\floatname{Crossing}{Crossing}
\newfloat{Vertex}{thp}{lop}
\floatname{Vertex}{Vertex}

\begin{document}

\preprint{FERMILAB-PUB-16-314-T}

\title{Study of weak corrections to Drell-Yan, top-quark pair and
  dijet Production at high energies with {\tt MCFM}}


\author{John M. Campbell}
\email[]{johnmc@fnal.gov}
\affiliation{Fermilab, PO Box 500, Batavia, IL 60510, USA}
\author{Doreen Wackeroth}
\email[]{dow@ubpheno.physics.buffalo.edu}
\affiliation{Fermilab, PO Box 500, Batavia, IL 60510, USA}
\affiliation{Department of Physics, University at Buffalo, 
The State University of New York, Buffalo, NY 14260, USA}
\author{Jia Zhou}
\email[]{jiazhou@buffalo.edu}
\affiliation{Department of Physics, University at Buffalo, 
The State University of New York, Buffalo, NY 14260, USA}

\begin{abstract}
\noindent
Electroweak (EW) corrections can be enhanced at high energies due to
the soft or collinear radiation of virtual and real $W$ and $Z$ bosons
that result in Sudakov-like corrections of the form
$\alpha_W^l\log^n(Q^2/M_{W,Z}^2)$, where $\alpha_W =\alpha/(4\pi\sin^2\theta_W)$
and $n\le 2l-1$. The inclusion of EW corrections
in predictions for hadron colliders is therefore especially important when searching for signals
of possible new physics in distributions probing the kinematic regime $Q^2
\gg M_V^2$. Next-to-leading order (NLO) EW corrections should also be taken into
account when their size ($\m{O}(\alpha)$) is comparable to that of QCD
corrections at next-to-next-to-leading order (NNLO) ($\m{O}(\alpha_s^2)$).
To this end we have implemented the
NLO weak corrections to the Neutral-Current
Drell-Yan process, top-quark pair production and di-jet production in
the parton-level Monte-Carlo program {\tt MCFM}. This enables a
combined study with the corresponding QCD corrections at NLO and NNLO. We
provide both the full NLO weak corrections and their Sudakov
approximation since the latter is often used for a fast evaluation of weak
effects at high energies and can be extended to higher orders. With both
the exact and approximate results at hand, the validity of the Sudakov approximation
can be readily quantified.
\end{abstract}


\maketitle


\section{Introduction}\label{sec:intro}

As the CERN Large Hadron Collider (LHC) is operating at an
unprecedented high energy and is reaching unrivalled precision, the
inclusion of electroweak (EW) corrections becomes increasingly
important.  This is equally true in tests of the Standard Model (SM) and
in searches for signals
of new physics, in particular in the high-energy and high-momentum
regimes of kinematic distributions (see, for example,
Ref.~\cite{Chiesa:2013yma} for a review).  
Electroweak corrections at high energies
may also play a significant role in the extraction of parton
distribution functions (PDFs), for instance in constraining the gluon
PDF at high momentum fraction in di-jet production (see,
for example, Refs.~\cite{Rojo:2014kta,Rojo:2015acz}).
The importance of weak corrections at high energies is due to the
occurrence of soft and collinear radiation of virtual and real $W$ and $Z$
bosons.  These give rise to Sudakov-like corrections that take the
form~\cite{Ciafaloni:1998xg},
\begin{equation}
\alpha_W^l\log^n(Q^2/M_{W,Z}^2) \qquad \mbox{where}~\;
\alpha_W = \frac{\alpha}{4\pi\sin^2\theta_W}
\quad \mbox{and}~\; n\le 2l-1 \; ,
\label{eq:Sudakovform}
\end{equation}
and $Q^2$ denotes a typical energy scale of the hard process.
Electroweak ${\cal O}(\alpha)$
corrections have been calculated for a number of
processes relevant to LHC physics, and are now becoming more widely
available, also in combination with QCD corrections, thanks to
automated tools such as {\tt RECOLA}~\cite{Actis:2016mpe}, {\tt
  SHERPA/MUNICH+OPENLOOPS}~\cite{Kallweit:2014xda,Cascioli:2011va},
{\tt GOSAM}~\cite{Chiesa:2015mya}, and {\tt
  MADGRAPH5\_aMC@NLO}~\cite{Alwall:2014hca,Frixione:2015zaa}. Recent progress in this
area is reviewed in Ref.~\cite{Badger:2016bpw}.  However, dedicated
and efficient computations for specific processes, including also
QCD corrections in the same way, is still highly desirable for LHC studies.

In this paper we present such calculations in the framework of the
widely used, publicly available parton-level Monte Carlo (MC) program
{\tt
  MCFM}~\cite{Campbell:1999ah,Campbell:2011bn,Campbell:2015qma,Boughezal:2016wmq}.
We will concentrate on the implementation of the weak one-loop
corrections to three key SM processes at the LHC: the Neutral-Current
(NC) Drell-Yan (DY) process, $pp \to \gamma, Z \to e^+ e^-,
\mu^+\mu^-$, and strong top-anti-top-quark pair ($t\bar t$) and di-jet
production. At leading order (LO) these processes are of ${\cal
  O}(\alpha^2)$ (NC DY) and ${\cal O}(\alpha_s^2)$ ($\ttb$ and di-jet
production), and we provide the cross sections due to the full set of
$W$ and $Z$ exchange diagrams at ${\cal O}(\alpha^3)$ (NC DY) and
${\cal O}(\alpha_s^2 \alpha)$ ($\ttb$ and di-jet production). These
contributions represent a gauge-invariant subset of the
next-to-leading-order (NLO) EW corrections and thus can be studied
separately. They provide the dominant EW effects in the Sudakov
kinematic regime, i.~e. when all Mandelstam invariants $\hat s_{ij}$
are of the same size and are much larger than the weak scale,
$|\hat{s}_{ij}|~\sim~\hat{s}\gg M_W^2$. Since here we are interested
in providing improved predictions with {\tt MCFM} in the Sudakov
regime, we leave the inclusion of the photonic ${\cal O}(\alpha)$
corrections to future work. It is important to note, however, that for
precision studies in the non-Sudakov regime, e.~g., around the $Z$
resonance of the NC DY process (see, e.~g., a recent status report in
Ref.~\cite{Alioli:2016fum} and references therein) and in the
forward-backward asymmetry in $\ttb$ production~\cite{Hollik:2011ps}, the consideration of
the full EW ${\cal}(\alpha)$ corrections is of the utmost importance.
Given the high relevance of these key SM processes at the LHC, they
have already been computed including exact NLO EW effects (NC
DY~\cite{Baur:2001ze,CarloniCalame:2007cd,Arbuzov:2007db,Dittmaier:2009cr,Li:2012wna},
$\ttb$~\cite{Beenakker:1993yr,Kuhn:2005it,Bernreuther:2005is,Kuhn:2006vh,Bernreuther:2006vg,Moretti:2006nf,Hollik:2007sw,Bernreuther:2008md,Bernreuther:2008aw,Kuhn:2009nf,Bernreuther:2010ny,Kuhn:2011ri,Hollik:2011ps,Bernreuther:2012sx,Kuhn:2013zoa,Pagani:2016caq,Denner:2016jyo}
and di-jet~\cite{Moretti:2006ea,Scharf:2009sp,Dittmaier:2012kx}) or at
next-to-next-to-leading-order (NNLO) QCD (NC
DY~\cite{Anastasiou:2003ds,Melnikov:2006kv,Catani:2009sm,Gavin:2010az,Boughezal:2016wmq},
$\ttb$~\cite{Czakon:2013goa,Czakon:2014xsa,Czakon:2015owf,Abelof:2015lna,Czakon:2016ckf}
and di-jet~\cite{Ridder:2013mf,Currie:2013dwa,Currie:2014upa}).
State-of-the-art fixed higher-order corrections have also been
implemented in, and matched to, parton-shower (PS) programs (NC DY at
NNLO QCD+PS~\cite{Karlberg:2014qua,Hoeche:2014aia} and NLO
EW+PS~\cite{Barze':2013yca}, $\ttb$ at NLO
QCD+PS~\cite{Frixione:2003ei,Frixione:2007nw} and di-jet at NLO
QCD+PS~\cite{Alioli:2010xa}) and improved by the analytic resummation
of logarithmically-enhanced corrections (NC DY at
NNLO+NNLL~\cite{Catani:2015vma} and $\ttb$ at
NNLO+NNLL~\cite{Czakon:2011xx,Kidonakis:2014isa,Kidonakis:2014pja,Kidonakis:2015ona}). For
the {\tt MCFM} implementation of the weak one-loop corrections to the
NC DY process we make use of the results provided in
Refs.~\cite{Hollik:1988ii,Beenakker:1991ca}, while in the case of
$\ttb$ production we implement the results of
Ref.~\cite{Kuhn:2005it,Kuhn:2006vh} for the virtual corrections. For
di-jet production we use results from the case of $\ttb$ production in
the limit $m_t \to 0$ and from $b$-jet production~\cite{Kuhn:2009nf},
where applicable, and re-calculate the remaining contributions.  The
${\cal O}(\alpha_s^2 \alpha)$ cross sections to $\ttb$ and di-jet
production also include real QCD radiation, whose effects have been
re-calculated and implemented using the {\tt MCFM} formulation of
the Catani-Seymour dipole subtraction
method~\cite{Catani:1996vz,Catani:2002hc}.  It is interesting to note
that this implementation of weak one-loop corrections to $\ttb$ and
di-jet production in {\tt MCFM} provides, for the first time, these
results in a readily available, fully flexible, public MC
code~\footnote{Weak corrections have been implemented in the publicly
  available {\tt HATHOR}~\cite{Aliev:2010zk} library, which provides
  predictions for total cross sections to $\protect\ttb$ production.}.
We validate the results of our implementation by comparing {\tt MCFM}
results for relative weak one-loop corrections to the total cross
sections and kinematic distributions with published
results in Ref.~\cite{Dittmaier:2012kx} (di-jet production) and
Ref.~\cite{Kuhn:2013zoa} ($t\bar t$ production), and by using the
publicly available MC program {\tt ZGRAD2}~\cite{Baur:2001ze}.  We
also compare the relative impact of weak one-loop and higher-order QCD
corrections and discuss two different approaches to combining these
corrections ({\em additive} and {\em multiplicative}). In the case of
the NC DY process, NNLO QCD predictions are also obtained with {\tt
  MCFM}~\cite{Boughezal:2016wmq}, while the NLO QCD predictions for
di-jet production are obtained with the MC program {\tt MEKS} (version
1.0)~\cite{Gao:2012he}, and the (N)NLO predictions for $\ttb$
production are taken from Refs.~\cite{Czakon:2015owf,Czakon:2016ckf}.

The important interplay of photon-induced processes and EW corrections
is illustrated in the case of the NC DY process.  At LO this already receives
a contribution from the tree-level photon-induced
process, $\gamma \gamma \to l^+ l^-$. We compare our {\tt MCFM}
results with the ones of Ref.~\cite{Dittmaier:2009cr} and discuss the
impact of this process on a number of interesting NC DY observables.
This is particularly interesting given the large uncertainty in the photon
PDF that is obtained in global PDF sets such as MRST2004QED~\cite{Martin:2004dh},
NNPDF3.0QED~\cite{Ball:2013hta,Ball:2014uwa}, and
CT14QED~\cite{Schmidt:2015zda}. A recent study of the combined impact of 
NLO EW effects and photon-induced processes in $t\bar t$ production can be 
found in Ref.~\cite{Pagani:2016caq}.

EW logarithmic corrections that take the form of Eq.~(\ref{eq:Sudakovform}) have been
studied at one-loop and beyond by several groups, e.~g., see
Refs.~\cite{Ciafaloni:1998xg,Kuhn:1999nn,Beccaria:1999fk,Fadin:1999bq,Ciafaloni:1999ub,Melles:2000ia,Denner:2000jv,Hori:2000tm,Melles:2000ed,Beenakker:2000kb,Melles:2000gw,Kuhn:2001hz,Denner:2001gw,Beccaria:2001yf,Melles:2001ye,Beenakker:2001kf,Denner:2003wi,Feucht:2004rp,Pozzorini:2004rm,Jantzen:2005az,Jantzen:2005xi,Denner:2006jr,Jantzen:2006jv,Chiu:2007yn,Denner:2008yn,Chiu:2008vv,Bell:2010gi,Manohar:2012rs,Becher:2013zua,Manohar:2014vxa,Bauer:2016kkv},
and references therein. As a first step to improving the predictions of
multi-purpose MC programs for the LHC at high energies, one could for instance
implement the Sudakov approximation of EW
corrections. Examples of such improvements are the implementation of
weak Sudakov corrections to $Z+ \le 3$~jets in {\tt
  ALPGEN}~\cite{Chiesa:2013yma} and in {\tt
  SHERPA}~\cite{Thompson:2016hzj}. Moreover, in cases where these EW
Sudakov corrections are indeed dominant and represent a good
approximation of the complete EW corrections, the known higher-order
EW Sudakov logarithms, i.~e. beyond one-loop order, 
could be used to further improve predictions in the
Sudakov regime.

The implementation of weak one-loop corrections in {\tt MCFM} includes
both the exact weak corrections as described above and their Sudakov
approximation based on the general algorithm of
Denner-Pozzorini~\cite{Denner:2000jv,Denner:2001gw}.  We compare these
predictions for observables in the NC DY process and for $\ttb$ and
di-jet production at high invariant masses of the leading pair
of final-state particles to provide insight into how well the approximation
works. Examples of similar studies can be found, for instance, in
Refs.~\cite{Zykunov:2005tc,Moretti:2006nf,Kuhn:2013zoa,Badger:2016bpw}.

In this paper we concentrate on the inclusion of virtual weak corrections.
This is because the masses of the weak gauge bosons provide a physical Infra-Red
(IR) cutoff so that in general there is no need for the inclusion of
real emission of weak gauge bosons. Moreover, the real emission of a
$W$ or $Z$ boson and their subsequent decays usually yields an
experimental signature that can easily be separated from the
no-emission case. Even in situations where the inclusive experimental
treatment of an observable requires the inclusion of both real and
virtual $W$ and $Z$ boson radiation, such EW Sudakov corrections can
still have a significant numerical impact due to an incomplete
cancellation of mass-singular EW logarithms between these two
contributions~\cite{Ciafaloni:2000df,Ciafaloni:2000gm,Ciafaloni:2006qu,Baur:2006sn,Bell:2010gi,Stirling:2012ak,Manohar:2014vxa}. This
is a consequence of not averaging over the initial-state isospin
degrees of freedom so that, unlike in QED and QCD, the Bloch-Nordsieck
theorem is violated~\cite{Ciafaloni:2000df,Ciafaloni:2000rp}. 
For example, a study in Ref.~\cite{Baur:2006sn} found that 
the inclusion of real $W$ and $Z$ boson radiation 
in NC DY production, $pp \to e^+ e^- V, \, V=W,Z$ with $V \to jj$ and $Z \to \bar \nu \nu$,  
in predictions for the invariant-mass distribution of the final-state $e^+ e^-$ pair ($M(e^+e^-)$) 
at the 14 TeV LHC can reduce
the impact of EW 1-loop corrections from about $-21\%$ to $-16\%$ of the LO cross section 
at $M(e^+e^-)=4$~TeV.
However, the necessity of
including real emission diagrams in a prediction as part of the EW
corrections, and the degree of their partial cancellation, strongly
depends on the details of the experimental analysis.  This therefore
requires careful consideration, ideally in consultation with the
experimentalists conducting the analysis. Therefore, we do not include
real $W/Z$ emission contributions in the predictions presented in this
paper, but rather concentrate on the implementation of virtual weak
corrections in {\tt MCFM}.  We note that a combined study of real and
virtual $W/Z$ emission to NC DY, $\ttb$ and di-jet production
can be conducted with {\tt MCFM} with realistic analysis cuts, where the former is based on the
tree-level processes $W/Zjj$, $l^+ l^- W/Z$ and $\ttb W/Z$. A recent discussion of the
resummation of EW Sudakov logarithms originating from 
real $W,Z$ radiation in the DY process can be found in Ref.~\cite{Bauer:2016kkv} 
(and references therein).

The paper is organized as follows. In Section~\ref{sec:implementation}
we provide the details for both the implementation of the exact
(Section~\ref{sec:exact}) and the Sudakov approximation
(Section~\ref{sec:sudakov}) of weak one-loop corrections to the NC DY
process, $\ttb$ and di-jet production.  We validate our implementation
by comparing with existing calculations and published results in
Section~\ref{sec:comparison} and provide a comparison of the exact
calculation with the Sudakov approximation at high invariant masses in
Section~\ref{sec:sudakovtest}. Section~\ref{sec:ncdycomp} also
includes a discussion of the impact of the photon-induced tree-level
process, $\gamma \gamma \to l^+ l^-$. Before we conclude in
Section~\ref{sec:conclusions}, we discuss the size of the weak one-loop
corrections relative to QCD corrections, and their combination,
in Section~\ref{sec:combination}. The details of the {\tt
  MCFM} implementation of real QCD radiation diagrams, which also
contribute at ${\cal O}(\alpha_s^2 \alpha)$ in $\ttb$ and di-jet
production, are provided in the appendix.

\section{Implementation of Weak Corrections in {\tt MCFM}}\label{sec:implementation}

The hadronic differential cross section $d\si$ for proton-proton
collisions at the LHC can be written as a convolution of a partonic
cross section $d\hat{\si}$ and PDFs
$f_i, f_j$ for partons $i,j$ carrying a fraction $x_{1,2}$ of the
protons' momenta $P_{1,2}$:
\begin{equation}\label{eq:hadronX}
  d\si\brc{P_1,P_2} = \frac{1}{1+\delta_{ij}}\sum_{i,j} \left[\int_0^1dx_1\int_0^1dx_2 
   f_i\brc{x_1,\mu_F^2}f_j\brc{x_2,\mu_F^2}d \hat{\si}_{ij} \brc{\mu_R^2}+
i \leftrightarrow j \right] \, ,
\end{equation}
where $\mu_F,\mu_R$ denote the factorization and renormalization scales
respectively.  Here we consider $2\to 2$ processes, $i(p_1)+j(p_2) \to k(p_3)+
l(p_4)$, where the partonic cross section $d \hat{\si}_{ij}$ can be expressed in terms of
the following Mandelstam variables:
\begin{equation}\label{eq:stu}
  \begin{aligned}
  &\hat{s} = \brc{p_1 + p_2}^2 = \brc{p_3 + p_4}^2, \\
  &\hat{t} = \brc{p_1 + p_3}^2 = \brc{p_2 + p_4}^2, \\
  &\hat{u} = \brc{p_1 + p_4}^2 = \brc{p_2 + p_3}^2, \\
  \end{aligned}
\end{equation}
where all momenta are assumed outgoing, so that $p_1 + p_2 + p_3 + p_4 = 0$.
The momenta of the
incoming partons are $-p_1 = x_1P_1$ and $-p_2 = x_2P_2$
and $p_3$,~$p_4$ are the outgoing momenta of the final-state particles
$k,l$. Up to one-loop electroweak corrections, $d\hat\si$ can be written in terms of the leading-order (LO)
amplitude, $\mm_0$, and the one-loop corrections, $\delta \mm$, as follows:
\begin{equation}
d\hat \si = dP_{kl} \overline{\sum}[|\mm_0|^2(\alpha^m \alpha_s^n)+2{\cal R}e (\delta \mm \times \mm_0^*)(\alpha^{m+1} \alpha_s^n)]
\end{equation}
where $dP_{kl}$ denotes the phase space of the final-state particles.
The barred summation indicates that we have
averaged (summed) over initial (final) state spin and color degrees of freedom.
The indices $m$,~$n$ are used to indicate the order in perturbation theory considered
here, by pulling out overall strong ($\alpha_s$) and weak ($\alpha$) coupling factors.
For the processes considered in this paper we have  $m=2, n=0$ for the NC DY process
and $m=0, n=2$ for top-quark pair and di-jet production. We provide a detailed
description of the {\tt MCFM} implementation of the exact ${\cal
  O}(\alpha^{m+1} \alpha_s^n)$ contributions to $d\hat \si$ in
Section~\ref{sec:exact} and of their Sudakov approximation in
Section~\ref{sec:sudakov}.

\subsection{Exact One-loop Corrections}\label{sec:exact}

\subsubsection{Neutral-current Drell-Yan production}
\label{sec:ncdy}

\begin{figure}[htbp]
\includegraphics[scale=1]{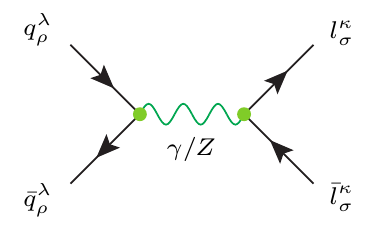}
\caption{Feynman diagrams for the NC DY process at LO.}
\label{fig:dy:lo}
\end{figure}

The LO parton-level process under consideration is
$q_\rho^\la\bar{q}_\rho^\la \ra \gamma,Z \ra
l_\si^\kappa\bar{l}_\si^\kappa$ shown in Fig. \ref{fig:dy:lo}, where
$q=u,d,s,c$ and $l=e,\mu$. The labels $\la=R,L,\kappa=R,L$ denote the chirality,
and $\rho=\pm, \si=\pm$ are the isospin indices. Note that we consider
all external fermions to be massless and that we do not include the
$b\bar b$-initiated process due to the smallness of the bottom-quark
PDF.

\begin{figure}[h]
\includegraphics[scale=1]{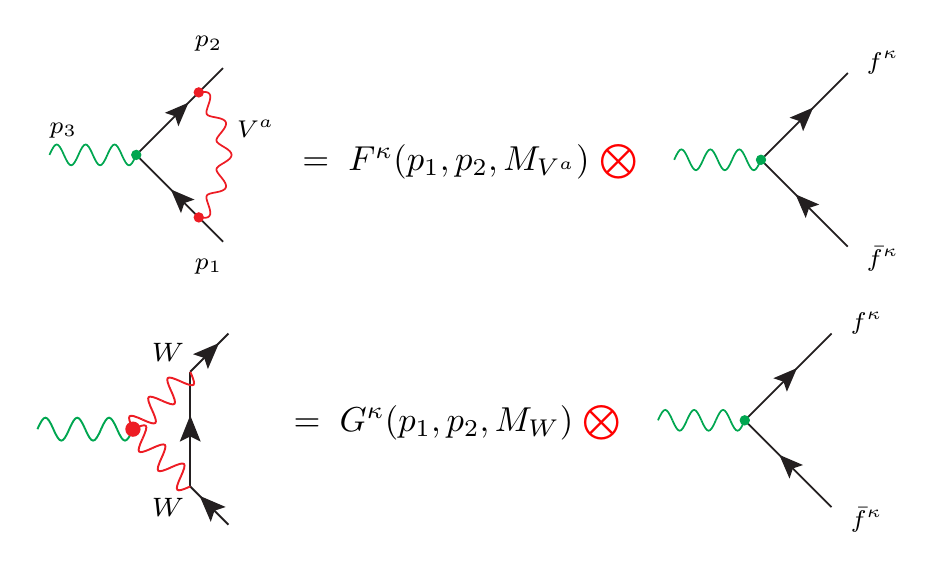}
\caption{Weak vertex corrections of $\m{O}(\alpha)$ to the NC DY
  process $(V^a=Z,W)$.\label{fig:dyvert}}
\end{figure}

\begin{figure}[h]
\includegraphics[scale=1]{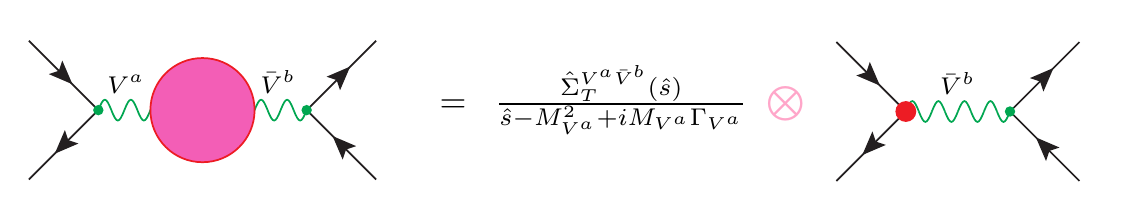}
\caption{Gauge-boson self-energy correction at ${\cal O}(\alpha)$,
where $\hat{\Sigma}_{T}^{V^a\bar{V}^b}$ is the renormalized vector
boson self-energy, and $\Gamma_{V^a}$ the width of the gauge boson
($V^a=\gamma,Z)$. The red dot 
denotes the tree-level coupling to the initial $\qqb$ pair, $I^{V^a}_{q\bar{q}}$,
and $I^{V^b}_{l\bar{l}}$ describes the coupling to the final lepton pair. 
\label{fig:dyself}}
\end{figure}

\begin{figure}[tbph]
\includegraphics[scale=1]{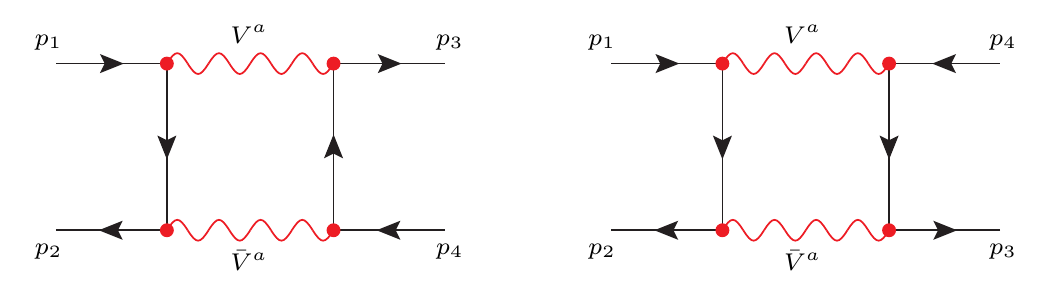}  
\caption{Feynman diagrams for weak one-loop box corrections to the NC
  DY process ($V^a = Z/W$).
\label{fig:z:box}}
\end{figure}

When we consider weak one-loop corrections to the parton-level LO NC
DY process, $q\bar q \to \gamma,Z \to l^+ l^-$, we refer to a
correction of $\m{O}(\alpha)$ involving only $W$ and $Z$ bosons in
UV-divergent vertex and self-energy corrections, and UV-finite box
corrections, as shown in Figs.~\ref{fig:dyvert}, \ref{fig:dyself} and
\ref{fig:z:box} respectively.  The weak one-loop vertex corrections
can be described by well-known form factors, $F^{\kappa}$ and
$G^\kappa$, which multiply the LO vertex as schematically shown in
Fig.~\ref{fig:dyvert}.
The gauge-boson self-energy correction at ${\cal O}(\alpha)$ can also
be factorized with respect to the LO amplitude as shown schematically
in Fig.~\ref{fig:dyself}.
We implemented the explicit expressions of Ref.~\cite{Hollik:1988ii}
for the unrenormalized vector boson self-energies (Appendix A), the
corresponding counterterms (Appendix B), and the renormalized form
factor $\hat F^{\kappa}$ (Appendix C.1, Eq.~(C.5)). For the renormalized
form factor $\hat G^{\kappa}$ we use Eq.~(3.13) of
Ref.~\cite{Beenakker:1991ca}.  Note that the counterterms are defined
in the on-shell renormalization scheme (see
Refs.~\cite{Hollik:1988ii,Beenakker:1991ca} for details).  
The {\tt MCFM} implementation of the
weak one-loop box contributions due to the exchange of $Z/W^{\pm}$ bosons,
shown in Fig.~\ref{fig:z:box}, represents our own calculation that
expresses the results in terms of scalar integrals, which are evaluated with
{\tt QCDLoop}~\cite{Ellis:2007qk}.

Finally, for the evaluation of the $Zf\bar f$ coupling at LO {\tt
  MCFM} provides three choices of the EW input scheme, i.e.  the
so-called $\alpha(0)$, $\alpha(M_Z^2)$, and $G_\mu$ schemes (see also
Ref.~\cite{Dittmaier:2009cr}).  Note that the corrections relative to
the LO cross section are always evaluated by using the fine-structure
constant $\alpha(0)$. Also, in all three schemes the cosine of the
weak mixing angle is defined via the physical $W$ and $Z$ masses as
$\cos\theta_W=M_W/M_Z$~\cite{Sirlin:1980nh}.  When the form factors of
the $Zf\bar{f}$ and $\gamma f\bar{f}$ vertices are renormalized in the
$\alpha(0)$-scheme, the corrections depend on the light-fermion masses
in a sensitive fashion due to terms proportional to $\alpha\log{m_f}$,
which enter through electric charge renormalization in the on-shell
scheme as~\cite{Denner:1991kt}
\[
\delta Z^{\mathrm{em}}_e=\mathrm{Re}\brc{\frac{1}{2}\frac{\ppar\Sigma^{\gamma}_T(k^2)}{\ppar k^2}\middle|_{k^2=0}}=\frac{2}{3}\frac{\alpha}{4\pi}\sum_{f\ne t}\NCf Q_f^2\log\brc{\frac{M_W^2}{m_f^2}}+\ldots
\]
The $\alpha(M_Z^2)$-scheme introduces a contribution
$\Delta\alpha(M_Z^2)=-\mathrm{Re}\hat{\Pi}^{\gamma}(M_Z^2)$~\cite{Denner:1991kt},
where $\hat{\Pi}^{\gamma}(k^2)=\Sigma_T^{\gamma}(k^2)/k^2-2\delta
Z^{\rm em}_e$ is the renormalized photon vacuum polarization, and the
LO coupling is evaluated at
$\alpha(M_Z^2)=\alpha(0)/[1-\Delta\alpha(M_Z^2)]$. Therefore, the
relative corrections in the $\alpha(M_Z^2)$-scheme absorb a term of
$2~\Delta\alpha(M_Z^2)$ resulting from the running of the
electromagnetic coupling from $q^2=0$ to $q^2=M_Z^2$. As a result the
logarithmic light-fermion terms are canceled at ${\cal O}(\alpha^3)$.

The $G_{\mu}$-scheme implies the replacement $\alpha(0) \to
\alpha_{G_\mu}$ with~\cite{Sirlin:1980nh}
\begin{equation}\label{eq:gmu}
\alpha_{G_\mu} = \sqrt{2}G_{\mu}\frac{M_W^2(M_Z^2-M_W^2)}{\pi M_Z^2}=\frac{\alpha(0)}{ (1-\Delta r)} \; ,
\end{equation}
where $G_{\mu}$ is the Fermi constant measured in muon decay.  The quantity
$\Delta r$ describes the radiative corrections to muon decay, which is
given at one-loop order by~\cite{Sirlin:1980nh,Marciano:1980pb,Hollik:1988ii},
\begin{equation}\label{eq:deltar}
  \Delta r^{\rm{1-loop}} = \Delta \alpha(M_Z^2)-\frac{c_w^2}{s_w^2}\Delta \rho+\Delta r_{rem}=\frac{\hat{\Sigma}_T^W(0)}{M_W^2}
  +\frac{\alpha}{4\pi\sw^2}\brc{6+\frac{7-4\sw^2}{4\sw^2}\log\cw^2} \; ,
\end{equation}
where $\hat{\Sigma}_T^W(0)$ is the renormalized $W$ boson self energy
evaluated at $q^2=0$ and we have introduced the short-hand notation,
$\cw=\cos\theta_W$ and $\sw=\sin\theta_W$.  Note that $\Delta r$
contains $\Delta\alpha$, so that the relative correction in the
$G_{\mu}$-scheme is also free of the logarithmic light-fermion mass
dependence. Moreover, it also contains corrections to the $\rho$
parameter. We therefore recommend use of the $G_{\mu}$-scheme for
obtaining precise predictions for the NC DY process, which is the
default scheme in {\tt MCFM}.

\subsubsection{Top-quark pair production}\label{sec:ttb}

\begin{figure}[thpb]
\includegraphics[scale=1]{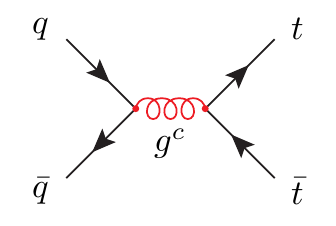}
\includegraphics[scale=1]{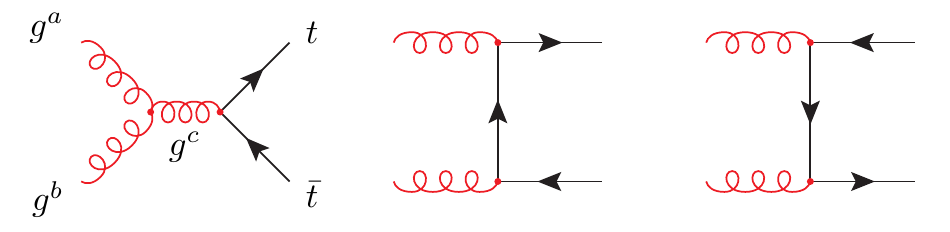}
\caption{Feynman diagrams for LO strong $\ttb$ production at ${\cal
    O}(\alpha_s^2)$.}
\label{fig:ttb:lo}
\end{figure}

\begin{figure}[thpb]
\includegraphics[scale=0.5]{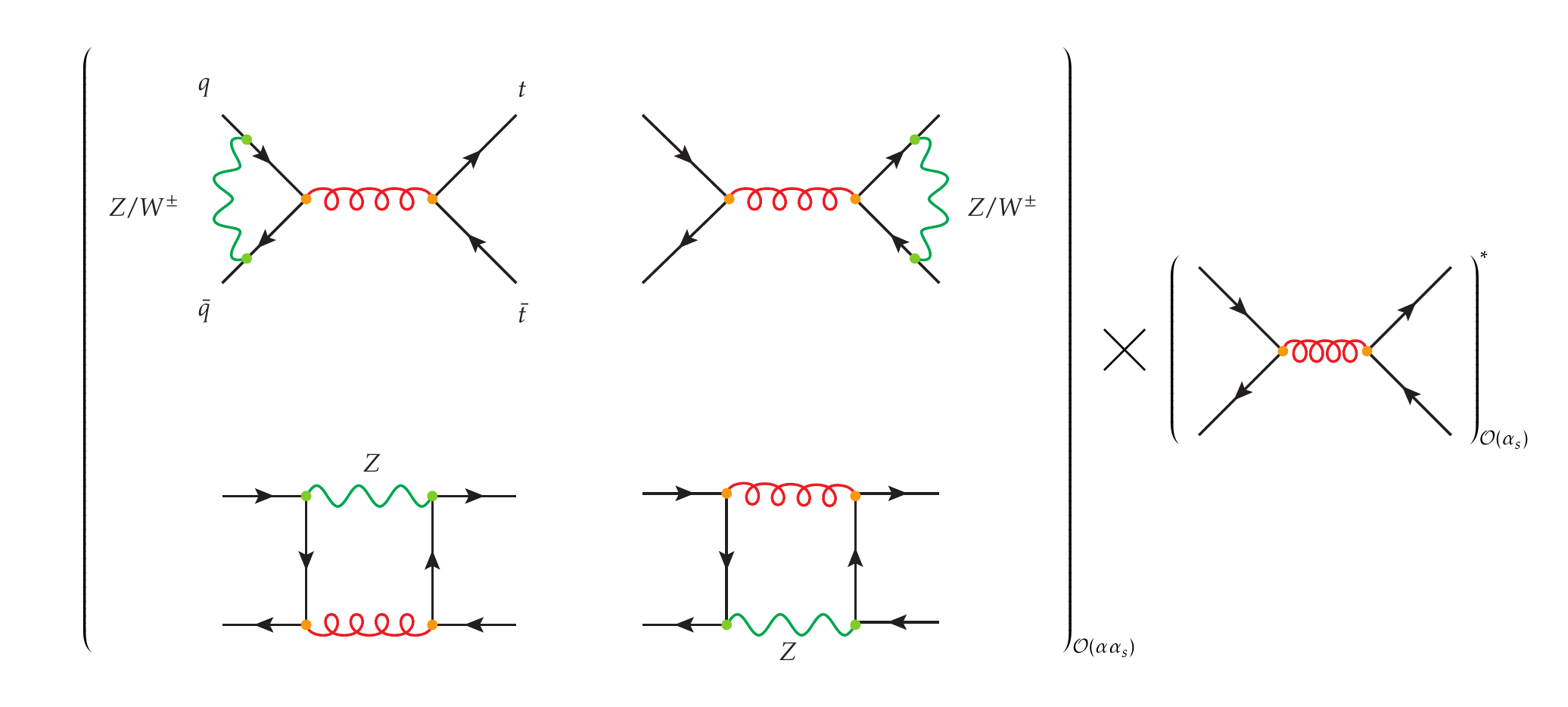}
\caption{Sample diagrams for one-loop weak virtual corrections to the
  quark-antiquark annihilation subprocess in strong $\ttb$ production,
  which consist of vertex and box corrections, respectively. The
  $\hat{u}$-channel box diagrams are not explicitly shown.}
\label{fig:ttb:nlo:qqb}
\end{figure}

\begin{figure}[thpb]
    \includegraphics[scale=0.5]{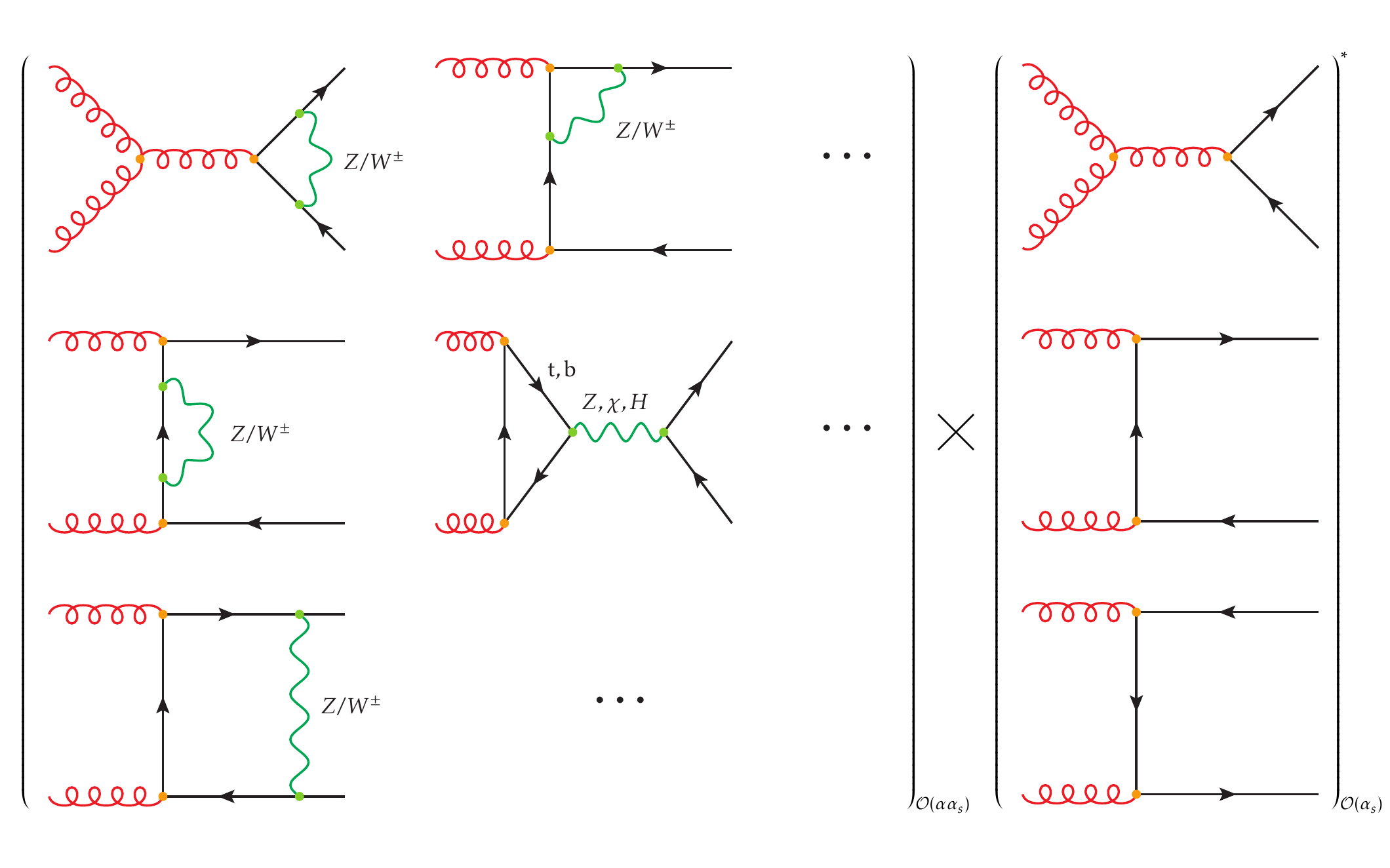}
    \caption{Sample diagrams for one-loop weak virtual corrections to
      the gluon fusion subprocess in strong $\ttb$ production, which
      consist of vertex, self-energy, and box corrections,
      respectively. The ellipses represent the vertex, self-energy and
      box diagrams which are not explicitly shown.}
    \label{fig:ttb:nlo:gg}
\end{figure}

Top-quark pairs are primarily produced through the strong
interaction, which occurs at ${\cal O}(\alpha_s^2)$ at LO.  The LO diagrams are shown
in Fig.~\ref{fig:ttb:lo}, with gluon fusion representing approximately
$90$\% of the rate at the LHC and quark-antiquark annihilation the remainder.
We consider NLO weak corrections to the strong $\ttb$
production processes, i.e. we include all contributions of ${\cal O}(\alpha_s^2 \alpha)$
to the cross sections of $\qqb$ annihilation
and gluon fusion. This includes weak one-loop
contributions of the form shown in Fig.~\ref{fig:ttb:nlo:qqb} (for $\qqb$ annihilation)
and Fig.~\ref{fig:ttb:nlo:gg} (for gluon fusion),
as well as $s$-channel $Z,H$ exchange
diagrams in the gluon fusion channel that are also shown in
Fig.~\ref{fig:ttb:nlo:gg}.  For the {\tt MCFM} implementation of the
renormalized weak one-loop corrections to $\qqb$ annihilation and
gluon fusion we have adopted the analytic expressions of
Ref.~\cite{Kuhn:2005it} and Ref.~\cite{Kuhn:2006vh}, respectively.  We have
re-calculated the contributions from the $s$-channel $Z,H$ exchange
diagrams. The UV poles in the vertex and self-energy corrections in
both the $\qqb$ annihilation and gluon fusion subprocesses are removed
by performing wave-function and top-mass renormalization in the
on-shell renormalization scheme (see
Refs.~\cite{Beenakker:1993yr,Kuhn:2005it,Kuhn:2006vh} for details).
We have numerically cross-checked the implementation of the pure weak
${\cal O}(\alpha_s^2\alpha)$ contribution to $\ttb$ production against
the calculation provided in Ref.~\cite{Beenakker:1993yr}.

\begin{figure}[thbp]
  \includegraphics[scale=0.55]{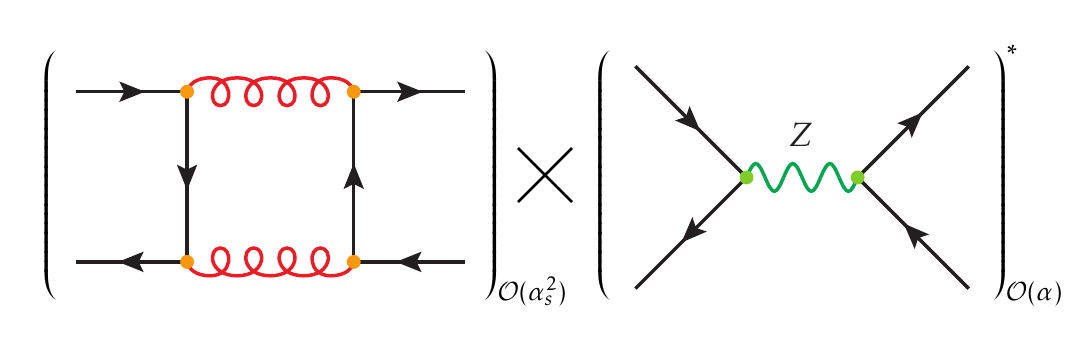}
  \caption{The contribution of the $\hat t$-channel two-gluon box
    diagram interfered with the $Z$-mediated Born diagram to the NLO
    cross section for $\ttb$ production at ${\cal O}(\alpha
    \alpha_s^2)$.  The contribution of the $\hat{u}$-channel two-gluon
    box diagram is not explicitly shown.}
  \label{fig:ttb:qcdbox}
\end{figure}

\begin{figure}[htpb]
\includegraphics[scale=1]{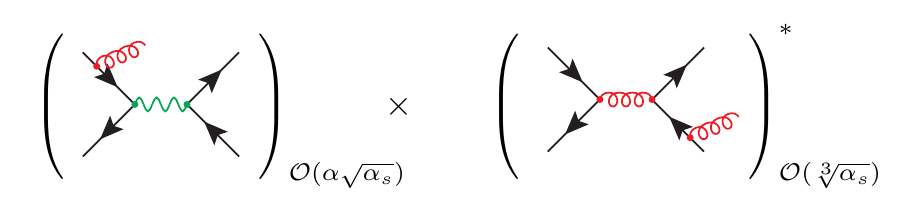}
\caption{Sample diagrams for real corrections to $\ttb$ production in
  the $\qqb$ annihilation channel contributing to the ${\cal O}(\alpha
  \alpha_s^2)$ cross section.}
\label{fig:ttb:real}
\end{figure}

In the case of $\qqb$ annihilation,  the ${\cal O}(\alpha_s^2 \alpha)$
corrections include box diagrams that contain a gluon in the loop, specifically
the gluon-$Z$ box diagram of
Fig.~\ref{fig:ttb:nlo:qqb} and the double-gluon box diagrams of
Fig.~\ref{fig:ttb:qcdbox}.  These contributions are UV finite but IR
divergent.  
The IR divergences are canceled by the corresponding real gluon
radiation contributions depicted in Fig.~\ref{fig:ttb:real}, as long as
IR-safe observables are considered. In {\tt MCFM} the extraction and
cancellation of the IR poles is performed by using the Catani-Seymour dipole
subtraction method~\cite{Catani:1996vz,Catani:2002hc}.  Note that the color
structure does not permit any contributions involving emitter and spectator
partons that are either both in the initial state or both in the final state.
The only dipole configurations that are present have one parton in the initial
state and one in the final state.  For completeness, the
explicit expressions for the real contribution to $\ttb$ production at
$\Oa$, as implemented in {\tt MCFM}, are provided in the
appendix.

\subsubsection{Di-jet production}\label{sec:dijet}

\begin{figure}[htbp]
 \includegraphics[scale=0.85]{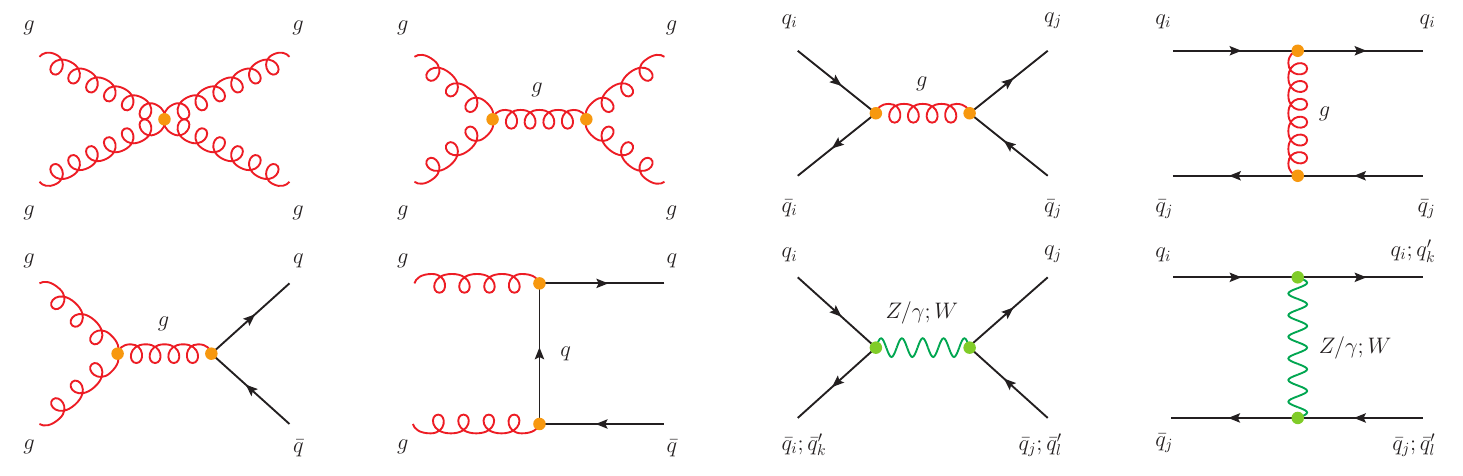}
 \caption{\label{fig:dijet:lo} Sample tree-level Feynman diagrams for
   di-jet production via QCD and EW interactions.}
\end{figure}

Di-jet production is a ${\cal O}(\alpha_s^2)$, ${\cal O}(\alpha^2)$ 
or ${\cal O}(\alpha_s \alpha)$ process at LO,
that is mediated by $2\to 2$
scattering processes involving light quarks and gluons, as shown in
Fig.~\ref{fig:dijet:lo}.  
The different subprocesses can be
categorized in terms of the number of external quarks and gluons:
four-quark, two-gluon-two-quark, and four-gluon subprocesses.
In Tables~\ref{tab:4q} and \ref{tab:2g2q} we list all
processes of the four-quark and two-gluon-two-quark category.  In practice
it is only necessary to perform explicit calculations of each
subprocess A listed in Tables~\ref{tab:4q} and \ref{tab:2g2q}, since all
other processes can be obtained via crossing symmetry. The crossing
relations are indicated in the tables.  The four-gluon subprocess does
not receive corrections at the order under consideration and thus only
contributes to the LO cross section for di-jet production. Note that
again we consider all external fermions to be massless and we do
not include the $b$-quark-initiated processes.

\begin{table}[H] 
\caption{\label{tab:4q} All four-quark subprocesses to di-jet
  production with the flavor indices $i,\, j$ so that $q_{i,j}\,
  \in\{\mathrm{u,d,c,s}\}$, where $i, \, j$ can be equal or
  different.}
\begin{center}
 \begin{tabular}{m{2cm}l}
\hline\hline
 A. &$q_i\bar{q}_i\ra q_j\bar{q}_j$, direct calculation \\
 B. &$q_iq_j\ra q_iq_j$, $(2\ra 3, \, 3\ra 4, \, 4\ra 2; \, s \ra t, \, t \ra u, \, u \ra s)$ \\
 C. &$\bar{q}_iq_i \ra \bar{q}_jq_j$, $(1\lra 2, 3\lra 4)$ \\
 D. &$\bar{q}_i\bar{q}_j \ra \bar{q}_i\bar{q}_j$, $(1 \ra 3, \, 3 \ra 2, \, 2 \ra 1; \, s \ra t, \, t \ra u, \, u \ra s)$ \\
 E. &$q_i\bar{q}_j \ra q_i\bar{q}_j$, $(2 \lra 3; \, s \lra t)$ \\
 F. &$\bar{q}_iq_j \ra \bar{q}_iq_j$, $(1 \ra 3, \, 3 \ra 4, \, 4 \ra 2, \, 2 \ra 1; \, s \lra t)$ \\
\hline\hline
 \end{tabular}
\end{center}
 \end{table}
\begin{table}[H] 
\caption{\label{tab:2g2q} All two-gluon-two-quark subprocesses to
  di-jet production, where $q \in\{\mathrm{u,d,c,s}\}$.  Note that the
  amplitude is multiplied by a minus sign when crossing a final/initial
  state quark to an initial/final state one.}
\begin{center}
\begin{tabular}{m{2cm}l}
\hline\hline
A. &$gg\ra q\bar{q}$, direct calculation \\
B. &$gq\ra gq$, $(2 \ra 3, 3 \ra 4, 4 \ra 2; \, s \ra t, t\ra u, u\ra s)$ \\
C. &$g\bar{q} \ra g\bar{q}$, $(2 \lra 3; \, s\lra t)$ \\
D. &$qg\ra qg$, $(1\lra 4; \, s \lra t)$ \\
E. &$\bar{q}g\ra \bar{q}g$, $(1\ra 2, \, 2\ra 4, 4\ra 3, \, 3\ra 1; \, s \lra t)$ \\
F. &$q\bar{q}\ra gg$, $(1\lra 3, \, 2\lra 4; \, t \lra u)$ \\
G. &$\bar{q}q \ra gg$ $(1\lra 4, \, 2\lra 3)$ \\
\hline\hline
\end{tabular}
\end{center}
\end{table}

The one-loop corrections to di-jet production at fixed $\Oa$ consist
of $\m{O}(\alpha)$ corrections to the QCD mediated processes
interfered with the LO ${\cal O}(\alpha_s)$ amplitudes and of
$\m{O}(\alpha_s)$ corrections to the QCD(weak) mediated processes interfered
with the LO $\m{O}(\alpha)(\m{O}(\alpha_s))$ amplitudes.  Their contributions to the
partonic di-jet cross section at $\Oa$ can be written symbolically as
\begin{eqnarray}\label{sig_nlo:propto}
d\hat{\sigma}(\alpha_s^2\alpha) & \propto &
2\mathrm{Re}\left[\delta\m{M}(\alpha_s\alpha)\cdot\m{M}_0^*(\alpha_s)
+ \delta \m{M}(\alpha_s^2)\cdot\m{M}_0^*(\alpha)\right]
\end{eqnarray}
where $\m{M}_0(\alpha_s)$ and $\m{M}_0(\alpha)$ denote the LO
amplitude with gluon and weak boson exchange, respectively.  $\delta
\m{M}(\alpha_s^2)$ denotes the QCD one-loop correction to the strong
LO amplitude while $\delta \m{M}(\alpha_s\alpha)$ represents both weak corrections
to the strong LO amplitude and QCD corrections to the weak LO
amplitude. As was the case for $\ttb$ production, we also need to take
into account real QCD corrections in order to cancel the IR
divergences stemming from the virtual QCD corrections. Explicit expressions for the
real corrections can be found in the appendix. In the following we
will present the virtual corrections to the
four-quark and two-gluon-two-quark subprocesses, $q_i\bar{q}_i \ra
q_j\bar{q}_j$ and $gg \ra q\bar{q}$.  All remaining subprocesses can
be obtained via the crossing relations listed in Table~\ref{tab:4q} and
\ref{tab:2g2q}.  The virtual corrections to the two-gluon-two-quark subprocess $gg \ra q\bar{q}$
consist of the same weak one-loop corrections as in $\ttb$ production, shown in Fig.~\ref{fig:ttb:nlo:gg},
with the top quark replaced by a massless quark.
For other subprocesses we have partially used the analytic expressions for the
weak corrections to $b$-jet production of
Ref.~\cite{Kuhn:2009nf}, where applicable to the case of di-jet production.

\begin{figure}[htbp]
 \includegraphics[scale = 0.5]{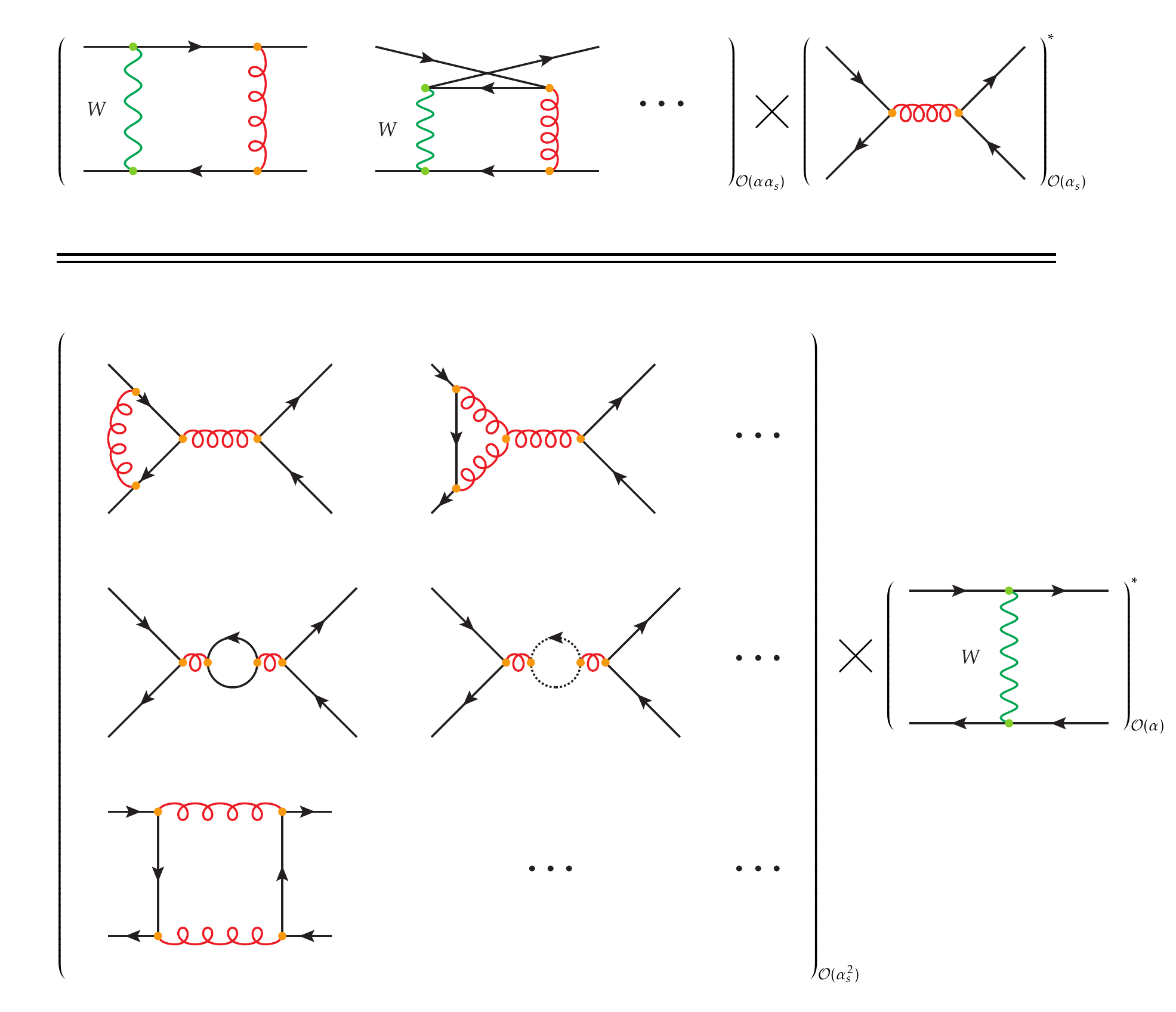}
 \caption{\label{fig:2j:4q:v2} Sample Feynman diagrams for virtual
   corrections to four-quark subprocesses of category 2 of
   Table~\ref{tab:cat:4q}, $u_i\bar{u}_i\ra d_j\bar{d}_j$, $d_i\bar{d}_i
   \ra u_j\bar{u}_j$, where $i,\,j$ denote the $i$th- or $j$th-generation
   of the light (anti)quarks, i.e, $i,\,j\in(1,2)$. The diagrams above
   the double line contribute to $\delta\m{M}(\alpha_s\alpha)$ and
   below the double line to $\delta\m{M}(\alpha_s^2)$.  Weak one-loop
   corrections similar to the ones shown in Fig.~\ref{fig:ttb:nlo:qqb} are
   not explicitly shown.}
\end{figure}

\begin{figure}[htbp]
 \includegraphics[scale = 0.5]{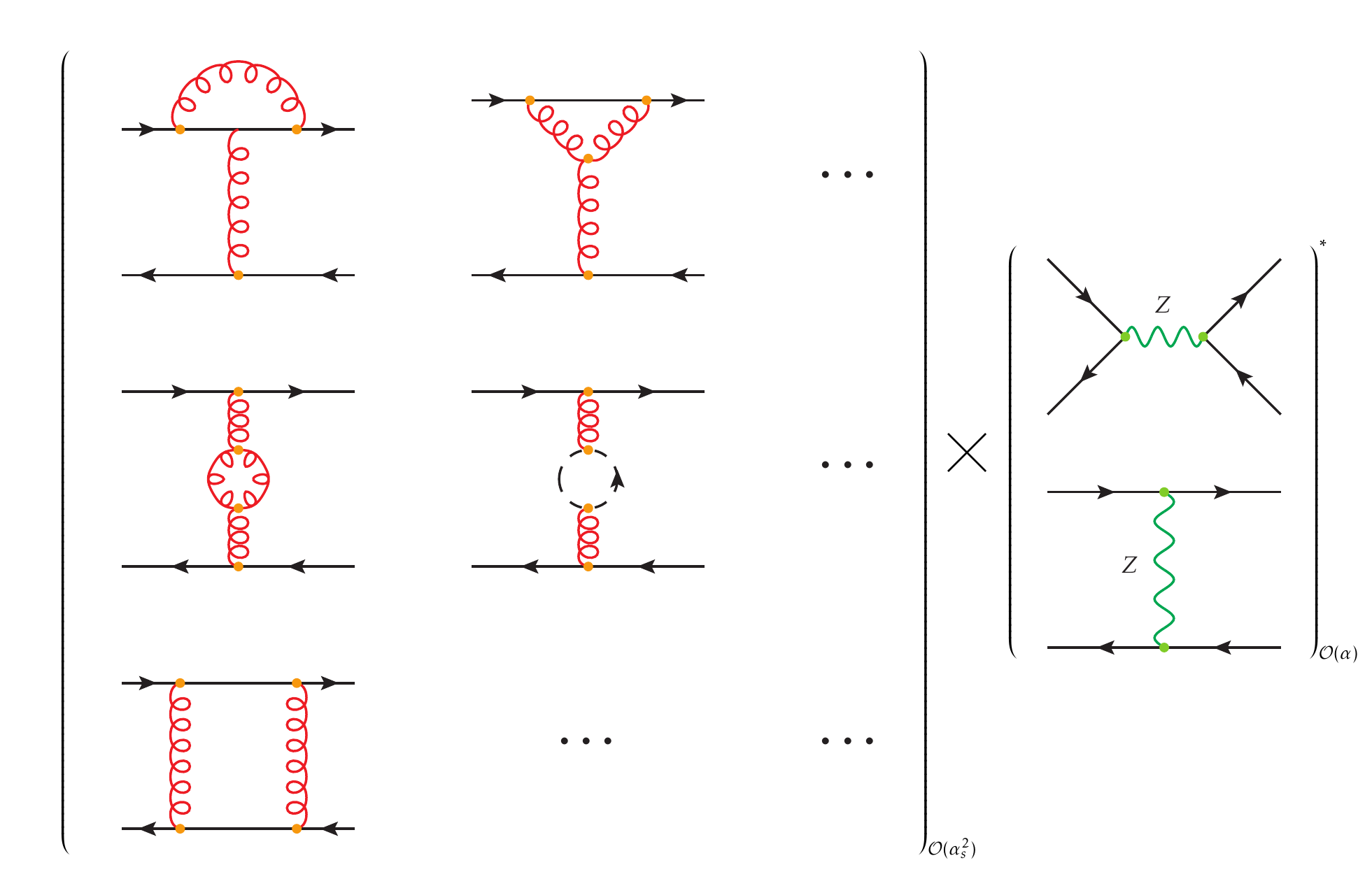}
 \caption{\label{fig:2j:4q:v3} Sample Feynman diagrams for virtual
   corrections to four-quark subprocesses of category 3 of
   Table~\ref{tab:cat:4q}, $q\bar{q}\ra q\bar{q}$, $q\in\{u_i,d_i\}$.
   Contributions similar to the ones shown in Fig.~\ref{fig:ttb:nlo:qqb}
   and \ref{fig:2j:4q:v2} are not explicitly shown.}
\end{figure}

\begin{table}[H]
\caption{\label{tab:cat:4q} The three categories of subprocesses that
  comprise the four-quark processes $q_i\bar{q}_i \ra q_j\bar{q}_j$ of
  di-jet production (with $u_{1,2}=u,c$ and $d_{1,2}=d,s$).}
\begin{center}
\begin{tabular}{m{2cm}l}
\hline\hline
category 1 & $u_i\bar{u}_i\ra u_j\bar{u}_j, d_i\bar{d}_i \ra d_j\bar{d}_j \, , \, \mbox{ for}~i\ne j$ \\ 
category  2 &  $u_i\bar{u}_i\ra d_j\bar{d}_j, d_i\bar{d}_i \ra u_j\bar{u}_j$ \\
category  3 &  $u_i\bar{u}_i\ra u_i\bar{u}_i, d_i\bar{d}_i \ra d_i\bar{d}_i$ \\
\hline\hline
\end{tabular}
\end{center}
\end{table}
In the case of the four-quark subprocesses $q_i\bar{q}_i \ra
q_j\bar{q}_j$ we further divide them into the three categories shown in
Table~\ref{tab:cat:4q} since, as discussed further shortly, they proceed
through different diagrams.
The virtual corrections to the four-quark subprocesses of category 1 of
Table~\ref{tab:cat:4q} can again be obtained from the weak corrections to
$\ttb$ production shown in  Fig.~\ref{fig:ttb:nlo:qqb}.
Sample diagrams for virtual corrections to the four-quark
subprocesses of category 2 and 3 of Table~\ref{tab:cat:4q} are shown in
Fig.~\ref{fig:2j:4q:v2} and Fig.~\ref{fig:2j:4q:v3}, respectively. While
the color structure ensures that
there is no contribution from the interference between one-loop QCD and LO
weak diagrams in category 1 (except for the mixed QCD-weak box
contribution), such corrections do survive in category 2
(diagrams below the double line in Fig.~\ref{fig:2j:4q:v2}) and
category 3 (Fig.~\ref{fig:2j:4q:v3}).

The UV poles in self-energy and vertex corrections are eliminated
after applying an appropriate renormalization procedure as described
below.  The IR poles originating from the soft and collinear virtual
gluon contributions in category 2 and 3 are canceled against their
counterparts from the real corrections and PDF counterterms as
described in the appendix. Note that these real QCD corrections yield
real corrections to the quark-gluon-initiated subprocesses of
Table~\ref{tab:2g2q} by crossing the emitted gluon to the initial
state.  These corrections exhibit a trivial initial-state collinear
singularity which is absorbed into the PDFs (as detailed in the appendix).

The weak one-loop vertex corrections in all three categories
involve $Z/W^{\pm}$ boson exchange in the $gq\bar{q}$ vertex, which can be
described in terms of the renormalized form
factor $f_1$ given in Eq.~(III.11) of Ref. \cite{Kuhn:2009nf} (or
(II.16) of Ref. \cite{Kuhn:2005it}), as
\begin{equation}\label{eq:2j:vert}
  \delta\m{M}\brc{\alpha_s\alpha}_y = -\frac{1}{2}\frac{\alpha}{4\pi}\m{M}_0\brc{\alpha_s}_y\left\{
    \sum_{f=q_i,q_j}\left[\brc{g_v^f}^2+\brc{g_a^f}^2\right]f_1\brc{\frac{M_Z^2}{y}} + \frac{1}{2s_w^2}f_1\brc{\frac{M_W^2}{y}} \; ,
    \right\}
\end{equation}
where $g_{v}^f(g_a^f)$ is the vector(axial) vector coupling of the fermion to the $Z$ boson, $g_v^f=1/(2s_wc_w) (T_3^f-2s_w^2 Q_f), g_a^f=1/(2s_wc_w) T^3_f$, and the subscript $y$ denotes the channel of the amplitude, while the variable $y$ in the 
function $f_1$ denotes the Mandelstam variable corresponding to that channel. 
The function $f_1$ is given by
\begin{equation}
f_1(x) = 1+2\left[
    \brc{1+\log(x)}\brc{2x+3}-2\brc{1+x}^2\brc{\Li2\brc{1+\frac{1}{x}}-\frac{\pi^2}{6}}\right].
\end{equation}

The renormalized contribution of the QCD vertex and self-energy
corrections in category 2 and 3 to $d\hat \si$ can be written as
\begin{equation}
  \delta\m{M}\brc{\alpha_s^2}_x = \frac{\alpha_s}{4\pi}\m{M}_0\brc{\alpha_s}_x
  \left[
    2\Lambda_1(x)+2\Lambda_2(x)+\Pi(x)+ \delta Z_{g_s}
  \right]
\end{equation}
where the subscript and variable $x$ have the same interpretation as $y$ in Eq.~(\ref{eq:2j:vert}),
and $\delta Z_{g_s}$ denotes the renormalization constant for the strong coupling
\begin{equation}
  \delta Z_{g_s} = \left[
    \brc{-\frac{11}{2} + \frac{n_F}{3}}\frac{1}{\epsilon} + \frac{1}{3}\brc{ \frac{1}{\epsilon}+ \log\frac{\mu^2}{m_t^2}}
    \right] \; ,
\end{equation}
where $m_t$ denotes the top-quark mass, $d=4-2\epsilon$ and we 
do not distinguish between IR and UV
poles. Consequently, the quark wave function renormalization constant
$\delta Z_q = -\alpha_s/(3\pi)B_0(0,0,0)\equiv0$.  The two form factors
$\Lambda_1$ and $\Lambda_2$, describing virtual gluon corrections to
$gq\bar{q}$ and $V^aq\bar{q}$ ($V^a=Z,\gamma,W^{\pm}$) vertices
respectively, read
\begin{eqnarray}
    \Lambda_1(x) &=& \frac{1-\Nc^2}{4\Nc}\left[\brc{d-7}~B_0\brc{x,0,0}-2~x~C_0\brc{0,x,0,0,0,0}\right] - \frac{\Nc\brc{\Nc^2-1}}{4} B_0\brc{x,0,0} \\
    \Lambda_2(x) &=& \frac{\brc{\Nc^2-1}^2}{4\Nc}\left[\brc{d-7}~B_0\brc{x,0,0}-2~x~C_0\brc{0,x,0,0,0,0}\right],
\end{eqnarray}
with the scalar integrals
\[
\begin{aligned}
& B_0(x,0,0) = \frac{1}{\epsilon}+\log\frac{\mu^2}{-x-i\varepsilon}+2 + \m{O}\brc{\epsilon} \\
  & C_0(0,x,0,0,0,0) = \frac{1}{x}\left[\frac{1}{\epsilon^2} + \frac{1}{\epsilon}\log\frac{\mu^2}{-x-i\varepsilon} + \frac{1}{2}\log^2\frac{\mu^2}{-x-i\varepsilon}\right] + \m{O}\brc{\epsilon} \,.
\end{aligned}
\]
Finally, the gluon self-energy correction reads
\begin{equation}
  \begin{aligned}
    \Pi(x) = \frac{1}{2(d-1)}&\big[4(d-2)~A_0\brc{m_t^2} + \brc{9d-6-2(d-2)n_F}~x~B_0\brc{x,0,0} \\
     & -2~\brc{4m_t^2+(d-2)x}~B_0\brc{x,m_t^2,m_t^2}\big],
  \end{aligned}
\end{equation}
with
\begin{align*}
  &A_0\brc{m_t^2} = m_t^2\left[\frac{1}{\epsilon}+1+\log\frac{\mu^2}{m_t^2-i\varepsilon}\right] + \m{O}\brc{\epsilon} \\
  &B_0\brc{x,m_t^2,m_t^2} = \frac{1}{\epsilon} + 2 + \log\frac{\mu^2}{m_t^2} - \sqrt{1-\frac{4m_t^2}{x}}\log\frac{1+\sqrt{1-\frac{4m_t^2}{x}}}{-1+\sqrt{1-\frac{4m_t^2}{x}}} + \m{O}\brc{\epsilon}
\end{align*}
Depending on the production channel considered, the variable $x$ could be $\hat{s}$, $\hat{t}$, or $\hat u$.

\def\uxs{\mathswitchr {u\times s}}
\def\uxt{\mathswitchr {u\times t}}
\def\sxs{\mathswitchr {s\times s}}
\def\sxt{\mathswitchr {s\times t}} 
\def\txs{\mathswitchr {t\times s}}
\def\txt{\mathswitchr {t\times t}}

The box contributions to $d\hat \si$ can be written in terms of
four contributions for all di-jet subprocesses, by taking advantage of appropriate crossing
relations.  In this way, $d\hat \si$ can be schematically written as
\begin{eqnarray}
d\hat\si =  4\pi\alpha\alpha_s^2 & \Big\{ 
\mathrm{prop}_{V^a}\brc{x} [cs_1 (\Box^{\txs}+
\Box^{\uxs})+cs_2 (\Box^{\txt}+\Box^{\uxt})](M_{V^a} \ne 0) \nonumber \\ & \qquad \qquad
+ (cs_1 (\Box^{\txs}+
\Box^{\uxs})+cs_2 \Box^{\uxt}+ cs_3 \Box^{\txt})(M_{V^a}=0) \Big\}
\end{eqnarray}
in terms of the three possible color factors
\[
cs_1 = \frac{\Nc^2-1}{4}, \quad cs_2 = \frac{-\Nc^2+1}{4\Nc}, \quad
cs_3 = \frac{\brc{\Nc^2-1}^2}{4\Nc}
\]
and the propagator function defined by
\begin{equation}\label{eq:2j:propx}
\mathrm{prop}_{V^a}\brc{x} = \frac{x\brc{x-M_{V^a}^2}}{\brc{x-M_{V^a}^2}^2+\Gamma_{V^a}^2M_{V^a}^2}
\end{equation}
The integral functions 
for the interference of the $t$-channel and $u$-channel
box diagrams with the $s$-channel and $t$-channel LO diagrams, $\Box^{\txs}$,
$\Box^{\uxs}$ and $\Box^{\txt}$, $\Box^{\uxt}$, respectively,
are given by
\begin{equation}\label{eq:2j:4q:bx:txs}
 \begin{aligned}
\Box^{\txs} = & \frac{8}{\hat{s}}  \bigg\{-\brc{g_a^f g_a^i+g_v^f g_v^i}\left[
    2\hat{u}\brc{B_0\brc{\hat{s},M_{V^a}^2} - B_0\brc{\hat{t},0}} 
     +\hat{t}(M_{V^a}^2+\hat{t}-\hat{u})\brc{C_0^1\brc{\hat{t},M_{V^a}^2} + C_0^2\brc{\hat{t},0}} \right] \\
  & + 2\brc{g_a^f g_a^i \brc{\hat{t}+\hat{u}}\brc{M_{V^a}^2-\hat{t}+\hat{u}}+g_v^f g_v^i\brc{M_{V^a}^2\brc{\hat{t}+\hat{u}}+3 \hat{t}^2+\hat{u}^2}} C_0^2\brc{\hat{s},M_{V^a}^2} \\ 
    & + \hat{t}\brc{-g_a^f g_a^i \brc{M_{V^a}^4+2 M_{V^a}^2 \hat{t}-\hat{t}^2+\hat{u}^2}-g_v^f g_v^i \brc{M_{V^a}^4+2 M_{V^a}^2 \hat{t}+3 \hat{t}^2+\hat{u}^2}}D_0\brc{\hat{s},\hat{t},M_{V^a}^2} \bigg\}
\end{aligned}
\end{equation}
\begin{equation}\label{eq:2j:4q:bx:uxs}
  \begin{aligned}
   \Box^{\uxs} = & \frac{8}{\hat{s}}  
\bigg\{ -\brc{g_a^f g_a^i-g_v^f g_v^i}\left[
      2 \hat{t}\brc{B_0\brc{\hat{s},M_{V^a}^2} - B_0\brc{\hat{u},0}}
      + \hat{u}\brc{M_{V^a}^2-\hat{t}+\hat{u}}\brc{C_0^1\brc{\hat{u},M_{V^a}^2} + C_0^2\brc{\hat{u},0}}\right] \\
  & +2\brc{g_a^f g_a^i \brc{\hat{t}+\hat{u}} \brc{M_{V^a}^2+\hat{t}-\hat{u}}-g_v^f g_v^i \brc{M_{V^a}^2 \brc{\hat{t}+\hat{u}}+\hat{t}^2+3 \hat{u}^2}}C_0^2\brc{\hat{s},M_{V^a}^2} \\  
  & +\hat{u}\brc{-g_a^f g_a^i \brc{M_{V^a}^4+2 M_{V^a}^2 \hat{u}+\hat{t}^2-\hat{u}^2}+g_v^f g_v^i \brc{M_{V^a}^4+2 M_{V^a}^2 \hat{u}+\hat{t}^2+3 \hat{u}^2}}D_0\brc{\hat{s},\hat{u},M_{V^a}^2} \bigg\}
  \end{aligned}
\end{equation}
\begin{equation}\label{eq:2j:4q:bx:txt}
  \begin{aligned}
    \Box^{\txt} = &  -\frac{8}{\hat{t}} \brc{g_a^f g_a^i+g_v^f g_v^i} \big[
      2\hat{u}\brc{B_0(\hat{s},M_{V^a}^2) - B_0(\hat{t},0)}
      + \hat{t} \brc{M_{V^a}^2+\hat{t}-\hat{u}}\brc{C_0^1(\hat{t},M_{V^a}^2) + C_0^2(\hat{t},0)} \\
      & -2 \brc{M_{V^a}^2 \brc{\hat{t}+\hat{u}}+\hat{t}^2+\hat{u}^2} C_0^2\brc{\hat{s},M_{V^a}^2} \\
      & +\hat{t} \brc{M_{V^a}^4+2 M_{V^a}^2 \hat{t}+\hat{t}^2+\hat{u}^2} D_0\brc{\hat{s},\hat{t},M_{V^a}^2}\big] 
  \end{aligned}
\end{equation}
\begin{equation}\label{eq:2j:4q:bx:uxt}
    \Box^{\uxt}  = \frac{16 \hat{u}^2}{\hat{t}} (g_a^f g_a^i+g_v^f g_v^i) (\hat{u} D_0(\hat{s},\hat{u},M_{V^a}^2)-2 C_0^2(\hat{s},M_{V^a}^2))
\end{equation}
In these expressions we have used the short-hand notation
\begin{equation*}
  \begin{aligned}
    & B_0\brc{x,y} = B_0\brc{x,y,0} \\
    & C_0^1\brc{x,y} = C_0\brc{x,0,0,0,0,y} \\
    & C_0^2\brc{x,y} = C_0\brc{x,0,0,0,y,0} \\
    & D_0\brc{x,y,z} = D_0\brc{0,0,0,0,x,y,z,0,0,0}
  \end{aligned}
\end{equation*}
for the scalar integrals, which read
\begin{subequations}\label{eq:2j:4q:int}
\begin{equation}\label{eq:2j:4q:int:b0}
  B_0\brc{x,y} = \left\{
  \begin{array}{lr}
    \frac{1}{\epsilon} + 2 + \log\frac{\mu^2}{-x} & (y = 0 \; || \; y = x) \\
    \frac{1}{\epsilon} + 2 + \frac{y-x}{x}\log\frac{y-x}{y} + \log\frac{\mu^2}{y} & \mathrm{otherwise}
  \end{array}
  \right.
\end{equation} \\
\begin{equation}\label{eq:2j:4q:int:c01}
  C_0^1\brc{x,y} = \left\{
  \begin{array}{lr}
    \frac{1}{x}\Li2\brc{1} = \frac{\pi^2}{6x} & (y = -x) \\
    C_0^2\brc{x,0} & (y = 0) \\
    \frac{1}{x}\left[\log\brc{-\frac{x}{y}}\log\brc{1+\frac{x}{y}} + \Li2\brc{-\frac{x}{y}}\right]
    & \mathrm{otherwise}
  \end{array}
  \right.
\end{equation} \\
\begin{equation}\label{eq:2j:4q:int:c02}
   C_0^2\brc{x,y} = \left\{
  \begin{array}{lr}
    \frac{1}{x}\left[
      \frac{1}{\epsilon^2} + \frac{1}{\epsilon}\log\frac{\mu^2}{-x}
      + \frac{1}{2}\log^2\frac{\mu^2}{-x}
      \right] & (y = 0)  \\
    -\frac{1}{2x}\left[
      \frac{1}{\epsilon^2} + \frac{1}{\epsilon}\log\frac{\mu^2}{y}
      + \frac{\pi^2}{6} + \frac{1}{2}\log^2\frac{\mu^2}{y}
      \right] & (y = x) \\
    \frac{1}{x}\left[
      \frac{1}{\epsilon}\log\frac{y}{y-x} + \log^2\frac{y}{y-x}
      + \log\frac{\mu^2}{y}\log\frac{y}{y-x} + \Li2\brc{\frac{x}{y}}
      \right] & \mathrm{otherwise}
  \end{array}
  \right.
\end{equation} \\
\begin{equation}\label{eq:2j:4q:int:d0}
   D_0\brc{x,y,z} = \left\{
  \begin{array}{ll}
    \frac{1}{xy}\left[
      \frac{4}{\epsilon^2} + \frac{2}{\epsilon}\brc{\log\frac{\mu^2}{-x}+\log\frac{\mu^2}{y}} + \log^2\frac{\mu^2}{-x} + \log^2\frac{\mu^2}{y} - \log^2\frac{x}{y} - \pi^2 \right]
       & (z = 0) \\
     \frac{1}{(x-z)y}\Big[
      \frac{1}{\epsilon^2} + \frac{1}{\epsilon}\brc{\log\frac{\mu^2}{y}
        + 2\log\frac{z}{z-x}}
      + \frac{1}{2}\log^2\frac{\mu^2}{-y} - \frac{1}{2}\log^2\frac{z}{-y} \\
      \qquad + 2\log\frac{\mu^2}{-y}\log\frac{z}{z-x} -4~\Li2\brc{\frac{x}{x-z}}- \Li2\brc{1+\frac{z}{y}}
      - \frac{\pi^2}{6}
      \Big] & (z \neq 0)
  \end{array}
  \right.
\end{equation}
\end{subequations}
Here $g_{v}^f,g_a^f$ parameterize both the coupling of the fermion to
the $Z$ and $W$ boson (with $g_{v}=g_a=1/(2 \sqrt{2} s_w)$ in the $W$ case).
It should be noted that the third expression in
Eq.~(\ref{eq:2j:4q:int:c01}) is exact only for $-|y|<x<0$, otherwise
there is a phase difference that we have omitted here. Since only the
real part contributes to $d\hat \si$ it would not alter the final
result.  For $M_{V^a} > 0 $, the expressions in
Eqs.~(\ref{eq:2j:4q:bx:txs})-(\ref{eq:2j:4q:bx:uxt}) describe box
diagrams with a gluon and a massive vector boson, while for $M_{V^a} =
0$, they describe the pure QCD box diagrams with two gluons. As in
$\ttb$ production, we differentiate between them as weak box and QCD
box contributions respectively.

\subsection{Leading and Subleading Logarithms in the Sudakov regime}
\label{sec:sudakov}

As discussed earlier, the NLO EW corrections at high energies are
dominated by logarithms of $\hat{s}_{ij}/M_{V^a}^2$, where
$\hat{s}_{ij}=(p_i+p_j)^2$ are Mandelstam variables of momenta
$p_i,~\;p_j$ associated with external particles $i, \, j$, and
$M_{V^a}$ is the mass of the weak gauge boson $V^a=Z,W$.
\begin{center}
\begin{figure}[htbp]
  \includegraphics[scale=0.22]{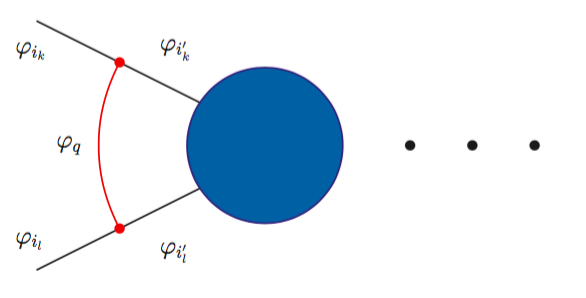} 
  \includegraphics[scale=0.22]{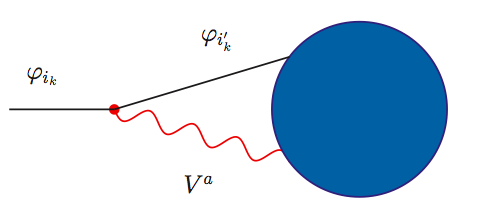}
  \includegraphics[scale=0.22]{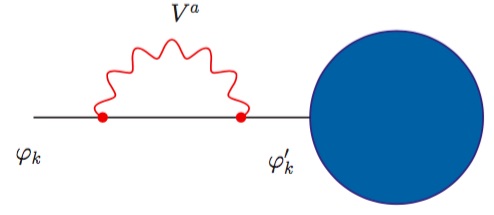}
  \caption{Feynman diagrams representing soft-collinear contributions
    resulting in double logarithms (left), collinear contributions
    (middle) and contributions from wave function renormalization
    (right) both resulting in single logarithms.}
  \label{fig:logs}
\end{figure}
\end{center}
For the implementation of the weak leading and subleading logarithms
at one-loop in {\tt MCFM} we adopt the formalism of
Refs.~\cite{Denner:2000jv,Denner:2001gw}.  As described in detail in
Ref.~\cite{Denner:2001gw}, the ${\cal O}(\alpha)$ corrections to a
$2\to 2$ process in logarithmic approximation (LA) in the Sudakov
regime, i.~e. when all Mandelstam variables are of the same size and
are much larger than the weak scale, $|\hat{s}_{ij}|~\sim~\hat{s}\gg
M_W^2$, factorize into the Born amplitude and double (DL) and single
logarithms (SL).  The double logarithms $\dl{\hat{s}_{kl}/M_{V^a}^2}$
originate from the soft-collinear contributions due to the exchange of
virtual EW gauge bosons between external legs $k,l$, as illustrated in
Fig.~\ref{fig:logs} (left).  The possible sources of single logarithms
in virtual EW corrections are collinear mass singularities and
wave-function and parameter renormalization when the UV singularities
are subtracted at $\mu_R \approx M_{V^a}$.  The contributions of
collinear radiation and wave-function renormalization are
schematically shown in Fig.~\ref{fig:logs} (middle and left).  Thus,
in the LA limit the ${\cal O}(\alpha)$ corrections to the LO amplitude
$\mm_0$ can be written as
\begin{equation}\label{eq:LA}
\delta\mm=\frac{\alpha}{4\pi}\brc{\delta^{\mathrm{LSC}}+\delta^{\mathrm{SSC}}+\delta^{\mathrm{C}}+
\delta^{\mathrm{PR}}} \mm_0 \; ,
\end{equation}
where $\delta^{\mathrm{LSC}}$ and $\delta^{\mathrm{SSC}}$ denote
respectively the leading and sub-leading logarithms of soft-collinear
origin, $\delta^{\mathrm{C}}$ the collinear logarithms and
$\delta^{\mathrm{PR}}$ the logarithms originating from parameter
renormalization.  For completeness, we provide in the following the
explicit expressions for $\delta^{\mathrm{LSC,SSC,C,PR}}$ and $\mm_0$
for the NC DY process, $t\bar t$ and di-jet production, as they are
implemented in {\tt MCFM}.  For the details of their derivation we
refer to Refs.~\cite{Denner:2000jv,Denner:2001gw}.

\subsubsection{Neutral-Current Drell-Yan Process}\label{sec:suda-ncdy}

The LO amplitude in LA for a given fermion chirality
$\tau,\lambda=\rL,\rR$ and isospin index $\rho,\si$ reads
\begin{equation}\label{eq:z:born}
  \mm^{q_\rho^\tau l_\si^\la}_0=4 \pi \alpha R_{q_{\rho}^{\tau}l_{\si}^{\lambda}}\frac{\m{A}_{\tau\lambda}}{\hat{s}}
\end{equation}
where
\[R_{q_{\rho}^{\tau}l_{\si}^{\lambda}}
=\sum_{N=Z,\gamma}I_{q_{\rho}^{\tau}}^NI_{l_{\si}^{\lambda}}^N
=\frac{Y_{q_{\rho}^{\tau}}Y_{l_{\si}^{\lambda}}}{4\cw^2}+\frac{T^3_{q_{\rho}^{\tau}}T^3_{l_{\si}^{\lambda}}}{\sw^2}
\]
and $Y$ and $T^3$ are the hypercharge and 3rd-component of the weak
isospin $T$, respectively, which are related to the electric charge
$Q$ via the Gell-Mann-Nishijima formula $Q = Y/2 + T^3$.  The LO
amplitude for each chirality combination $\m{A}_{\tau\lambda}$ is
\(\m{A}_{\rL\rL}=\m{A}_{\rR\rR} =\hat{u}\), and
\(\m{A}_{\rL\rR}=\m{A}_{\rR\rL}=\hat{t}\).  After removing the
photonic virtual corrections from the expressions provided in
Ref.~\cite{Denner:2001gw}, the different contributions in
Eq.~(\ref{eq:LA}) read,
\begin{equation}\label{eq:z:dl}
  \begin{aligned}
   & \delta^{\mathrm{LSC}}_{q_\rho^\tau l_\si^\la} = - \brc{\cwk_{q_{\rho}^{\tau}} + \cwk_{l_{\si}^{\lambda}}}\log^2\brc{\frac{\hat{s}}{M_W^2}}
    + 2\Lm\left[\brc{I^Z_{q_{\rho}^{\tau}}}^2+\brc{I^Z_{l_{\si}^{\lambda}}}^2\right]\log\brc{\frac{\hat{s}}{M_W^2}}, \\
   & \begin{aligned}
       \delta^{\mathrm{SSC}}_{q_\rho^\tau l_\si^\la} =
       &-4\left[\log\brc{\frac{\hat{s}}{M_W^2}}-\Lm\right]
R_{q_{\rho}^{\tau}l_{\si}^{\lambda}}
\brc{I^Z_{q_{\rho}^{\tau}}I^Z_{l_{\si}^{\lambda}}}\log\brc{\frac{\hat{t}}{\hat{u}}} \\
       &- \frac{\delta_{\tau\rL}\delta_{\lambda\rL}}{\sw^4R_{q_{\rho}^{\tau}l_{\si}^{\lambda}}}\log\brc{\frac{\hat{s}}{M_W^2}}\left[\delta_{\rho\sigma}\log\brc{\frac{|\hat{t}|}{\hat{s}}} - \delta_{-\rho\sigma}\log\brc{\frac{|\hat{u}|}{\hat{s}}}\right].
      \end{aligned}
  \end{aligned}
\end{equation}
\begin{equation}\label{eq:z:cl}
  \delta^{\mathrm{C}}_{q_\rho^\tau l_\si^\la}=
    3\brc{\cwk_{q_{\rho}^{\tau}} + \cwk_{l_{\si}^{\lambda}}}\log\brc{\frac{\hat{s}}{M_W^2}}
\end{equation}
\begin{equation}\label{eq:z:pr}
  \begin{aligned}
    &\delta^{\mathrm{PR}}_{q_\rho^\tau l_\si^\la}= 
    \brc{\frac{\sw}{\cw}\bew_{\mathrm{AZ}}\Delta_{q^{\tau}_{\rho}l^{\lambda}_{\si}} - \bew_{\mathrm{AA}}}\log\brc{\frac{\hat{s}}{M_W^2}}  \\
{\rm with} \,    &\Delta_{q^{\tau}_{\rho}l^{\lambda}_{\si}}:=\frac{-\frac{1}{4\cw^2}Y_{q^{\tau}_{\rho}}Y_{l^{\lambda}_{\si}} + \frac{\cw^2}{\sw^4}T^3_{q^{\tau}_{\rho}}T^3_{l^{\lambda}_{\si}}}{R_{q_{\rho}^{\tau}l_{\si}^{\lambda}}} , \; 
  \bew_{AZ} = -\frac{19+22\sw^2}{6\sw\cw}, \;\;\; \bew_{AA}=-\frac{11}{3}.
  \end{aligned}
\end{equation}
$\cwk_{f^\kappa}$ is defined in terms of the electroweak Casimir
operator as $\cwk_{f^\kappa}=C^{ew}_{f^\kappa}-Q_{f^\kappa}^2$ and
\[
I^Z_{f^{\kappa}} = \frac{T^3_{f^{\kappa}}-\sw^2Q_{f^{\kappa}}}{\sw\cw}.
\]
Explicit expressions for $C^{ew}_{f^\kappa}$ can be found
in Appendix~B of Ref.~\cite{Denner:2001gw}.

\subsubsection{Top-quark Pair Production}\label{sec:suda-ttb}

The weak ${\cal O}(\alpha)$ corrections in LA to the LO amplitudes for
the quark-antiquark annihilation and gluon fusion channels again
consist of the different contributions in Eq.~(\ref{eq:LA}) which read
\begin{equation}\label{eq:ttb:dl}
  \begin{aligned}
   & \delta^{\mathrm{LSC}}_{\qi\qf} = - \brc{\cwk_{\qi} + \cwk_{\qf}}\log^2\brc{\frac{\hat{s}}{M_W^2}}
    + 2\Lm\left[\brc{I^Z_{\qf}}^2+\brc{I^Z_{\qf}}^2\right]\log\brc{\frac{\hat{s}}{M_W^2}}, \\
   & \begin{aligned}
       \delta^{\mathrm{SSC}}_{\qi\qf} = 
       &-4\left[\log\brc{\frac{\hat{s}}{M_W^2}}-\Lm\right]
\brc{I^Z_{\qi}I^Z_{\qf}}\log\brc{\frac{\hat{t}}{\hat{u}}} \, \delta_{q_1 q} 
      \end{aligned}
  \end{aligned}
\end{equation}
\begin{equation}\label{eq:ttb:cl}
  \delta^{\mathrm{C}}_{\qi\qf}=
    3\brc{\cwk_{\qi} + \cwk_{\qf}}\log\brc{\frac{\hat{s}}{M_W^2}}
   -\frac{1}{4\sw^2}\bigg[\delta_{q_1 t}\brc{1+\delta_{\tau R}} 
+ \delta_{q_2 t}\brc{1+\delta_{\lambda R}}\bigg]\frac{m_t^2}{M_W^2} \log\brc{\frac{\hat{s}}{m_t^2}}
\end{equation}
The subscripts $\tau, \lambda$ denote the chiralities of the
initial-state light quarks ($q_1=q$) and final-state top quarks
($q_2=t$) in the quark-antiquark channel and of the top and anti-top
quark ($q_{1,2}=t,t$) in the gluon-fusion channel, respectively.  
Note that $\delta^{\rm PR}_{\qi\qf}=0$, since there is no need for
the renormalization of the electric charge, weak mixing angle, Yukawa and scalar-self coupling in strong $\ttb$ production. In
the case of top-pair and di-jet production we implemented in {\tt
  MCFM} the expressions for the amplitude squared, averaged(summed)
over initial(final)-state spin and color degrees of freedom as
follows: \def\hhs{\hat{s}} \def\hht{\hat{t}} \def\hhu{\hat{u}}
\begin{equation}\label{eq:ttb:suda:1}
  \begin{aligned}
\sum_{\tau=\rL,\rR}\sum_{\lambda=\rL,\rR}    &\brc{\delta\mm^{q_1q_2}_{\tau\la}}\cdot\brc{\mm_{0,\tau \la}^{q_1q_2}}^* = \frac{1}{4} \frac{1}{N^2_{q_1 q_2}}\frac{\alpha}{4\pi}\sum_{\tau=\rL,\rR}\sum_{\lambda=\rL,\rR}\brc{\delta^{\mathrm{LSC}}_{\qi\qf}+\delta^{\mathrm{SSC}}_{\qi\qf}+\delta^{\mathrm{C}}_{\qi\qf}}|\mm^{q_1q_2}_{0,\tau\lambda}|^2
  \end{aligned}
\end{equation}
with $N_{qt} = \Nc = 3$ for $q\bar q$ annihilation and $N_{tt} = \Nc^2-1 = 8$ for gluon fusion.  The LO amplitudes squared for
 $q\bar q$  annihilation for each combination of quark chiralities are
\begin{equation}\label{eq:ttb:qqb}
  \begin{aligned}
    &|\mm^{qt}_{0,\rL\rL}|^2 = |\mm^{qt}_{0,\rR\rR}|^2 = \brc{4\pi\alpha_s}^2~2\brc{\Nc^2-1}\frac{\brc{\hat{t}^2+m_t^2\hat{s}}}{\hat{s}^2} \\
    &|\mm^{qt}_{0,\rL\rR}|^2 = |\mm^{qt}_{0,\rR\rL}|^2 = \brc{4\pi\alpha_s}^2~2\brc{\Nc^2-1}\frac{\brc{\hat{u}^2+m_t^2\hat{s}}}{\hat{s}^2},
  \end{aligned}
\end{equation}
and for gluon fusion
\begin{equation}\label{eq:ttb:gg}
  \begin{aligned}
    |\mm^{tt}_{0,\tau\lambda}|^2
    & =
  \left[(T^aT^b)(T^aT^b)^* - (T^aT^b)(T^bT^a)^*\right]\cdot\brc{|\m{A}|^2 + |\m{B}|^2}_{\tau\lambda} + (T^aT^b)(T^bT^a)^*\cdot|\m{C}|^2_{\tau\lambda}, \\
  & = \frac{\Nc(\Nc^2-1)}{4}\cdot\brc{|\m{A}|^2 + |\m{B}|^2}_{\tau\lambda}
  + \frac{-\Nc^2+1}{4\Nc}\cdot|\m{C}|^2_{\tau\lambda}, 
  \end{aligned}
\end{equation}
with
\begin{equation}\label{eq:ttb:gg2}
  \begin{aligned}
   & \m{A}_{\rL\rL} = \m{A}_{\rR\rR} = \brc{4\pi\alpha_s}^2\frac{4\brc{\hhs^2~\hht~\hhu - 2 \hht^2~\hhu^2 + 6 m_t^2~\hhs~\hht~\hhu - m_t^2~\hhs^3 - 2 m_t^4~\hhs^2}}{\hhs^2~\hht^2}, \\
   & \m{A}_{\rL\rR} = \m{A}_{\rR\rL} = \brc{4\pi\alpha_s}^2\frac{4 m_t^2\brc{\hhs^2 - 2 \hht~\hhu - 2 m_t^2~\hhs}}{\hhs~\hht^2}, \\
   & \m{B}_{\rL\rL} = \m{B}_{\rR\rR} = \brc{4\pi\alpha_s}^2\frac{4\brc{\hhs^2~\hht~\hhu - 2 \hht^2~\hhu^2 + 6 m_t^2~\hhs~\hht~\hhu - m_t^2~\hhs^3 - 2 m_t^4~\hhs^2}}{\hhs^2~\hhu^2}, \\
   & \m{B}_{\rL\rR} = \m{B}_{\rR\rL} = \brc{4\pi\alpha_s}^2\frac{4 m_t^2\brc{\hhs^2 - 2\hht~\hhu - 2 m_t^2~\hhs}}{\hhs~\hhu^2}, \\
    & \m{C}_{\rL\rL} = \m{C}_{\rR\rR} = \brc{4\pi\alpha_s}^2\frac{4\left[\hht~\hhu~\brc{\hht^2 + \hhu^2} - m_t^2~s~\brc{\hht^2 - 4\hht~\hhu + \hhu^2} - 2 m_t^4~\hhs^2\right]}{\hht^2~\hhu^2}, \\
   & \m{C}_{\rL\rR} = \m{C}_{\rR\rL} = \brc{4\pi\alpha_s}^2\frac{4 m_t^2~\hhs~\brc{\hht^2 - 2 m_t^2~\hhs + \hhu^2}}{\hht^2 + \hhu^2}.
  \end{aligned}
\end{equation}

\subsubsection{Di-jet Production}\label{sec:suda-dijet}

The weak one-loop Sudakov corrections to the $gg \to q \bar q$ subprocess of
Table~\ref{tab:2g2q} (process A) and the four-quark subprocess of
category 1 of Table~\ref{tab:cat:4q} (shown as the pure weak contribution in
Fig.~\ref{fig:ttb:nlo:qqb}) can be directly obtained from the results for
$\ttb$ production of Section~\ref{sec:suda-ttb} by taking the
limit $m_t \to 0$.  There are, however, additional soft-collinear
contributions of ${\cal O}(\alpha \alpha_s^2)$ to the four-quark
subprocesses of categories 2 and 3 of Table~\ref{tab:cat:4q}, which
originate from the pure weak contribution shown in
Fig.~\ref{fig:2j:4q:v2}. The resulting contribution to the partonic
cross section reads 

\begin{equation}
  \begin{aligned}
  \brc{\delta\mm}_W\cdot\brc{\mm_0}^* =
  &-\frac{\alpha}{2\pi\sw^2}\left[\log\brc{\frac{\hat{s}}{M_W^2}}-\log\frac{M_Z^2}{M_W^2}\right]
  \bigg[
    \log\brc{\frac{-\hat{t}}{\hat{s}}}\delta_{q_iq_j}\brc{|\mm^{\qqb}_{\rL\rL}|^2_{\txt}+|\mm^{\qqb}_{\rL\rL}|^2_{\txs}} \\
    &-\log\brc{\frac{-\hat{u}}{\hat{s}}} |V_{q_i q_j}|^2 |\mm^{\qqb}_{\rL\rL}|^2_{\txs}
    \bigg].
  \end{aligned}
\end{equation}
where we assume a diagonal CKM matrix with $V_{ud}=V_{cs}=1$.  The
Born matrix elements squared read:
\begin{equation}
  \begin{aligned}
    &|\mm^{\qqb}_{\rL\rL}|^2_{\sxt} = |\mm^{\qqb}_{\rR\rR}|^2_{\sxt}=|\mm^{\qqb}_{\rL\rL}|^2_{\txs} = |\mm^{\qqb}_{\rR\rR}|^2_{\txs} = -\brc{4\pi\alpha_s}^2\frac{2(\Nc^2-1)}{\Nc}\frac{\hhu^2}{\hhs~\hht}, \\
    &|\mm^{\qqb}_{\rL\rL}|^2_{\txt} = |\mm^{\qqb}_{\rR\rR}|^2_{\txt} = \brc{4\pi\alpha_s}^2~2\brc{\Nc^2-1}\frac{\hhu^2}{\hht^2} .
  \end{aligned}
\end{equation}

\section{Impact of weak one-loop corrections and comparison with existing results}\label{sec:comparison}

In this section we will validate our calculation of the full weak
one-loop corrections described in Section~\ref{sec:exact} and their
implementation in {\tt MCFM} by cross-checking with existing results in the
literature.  In order to do so we will compare with the results
provided in Ref.~\cite{Dittmaier:2012kx} (di-jet production),
Ref.~\cite{Kuhn:2013zoa} ($t\bar t$ production), and by using the
publicly available MC program {\tt ZGRAD2}~\cite{Baur:2001ze}
(Neutral-Current Drell-Yan production). In the cases of $\ttb$ and di-jet
production we adopt the particular setup used in the publications.
Predictions with more up-to-date theoretical inputs can of course be
computed with {\tt MCFM}, and, in general, are not expected to differ
much from the ones presented here. This validation also provides an
opportunity to discuss the impact of the full weak one-loop
corrections on a variety of LHC observables, especially in the
high-energy regime.

\subsection{Neutral-current Drell-Yan production}\label{sec:ncdycomp}

We perform a tuned comparison of our {\tt MCFM} implementation of the
full weak one-loop corrections to the Neutral-Current Drell-Yan (NC
DY) process as described in Section~\ref{sec:ncdy} with the
calculation implemented in {\tt ZGRAD2}~\cite{Baur:2001ze}.  We
present results for the relative weak one-loop correction defined as
\begin{equation}\label{eq:relcorr}
\delta_{\wk} = \frac{d\sigma_{NLO}^{\wk} - d\sigma_{LO}}{d\sigma_{LO}},
\end{equation}
where $d\sigma_{LO}$ denotes the LO cross section and $d\sigma_{NLO}^{\wk}$
the NLO cross section including weak one-loop corrections.  The relative
correction may be defined after integration over the entire phase space, or
bin-by-bin in a differential distribution.

Our choices for the particle masses and widths, together with the
relevant electroweak couplings, are shown in
Table~\ref{table:parameters_DY}.  Results are obtained in the on-shell
renormalization scheme and by using a fixed-width scheme.  When using
the fixed-width scheme the values for the weak gauge boson masses,
$M_W$ and $M_Z$, and their total widths, $\Gamma_W$ and $\Gamma_Z$,
differ from those recommended by the PDG~\cite{Agashe:2014kda}, since
the PDG values have been extracted assuming a running gauge boson
width (see, for example, Refs.~\cite{Dittmaier:2009cr,Alioli:2016fum} for
details). As EW input scheme we use the $G_\mu$ scheme as described in
Section~\ref{sec:ncdy}.  Note that we only retain lepton and quark
masses in closed fermion loops and treat external fermions as massless
particles.  As a result of using the $G_\mu$ scheme the dependence on
the light quark masses cancels in the weak one-loop corrections. We
use the MSTW2008NLO~\cite{Martin:2009iq} set of Parton Distribution
Functions (PDF), which corresponds to a strong coupling of
$\alpha_s(M_Z)=0.12018$, and choose $\mu_F = \mu_R = M_Z$.
\begin{table}[htpb]
  \begin{tabular}{|l|l|}
  \hline
  $M_W$ = 80.3695\, GeV &
  $\Gamma_W$ = 2.1402\, GeV \\

  $M_Z$ = 91.1535\, GeV &
  $\Gamma_Z$ = 2.4943\, GeV \\

  $M_H$ = 126\, GeV & 
  $m_t$ = 172.5\, GeV \\
  $m_b$ = 4.82\, GeV & 
  $m_c$ = 1.2\, GeV \\

  $m_s$ = 150\, MeV &
  $m_u$ = 66\, MeV \\

  $m_d$ = 66\, MeV &
  $m_e$ = 0.51099892\, MeV \\

  $m_{\mu}$ = 105.658369\, MeV &
  $m_{\tau}$ = 1.777\, GeV \\

  $G_{\mu}$ = 1.16637~$\times~10^{-5}\;\GeV^{-2}$ &
  $\alpha_{G_\mu}$ = 1/132.4525902 \\

  $\sin^2\theta_W$ = $1 - M_W^2/M_Z^2$ & \\
  \hline
\end{tabular}
\caption{Input parameters used in the calculation of the Neutral-Current
  Drell-Yan process and of di-jet production. \label{table:parameters_DY}}
\end{table}

For the results presented here we concentrate on the LHC operating at
$\sqrt{S} =13$~TeV and apply a simple set of acceptance cuts for the
charged leptons.  These constrain the transverse momenta of the
leptons ($p_{T}(l^\pm)$), their pseudorapidities ($\eta(l^\pm)$) and
the invariant mass of the lepton-pair ($M(l^+l^-)$, $l=e, \mu$),
\begin{equation}
\label{eq:cutsDY}
 p_T(l^\pm) >  25~\mbox{GeV}, \quad |\eta(l^\pm)| <  2.5 \;, \quad M(l^+ l^-) > 60~\mbox{GeV} \,.
\end{equation}
With this setup and cuts, {\tt MCFM} yields a total cross section for 
$pp \to \gamma, Z \to l^+ l^-$ ($l=e$ or $\mu$) at LO of
\begin{equation}
\si_{\mathrm{LO}} = 712.44(2)~\pb,
\end{equation}
and a relative one-loop weak correction of
\begin{equation}
\delta_{\wk} = \frac{-4.474(3)~\pb}{712.44(2)~\pb} = -0.628 \%. 
\end{equation}
This is in excellent agreement with the {\tt ZGRAD2} results, which give
$\si_{\mathrm{LO}} = 712.41(2)~\pb$ and $\delta_{\wk} =
\frac{-4.483(3)~\pb}{712.41(2)~\pb} = -0.629 \%$.  

\begin{figure}[htpb]
    \includegraphics[scale=0.29]{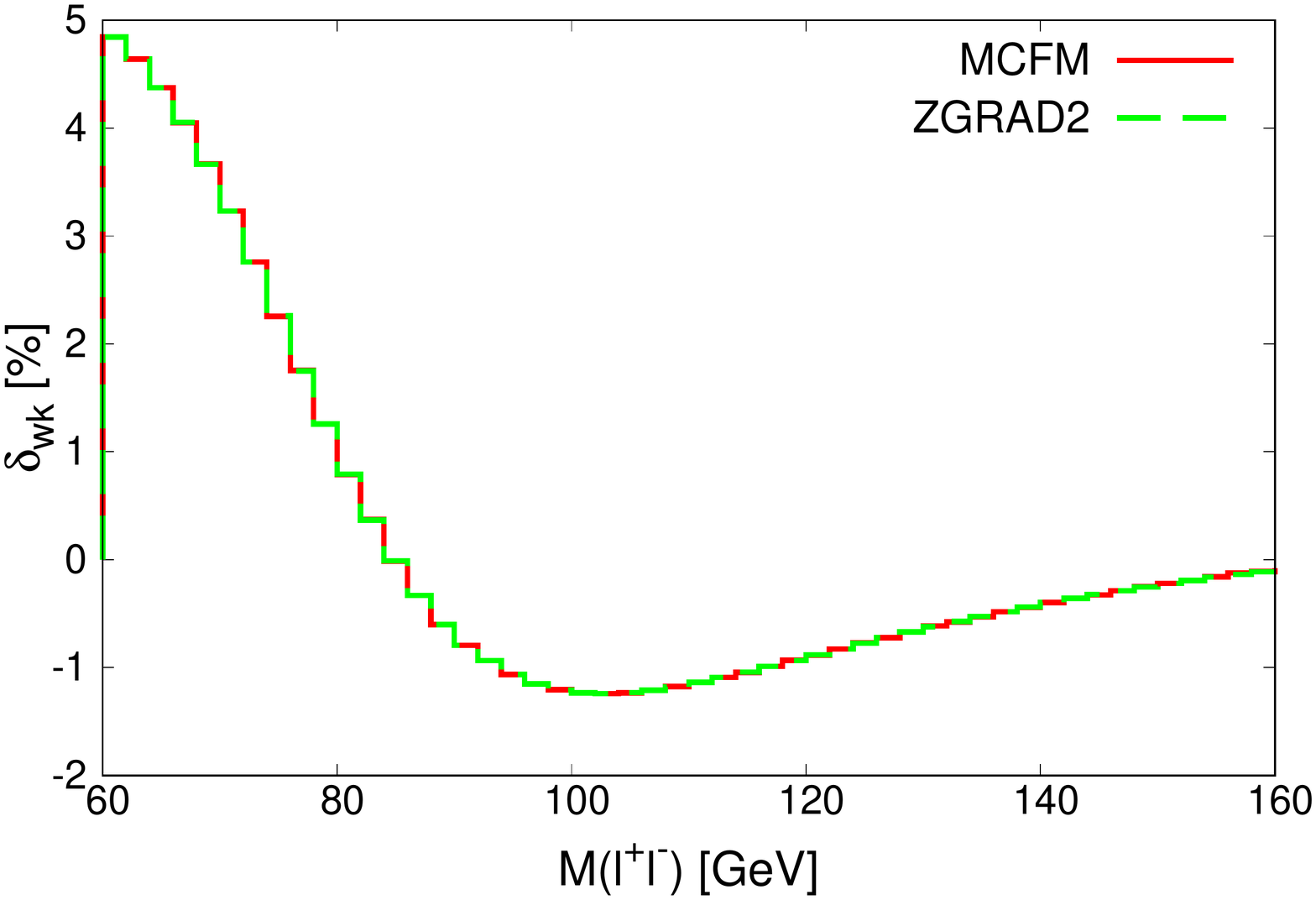}
    \includegraphics[scale=0.29]{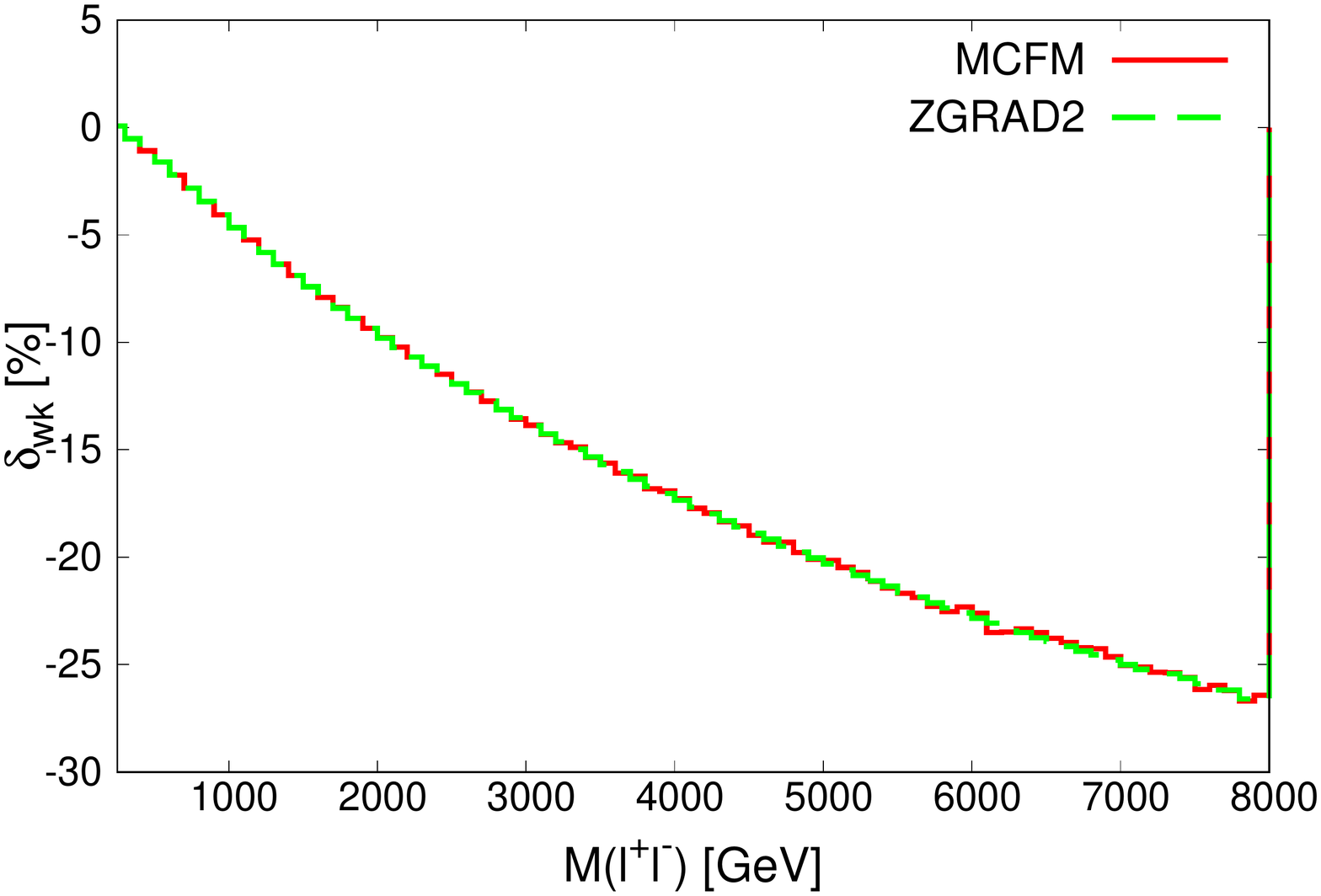} \\
    \includegraphics[scale=0.29]{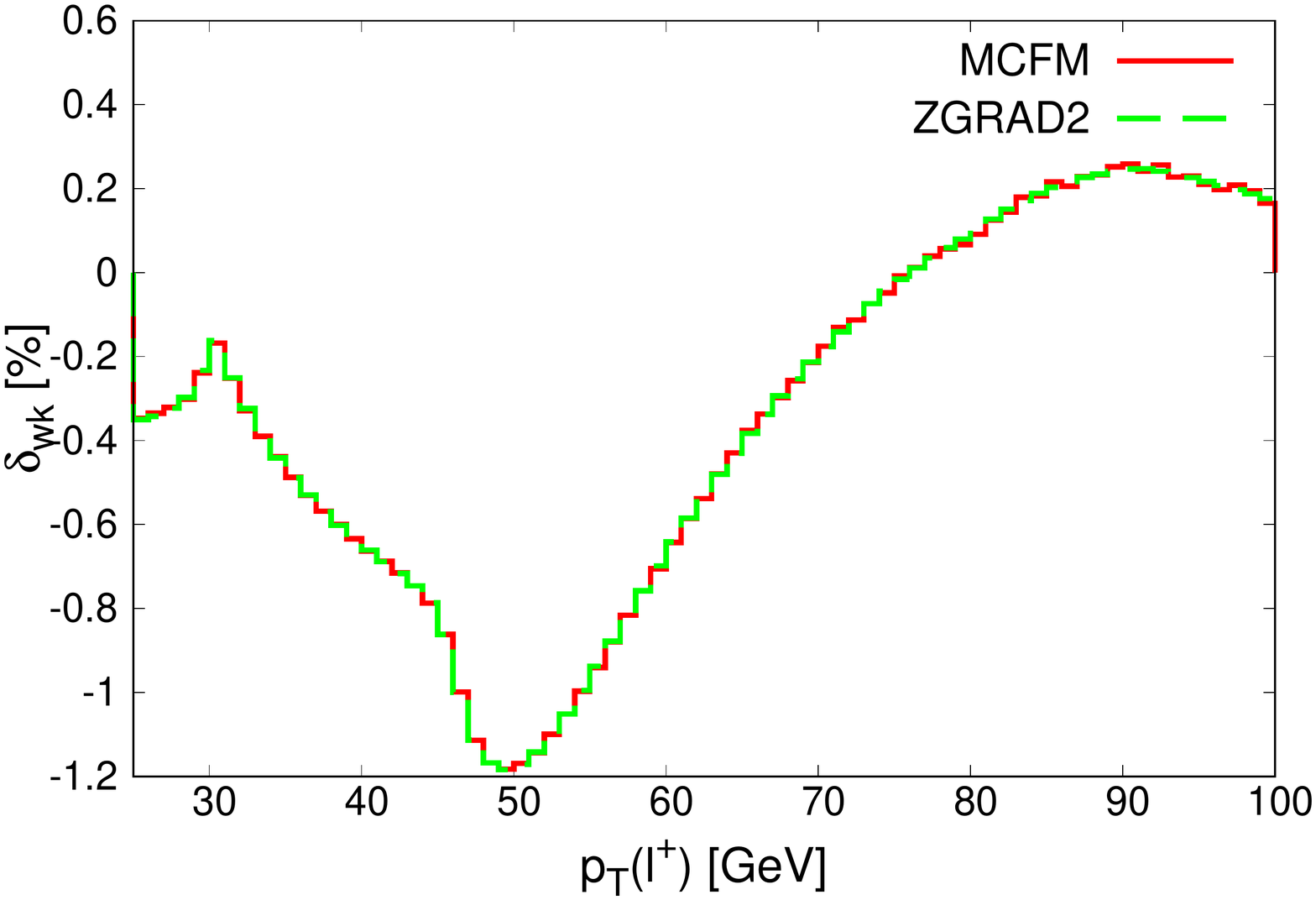} 
    \includegraphics[scale=0.29]{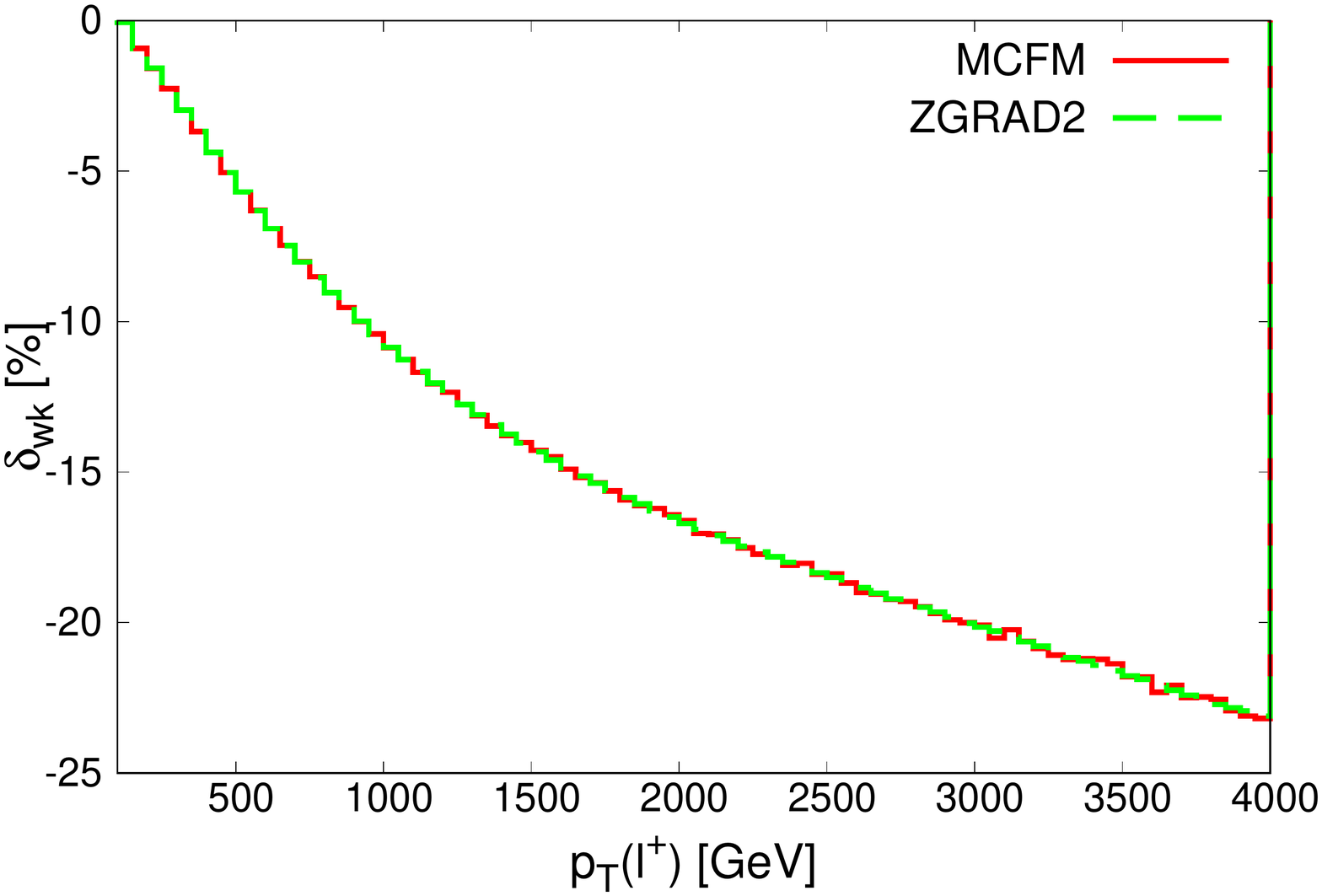} \\
    \includegraphics[scale=0.29]{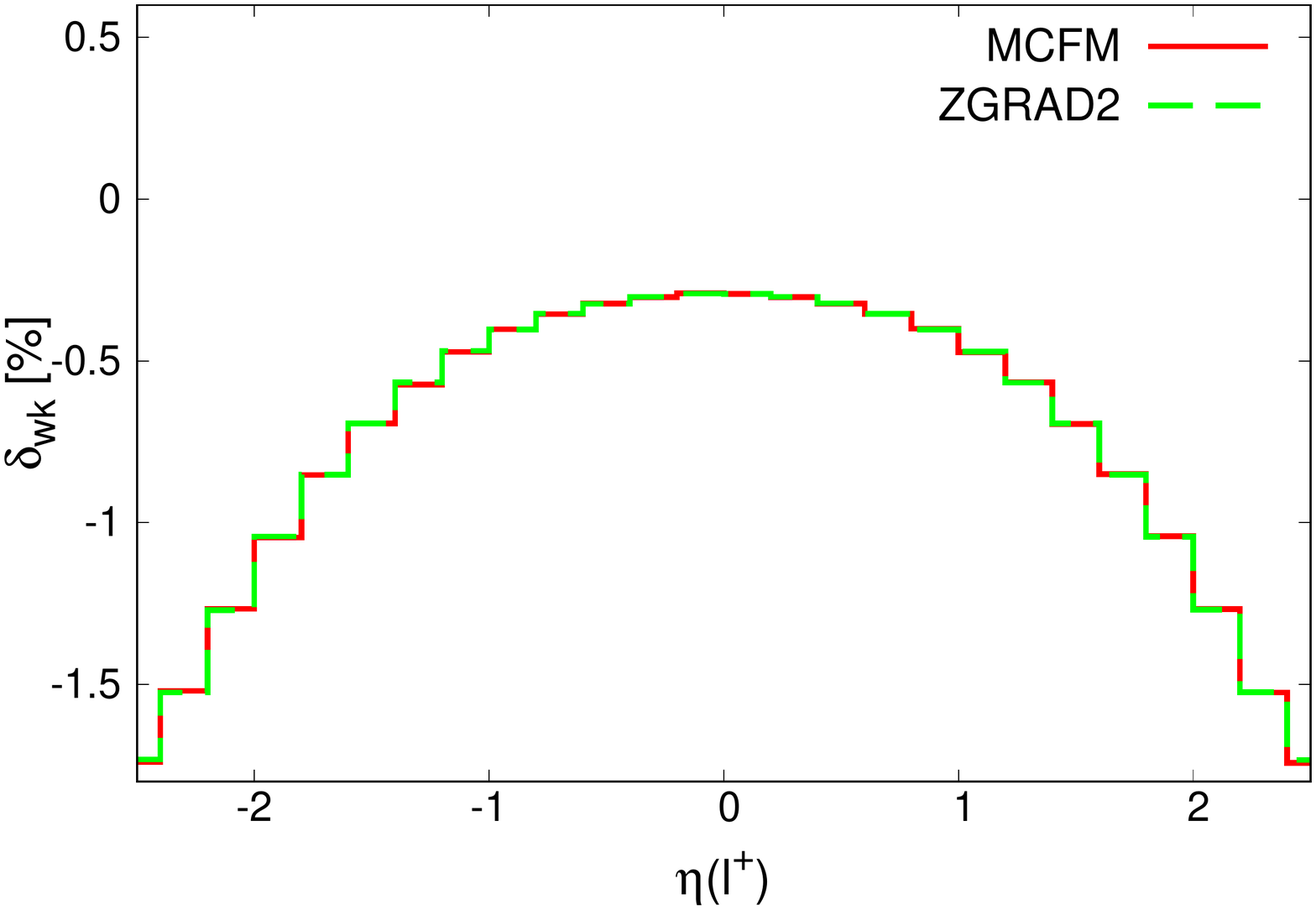} 
    \includegraphics[scale=0.29]{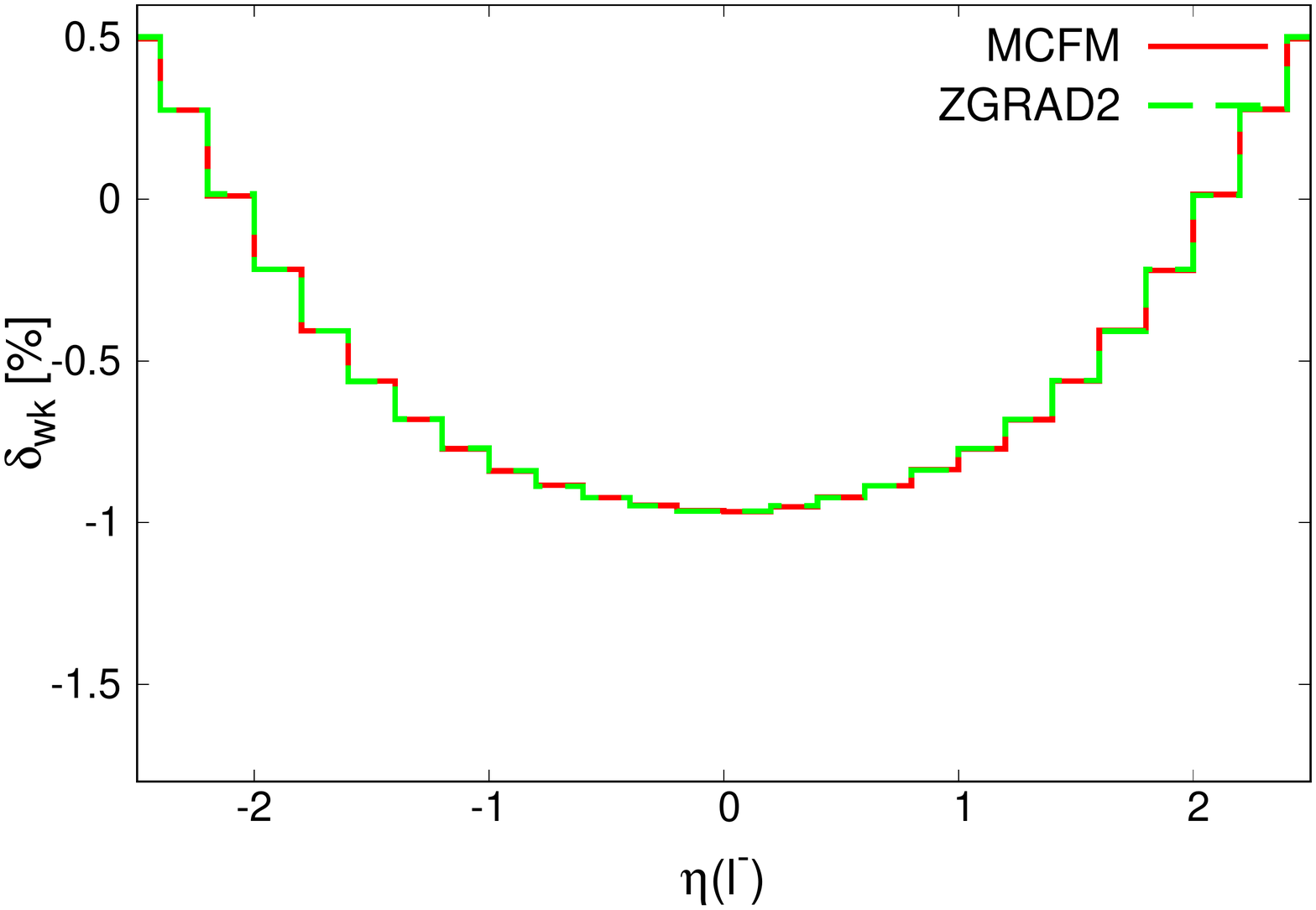} 
    \caption{Comparison of {\tt MCFM} (red, solid) and {\tt
        ZGRAD2} (green, dashed) predictions for the weak one-loop
      relative correction $\delta_{\wk}$ to the invariant mass of the
      lepton-pair ($M(l^+l^-)$, top), lepton transverse momentum
      ($p_{T}(l^+)$, middle), and lepton rapidities ($\eta(l^\pm)$,
      bottom) distributions in NC DY production at the 13 TeV LHC. The
      correction is expressed as a percentage of the LO result in each
      bin, according to Eq.~(\ref{eq:relcorr}).
    \label{fig:z:comp}}
\end{figure}

A comparison of {\tt MCFM} and {\tt ZGRAD2} results for the relative
one-loop weak corrections to the distributions of $M(l^+l^-)$,
$p_{T}(l^+)$ and $\eta(l^\pm)$ (for $l=e$ or $\mu$) is shown in
Fig.~\ref{fig:z:comp}. As can been seen, all {\tt MCFM} results for NC
DY production at the LHC are in excellent agreement with the {\tt
  ZGRAD2} predictions.

At ${\cal O}(\alpha^2)$ the NC DY process also receives a
contribution from the tree-level photon-induced process, $\gamma
\gamma \to l^+ l^-$. In Table~\ref{tab:phind} we compare the {\tt
  MCFM} result and the results presented by Dittmaier and Huber in
Ref.~\cite{Dittmaier:2009cr} (denoted as DH in the following) for the
total tree-level cross sections for the $q\bar q$- ($\sigma_0^{MCFM},
\sigma_0^{DH}\left|\right._{FS/PS}$) and $\gamma \gamma$-initiated
($\sigma_{\gamma\gamma,0}^{MCFM}$) processes for various $M(l^+l^-)$
regions at the 14 TeV LHC. The DH LO cross section 
$\sigma_0^{DH}\left|\right._{FS/PS}$ is obtained in the so-called
Factorized Scheme (FS) or Pole Scheme (PS), which differ from the
Complex-Mass Scheme (CMS) in the treatment of the $Z$ resonance (see
Ref.~\cite{Dittmaier:2009cr} for details).  For the comparison we
adopt the setup of Ref.~\cite{Dittmaier:2009cr} and use the
MRST2004QED~\cite{Martin:2004dh} PDF set. Note that the MRST2004QED
PDF set is by now outdated and up-to-date PDF sets, such as
NNPDF3.0QED~\cite{Ball:2013hta,Ball:2014uwa}, and
CT14QED~\cite{Schmidt:2015zda}, should be used. In
Table~\ref{tab:phind} we also compare the contribution of the LO
photon-induced process relative to the $\qqb$-initiated process,
$\delta_{\gamma \gamma,0}=\sigma_{\gamma
  \gamma,0}/\sigma_0$.  We find that the LO cross
sections $\sigma_0^{DH}$ and $\sigma_0^{MCFM}$ agree at the 0.15\%
level (and better for small $M(l^+l^-)$), and that there is excellent
agreement in $\delta_{\gamma\gamma,0}$. Note that in Ref.~\cite{Dittmaier:2009cr} 
also the EW ${\cal O}(\alpha)$ 
corrections to $\gamma \gamma \to l^+l^-$ have been calculated and found to be negligible. 

\begin{figure}[htpb]
\includegraphics[scale=0.31]{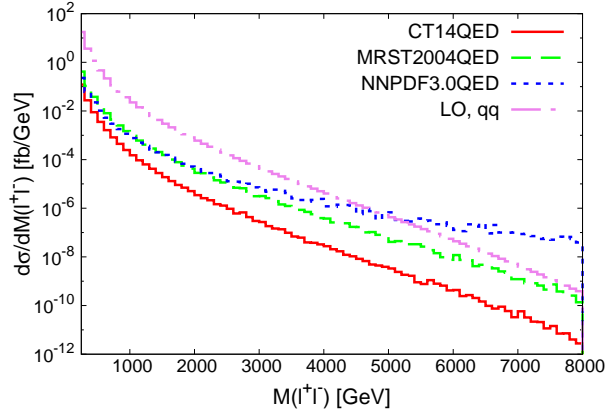}
\caption{LO predictions for the invariant lepton-pair mass
  distribution for the photon-induced process and the $q\bar
  q$-initiated (pink, long-dashed-dotted) NC DY process at the 13 TeV
  LHC. The photon-induced LO prediction is obtained with different
  photon PDFs as provided by CT14QED (red, solid), MRST2004QED (green,
  dashed), and NNPDF3.0QED (blue, dotted).\label{fig:phind}}
\end{figure}

In Fig~\ref{fig:phind} we show {\tt MCFM} predictions for the
$M(l^+l^-)$ distribution ($l=e$ or $\mu$) for the photon-induced tree-level
production process at the 13 TeV LHC when using a variety of photon PDFs,
compared to the $q\bar q$-induced NC DY distribution at LO (the setup
of Table~\ref{table:parameters_DY} is used with
$\mu_F=\mu_R=M_Z$). The spread of predictions, especially at high
invariant masses, indicates the large uncertainties associated with the
photon PDF of current global PDF sets (see
Refs.~\cite{Ball:2013hta,Ball:2014uwa,Schmidt:2015zda} for a detailed
discussion).  Dedicated efforts to improve the knowledge of the
photon PDF are under
way~\cite{Martin:2014nqa,Harland-Lang:2016kog,Manohar:2016nzj}.  In view of
the situation presented in Fig.~\ref{fig:phind}, we cannot conclusively assess by 
how much the positive photon-induced contribution
affects the impact of the negative weak one-loop corrections on NC DY observables until these more
precise determinations of photon PDFs become readily available.

\begin{table}
  \centering
  \begin{tabular}{c|*{6}{c}}
    \hline\hline
    $M(l^+l^-)$ [GeV] &50-$\infty$ &100-$\infty$ &200-$\infty$ &500-$\infty$ &1000
-$\infty$ &2000-$\infty$ \\
    \hline
    $\sigma_{\gamma\gamma,0}^{MCFM}$ [fb] &1287.98(7) &377.77(5) &63.88(1) &3.9809(7) &0.35407(7) &0.018759(4) \\
    \hline
    $\sigma_0^{DH}\left|\right._{FS/PS}$ [fb] &738773(6) &32726.8(3) &1484.92(1) &80.9489(6) &6.80008(3) &0.303767(1) \\
    \hline
    $\sigma_0^{MCFM}$ [fb] &739272(13) &32881.5(6) &1484.37(30) &81.0745(16) &6.8103(1) &0.304209(5) \\
    \hline
    $\delta_{\gamma\gamma,0}^{DH}$[\%] &0.17 &1.15 &4.30 &4.92 &5.21 &6.17 \\
    \hline
    $\delta_{\gamma\gamma,0}^{MCFM}$[\%] &0.17 &1.15 &4.30 &4.91 &5.20 &6.17 \\
    \hline\hline
  \end{tabular}
  \caption{{\tt MCFM} cross sections for the tree-level
    photon-induced process, $\sigma_{\gamma \gamma,0}^{MCFM}$, for various ranges of 
    the invariant di-lepton mass ($M(l^+l^-)$)
    obtained with MRST2004QED at the 14 TeV LHC. We also show a
    comparison of the LO $q\bar q$-initiated NC DY cross section,
    $\sigma_0$, and of the ratio
    $\delta_{\gamma\gamma,0}=\sigma_{\gamma \gamma,0}/\sigma_0$ from
    {\tt MCFM} and Table~1 of Ref.~\cite{Dittmaier:2009cr} (labeled as
    DH).  \label{tab:phind}}
\end{table}

\subsection{Top-quark pair production}\label{sec:ttbcomp}

We perform a tuned comparison of our {\tt MCFM} implementation of weak
one-loop corrections to $t \bar t$ production at the LHC as presented
in Section~\ref{sec:ttb} with the one presented by K{\"u}hn, Scharf
and Uwer in Ref.~\cite{Kuhn:2013zoa} (denoted as KSU in the
following).  We adopt the setup used therein, which corresponds to the
masses and couplings shown in Table~\ref{table:parameters_top}.
Furthermore, we set the renormalization and factorization scales equal
to the mass of the top quark, $\mu_F = \mu_R = m_t$ and we employ the
MSTW2008NNLO PDF set~\cite{Martin:2009iq} that specifies
$\alpha_s(m_Z)=0.11707$.  With this setup the value of the strong
coupling used in the calculation is $\alpha_s(m_t)$, as given in the
table.
\begin{table}[htpb]
  \begin{tabular}{|l|l|}
  \hline
  $M_W = 80.385\,\GeV$ &
  $M_Z = 91.1876\,\GeV$ \\  
  $M_H = 126\,\GeV$ &
  $m_t = 173.2\,\GeV$ \\ 
  $m_b = 4.82\,\GeV$ &
  $\alpha(m_t) = 1/127$ \\
  $\sin^2\theta_W = 1 - M_W^2/M_Z^2$ &  
  $\alpha_s(m_t) = 0.106823$ \\ 
  \hline
\end{tabular}
\caption{Input parameters used in the validation of the weak one-loop
  corrections to the $t\bar t$ production process.  These parameters
  are chosen in order to facilitate a comparison with the results of
  Ref.~\cite{Kuhn:2013zoa}.
\label{table:parameters_top}}
\end{table}

\begin{figure}[htpb]
\centering
  \includegraphics[scale=0.29]{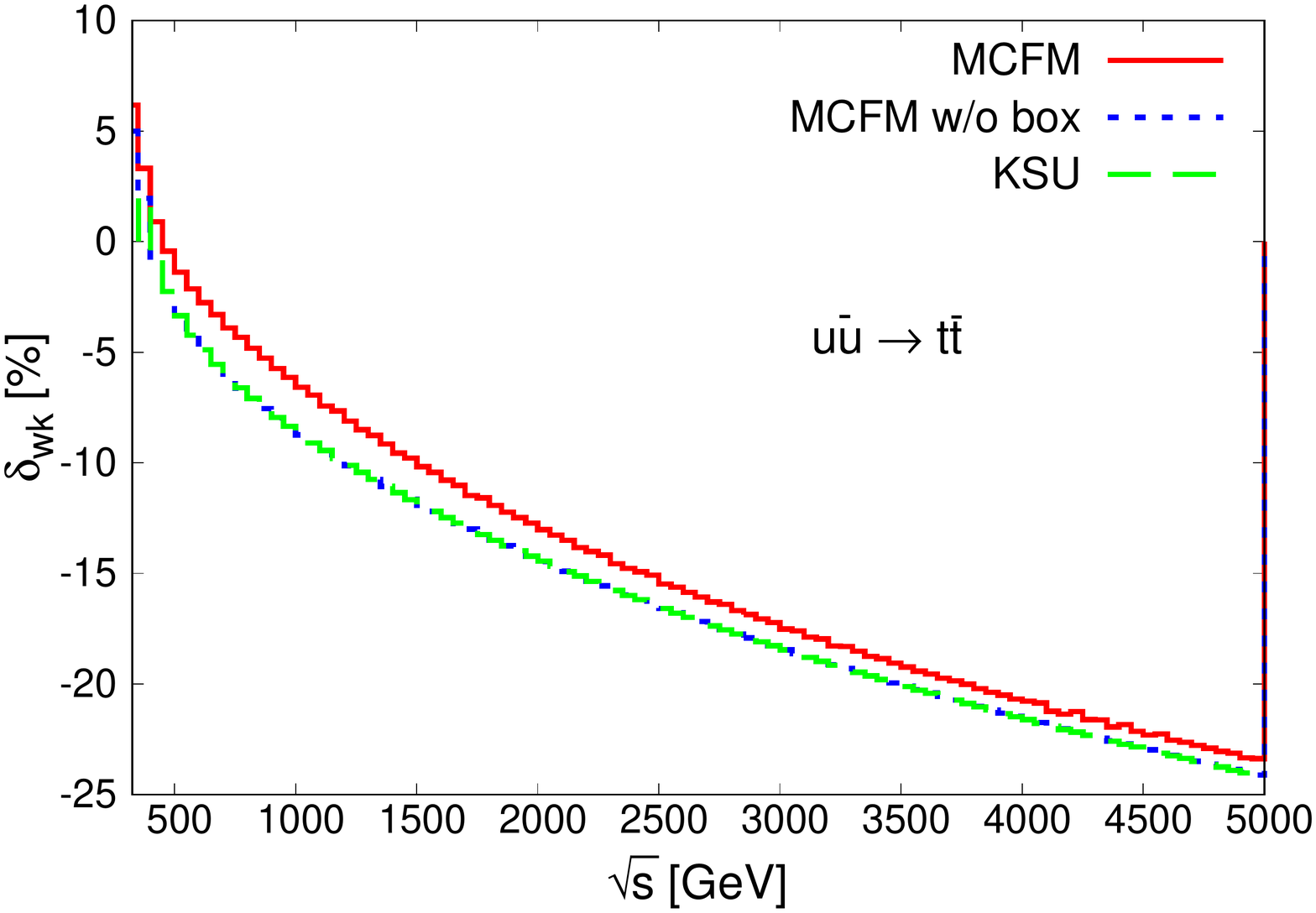}
  \includegraphics[scale=0.29]{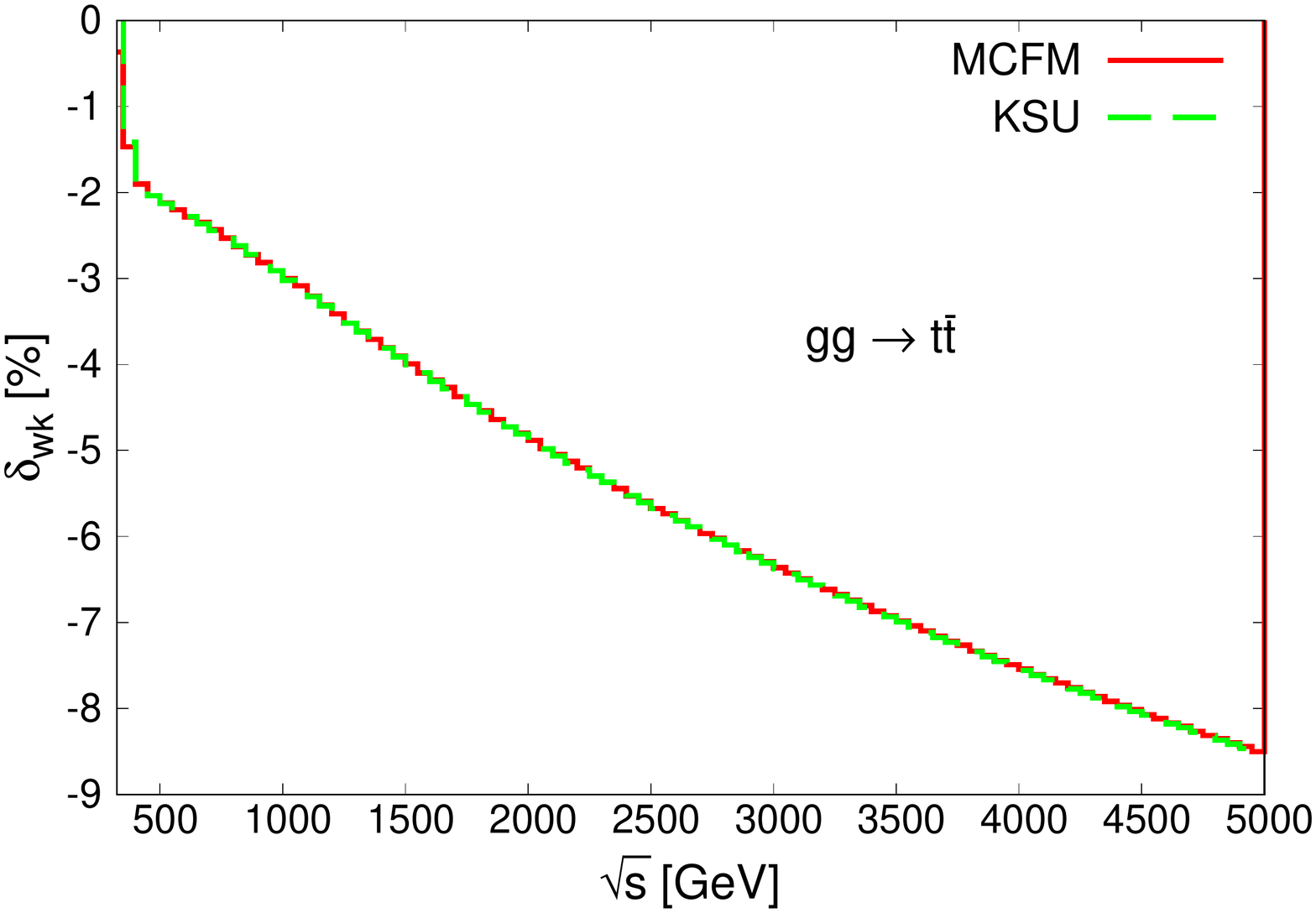}
    \caption{Comparison of relative corrections $\delta_{\wk}$ to the
      parton-level cross sections for $u \bar u \to t \bar t$ (left)
      and $gg \to t \bar t$ (right) from {\tt MCFM} and the results
      from Fig.~8 of Ref.~\cite{Kuhn:2013zoa} (KSU) (green, dashed).
      The correction is expressed as a percentage of the LO ${\cal
        O}(\alpha_s^2)$ cross section according to
      Eq.~(\ref{eq:relcorr}).  In case of the $u \bar u \to t \bar t$
      process the KSU results shown here only include weak one-loop
      vertex corrections, while {\tt MCFM} results are provided for
      both the weak one-loop vertex corrections (blue, dotted) and the
      full ${\cal O}(\alpha_s^2 \alpha)$ contribution (red, solid).
    \label{fig:partonic}}
\end{figure}

In Fig.~\ref{fig:partonic} we present a comparison of the relative
corrections to the parton-level processes, $u \bar u \to t \bar t$ and
$gg \to t \bar t$~\footnote{Throughout this paper we have extracted
  the numerical values from distributions in published figures by
  using the tool {\tt EasyNdata}~\cite{Uwer:2007rs}.}.  In the case of
the $\qqb$-initiated process we show results for $\delta_{\wk}$ of
Eq.~(\ref{eq:relcorr}) separately for the weak one-loop vertex
corrections (diagrams shown in the upper part of
Fig.~\ref{fig:ttb:nlo:qqb}) and for the full ${\cal O}(\alpha_s^2
\alpha)$ contribution (which now also includes the box diagrams in
Fig.~\ref{fig:ttb:nlo:qqb} and the virtual and real contributions of
Fig.~\ref{fig:ttb:qcdbox} and Fig.~\ref{fig:ttb:real}, respectively).
The parton-level results presented in Fig.~8 of
Ref.~\cite{Kuhn:2013zoa} only include weak one-loop vertex
corrections, since the remaining ${\cal O}(\alpha_s^2 \alpha)$
contributions were studied in detail and found to be very small.  This
is also supported by the results of our calculation shown in
Fig.~\ref{fig:partonic}.  Moreover, we observe that the agreement
between the results of {\tt MCFM} and those of KSU is excellent.

At the hadron level we perform the comparison for the LHC operating at
13~TeV, with no cuts applied to the top quarks except where noted
specifically below. Note that the hadron-level results of
Ref.~\cite{Kuhn:2013zoa} now also include the full ${\cal
  O}(\alpha_s^2 \alpha)$ contributions, i.e. also including box diagrams and real corrections.  For this set-up we find a LO
cross section for $\ttb$ production at ${\cal O}(\alpha_s^2)$ of,
\begin{equation}
\si_{\mathrm{LO}} = 474.60(4)~\pb.
\end{equation}
The overall effect of the full ${\cal
  O}(\alpha_s^2 \alpha)$  contribution on the total cross
section is rather small and results in a relative correction,
\begin{equation}
\delta_{\wk} = \frac{-9.509(1)~\pb}{474.60(4)~\pb} = -2.00 \%. 
\end{equation}
This is in perfect agreement with the results from
Ref.~\cite{Kuhn:2013zoa}, which gives $\delta_{\wk} = -2.00$\%.

\begin{figure}[htpb]
\centering
  \includegraphics[scale=0.29]{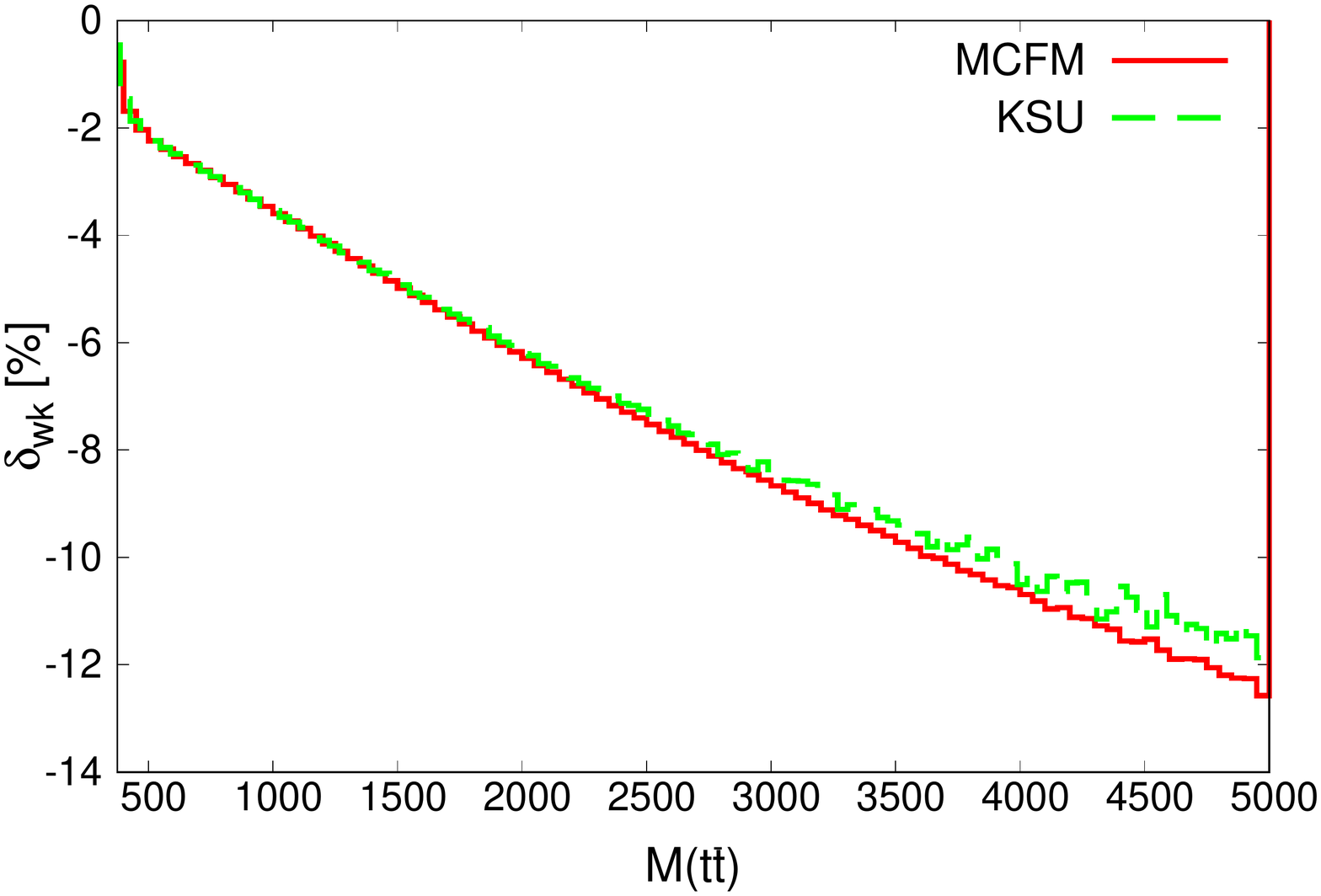}\\
  \includegraphics[scale=0.29]{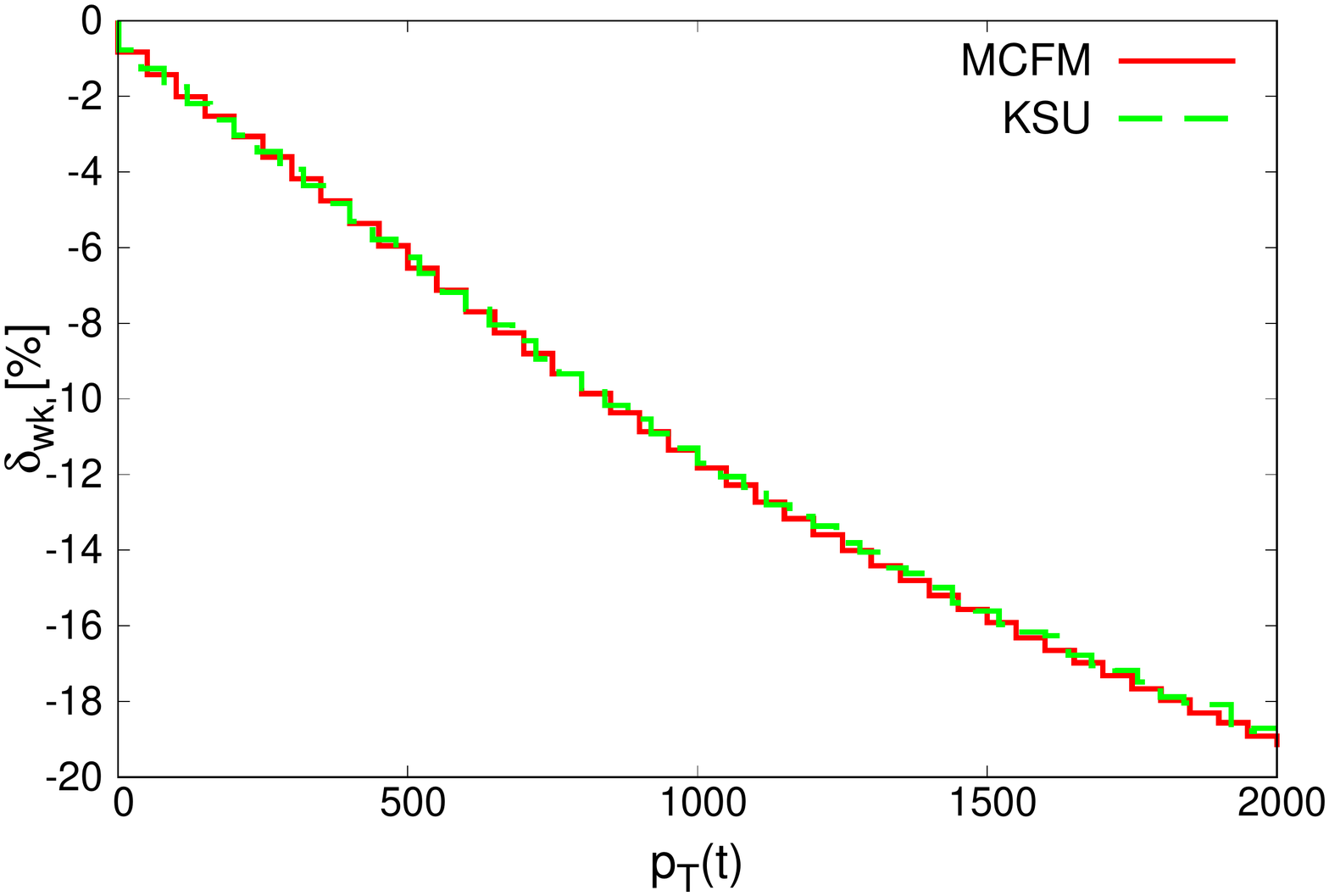}
  \includegraphics[scale=0.29]{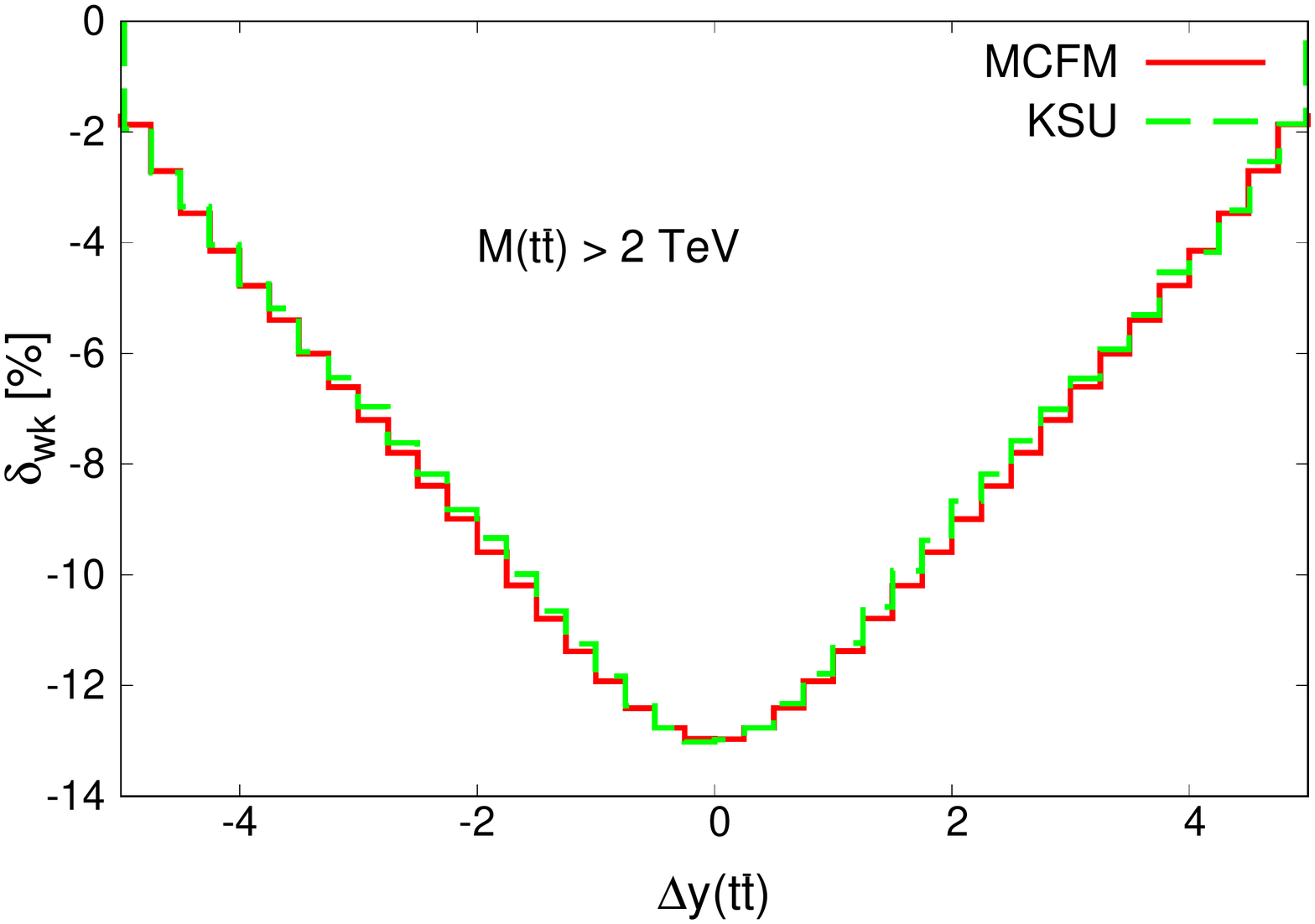}
    \caption{Comparison of relative corrections $\delta_{\wk}$ from
      {\tt MCFM} (red, solid) and the results from Fig.~20 and Fig.~22 of
      Ref.~\cite{Kuhn:2013zoa} (KSU) (green, dashed) to the invariant mass
      distribution of the top-antitop pair ($M(t{\bar t})$) (top center), the
      transverse momentum of the top quark ($p_T(t)$) (left), and the
      rapidity difference between the top and anti-top quark ($\Delta
      y(t\bar t)$) (right), in $\ttb$ production at the 13~TeV LHC.  In the calculation of the
      $\Delta y(t\bar t)$ distribution a cut of $M(t{\bar t})>2$~TeV
      is applied.  The correction is expressed as a percentage of the
      LO ${\cal O}(\alpha_s^2)$ cross section according to Eq.~(\ref{eq:relcorr}).
    \label{fig:ttbdistri}}
\end{figure}

We now turn to the comparison of results for differential
distributions, in particular for the top-pair invariant mass
distribution ($M(t\bar t)$), the transverse momentum of the top quark
($p_T(t)$) and the rapidity difference between the top and anti-top
quarks, $\Delta y(t\bar t) = y(t) - y(\bar t)$, where $y(t)$ and
$y(\bar t)$ are the rapidities in the lab frame.  This comparison is
presented in Fig.~\ref{fig:ttbdistri} where, for the $\Delta y(t\bar
t)$ distribution, a cut $M(t \bar t) > 2$~TeV has been applied.  The
KSU results are taken from Figs.~20 and~22~(right) of
Ref.~\cite{Kuhn:2013zoa}.  While the parton-level results are in
excellent agreement, we observe a small difference in the $M(t\bar t)$
distribution at the 0.5\% level at $M(t\bar t) \approx 5$~TeV.  We
have not been able to trace the origin of this discrepancy, which may
simply be due to the fact that different approaches for the treatment
of IR singularities have been used, namely dipole subtraction and
phase-space slicing. Each of these methods has its own challenges in
obtaining precise numerical results at such large invariant masses.

\subsection{Di-jet production}\label{sec:dijetcomp}

We perform a tuned comparison of our {\tt MCFM} implementation of di-jet production at ${\cal O}(\alpha_s^2 \alpha)$ 
as described in Section~\ref{sec:dijet} with the results of Dittmaier, Huss, and Speckner in 
Ref.~\cite{Dittmaier:2012kx} (denoted as DHS in the following), and
thus adopt the setup used therein, which corresponds to the masses and
couplings shown in Table~\ref{table:parameters_DY}.  The factorization
scale ($\mu_F$) and renormalization scale ($\mu_R$) are set equal,
$\mu_F= \mu_R = k_{T}(j_1)$, where $k_{T}(j_1)$ is the transverse
momentum of the leading jet. The CTEQ6L1 set of
PDFs~\cite{Pumplin:2002vw} is used, which corresponds to a strong
coupling of $\alpha_s(M_Z)=0.129783$.  To identify the jets the
anti-$k_T$ jet clustering algorithm is used with a pseudo-cone size of
$R = 0.6$, and the following jet cuts are applied:
\begin{equation}\label{jetcut}
  k_T(j) > 25\; \GeV, \;\;\; |y(j)| < 2.5.
\end{equation}
A comparison of {\tt MCFM} and DHS results for relative corrections
in various ranges of the invariant mass of the two leading jets ($M(j_1j_2)$)
at the 14 TeV LHC is shown in Table~\ref{tbl:m12:14tev}.  A similar
comparison, for various ranges of the transverse momentum of the
leading jet ($k_{T}(j_1)$), is given in Table~\ref{tbl:kt1:14tev}.  In
both cases we compare the relative one-loop weak corrections
($\delta_{\wk}$ of Eq.~(\ref{eq:relcorr})) and the effect of additional
tree-level contributions mediated by EW interactions
($\delta_{\EW}^{\tree}$).  The latter correction is defined by,
\begin{equation}\label{eq:relcorr_ew}
\delta_{\EW}^{\tree} = \frac{d\sigma_{LO+ew} - d\sigma_{LO}}{d\sigma_{LO}}
\end{equation}
where $\sigma_{LO}$ represents the QCD-mediated LO cross section of ${\cal O}(\alpha_s^2)$, while
$\sigma_{LO+ew}$ also contains the additional ${\cal O}(\alpha_s
\alpha)$ and ${\cal O}(\alpha^2)$ contributions due to the $Z,\gamma$ and $W$-exchange diagrams shown in Fig.~\ref{fig:dijet:lo}. As can be seen from
the tables, the inclusion of these terms partially cancels the effect
of the weak one-loop corrections.

\begin{table}[htpb]
  \small
  \centering
  \begin{tabular}{c|*{8}{p{1.6cm}}}
$M(j_1j_2)$ [GeV] & & $50-\infty$ & $100-\infty$ & $200-\infty$ & $500-\infty$ & $1000-\infty$ & $2000-\infty$ & $5000-\infty$ \\
\hline
\multirow{2}{*}{$\delta_\wk [\%]$}
  &DHS &-0.02 &-0.03 &-0.07 &-0.31 &-0.88 &-2.20 &-5.53 \\
  &MCFM &-0.02 &-0.03 &-0.07 &-0.31 &-0.88 &-2.23 &-5.57 \\
  \hline
  \multirow{2}{*}{$\delta^{\tree}_\EW [\%]$}
  &DHS &\ 0.03 &\ 0.01 &\ 0.02 &\ 0.10 &\ 0.34 &\ 1.00 &\ 2.56 \\
  &MCFM &\ 0.03 &\ 0.01 &\ 0.02 &\ 0.08 &\ 0.30 &\ 0.96 &\ 2.61 \\
  \hline
\end{tabular}
\caption{Comparison of relative corrections $\delta_\wk$ of
  Eq.~(\ref{eq:relcorr}) and $\delta_{\EW}^{\tree}$ of
  Eq.~(\ref{eq:relcorr_ew}) from {\tt MCFM} and Table~1 of
  Ref.~\cite{Dittmaier:2012kx} (DHS) for various ranges of the
  invariant di-jet mass ($M(j_1j_2)$) in di-jet production at the 14
  TeV LHC.
  \label{tbl:m12:14tev}}
\end{table}

\begin{table}
  \small
  \centering
  \begin{tabular}{c|*{8}{p{1.6cm}}}
$k_{T}(j_1)$ [GeV]  && $25-\infty$ & $50-\infty$ & $100-\infty$ & $200-\infty$ & $500-\infty$ & $1000-\infty$ & $2500-\infty$ \\
\hline
  \multirow{2}{*}{$\delta_\wk [\%]$}
  &DHS &-0.02 &-0.08 &-0.28 &-0.84 &-2.72 &-5.48 &-10.49 \\
  &MCFM &-0.02 &-0.08 &-0.28 &-0.83 &-2.75 &-5.64 &-10.41 \\
  \hline
  \multirow{2}{*}{$\delta^{\tree}_\EW [\%]$}
  &DHS &\ 0.03 &\ 0.03 &\ 0.12 &\ 0.36 &\ 1.44 &\ 4.62 &\ 18.28 \\
  &MCFM &\ 0.03 &\ 0.03 &\ 0.11 &\ 0.33 &\ 1.42 &\ 4.72 &\ 18.88 \\
  \hline
\end{tabular}
\caption{Comparison of relative corrections $\delta_\wk$ of Eq.~(\ref{eq:relcorr}) and $\delta_{\EW}^{\tree}$ of Eq.~(\ref{eq:relcorr_ew}) from {\tt MCFM} and
  Table~2 of Ref.~\cite{Dittmaier:2012kx} (DHS) for various ranges of
  the transverse momentum of the leading jet ($k_T(j_1)$) in di-jet
  production at the 14 TeV LHC.
\label{tbl:kt1:14tev}}
\end{table}

\begin{figure}[htpb]
\includegraphics[scale=0.29]{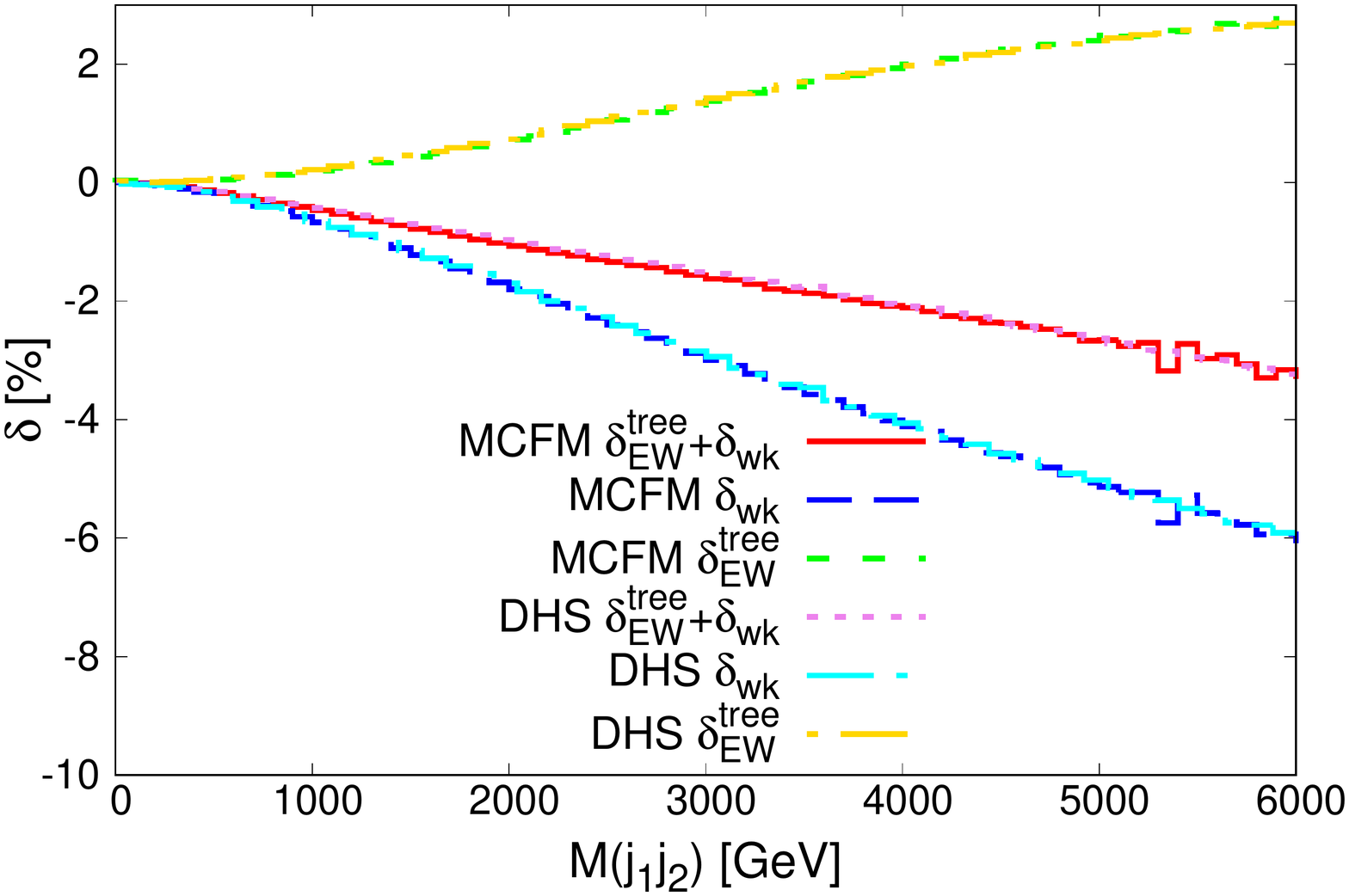} 
\includegraphics[scale=0.29]{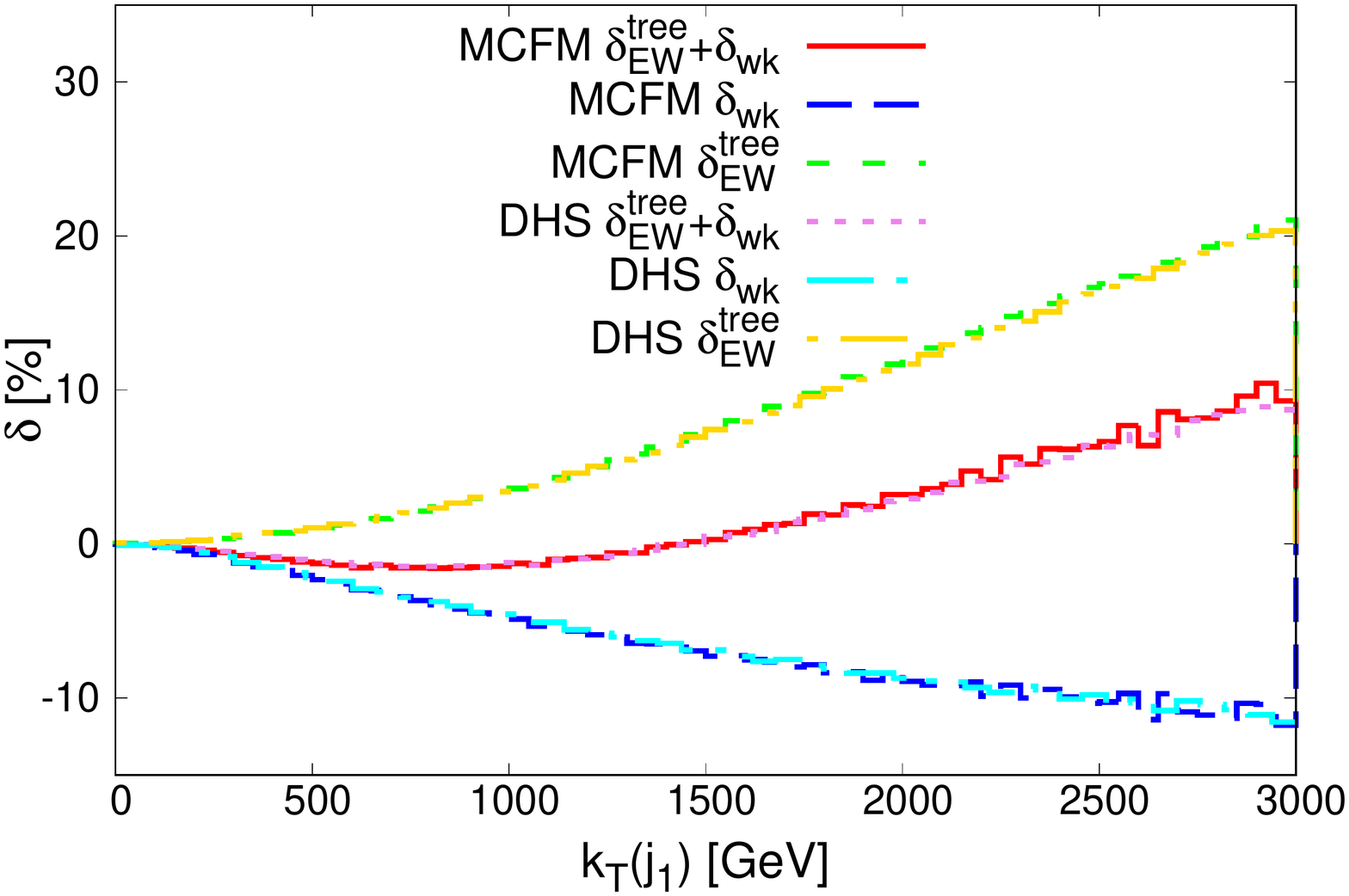} 
\caption{Comparison of relative corrections $\delta_{\wk}$ of Eq.~(\ref{eq:relcorr})  (blue, dashed and light blue, long-dashed-dotted),
  $\delta_{\EW}^{\tree}$ of Eq.~(\ref{eq:relcorr_ew}) (green, short-dashed and yellow, dotted-long-dashed), 
 and their sum (red, solid and pink, dotted) from {\tt MCFM} and the results
  of Figures~9 and 12 of Ref.~\cite{Dittmaier:2012kx} (DHS). Results are shown for the
  invariant di-jet mass ($M(j_1j_2)$) (left)) and transverse jet momentum
  distributions  for the leading jet ($k_{T}(j_1)$) (right)
  in di-jet production at the 14 TeV LHC.  
    \label{fig:jj:14tev}}
\end{figure}

A comparison of the relative corrections to the di-jet invariant mass
($M(j_1j_2)$) and the transverse jet momentum distributions of the
leading jet ($k_{T}(j_1)$) at the 14 TeV LHC is shown in
Fig.~\ref{fig:jj:14tev}. As can been seen, all of the {\tt MCFM}
results for di-jet production at the LHC are in good agreement with
those presented by DHS in Ref.~\cite{Dittmaier:2012kx}, with only small differences 
at large values of $k_T(j_1)$ of at most $3\%$ of the relative correction.

\section{Effectiveness of the Sudakov approximation}\label{sec:sudakovtest}

Using the two implementations of weak one-loop corrections in {\tt
  MCFM}, i.~e. the full corrections and their Sudakov approximation as
described in Section~\ref{sec:sudakov}, we can now easily assess the
effectiveness of the Sudakov approximation of
Ref.~\cite{Denner:2001gw} in the tails of kinematic distributions by
comparing with the exact results.  As pointed out
earlier~\cite{Moretti:2006nf,Kuhn:2013zoa}, the Sudakov approximation
is expected to have only limited application in $\ttb$ and di-jet
production.  The Sudakov logarithms are only dominant when all
invariants are much larger than the weak gauge boson mass and, in
general, these terms fail to capture the correct angular distribution
of particles in the final state. Nevertheless, a comparison of the
exact and Sudakov results may serve as a guide for cases in which a
full, exact calculation is infeasible and the Sudakov approximation is
the only available recourse.

If not mentioned otherwise, all results in this section are obtained
with {\tt MCFM} using the setup and cuts described in
Section~\ref{sec:comparison}.  For the sake of definiteness, we define
a relative weak Sudakov correction in direct analogy to
Eq.~(\ref{eq:relcorr}) through,
\begin{equation}\label{eq:relcorr_suda}
\delta_{\Sudakov} = \frac{d\sigma_{NLO}^{\Sudakov} - d\sigma_{LO}}{d\sigma_{LO}},
\end{equation}
where $\sigma_{NLO}^{\Sudakov}$ includes the NLO Sudakov corrections
described in Section~\ref{sec:sudakov}.

\subsection{Neutral-Current Drell-Yan process}\label{sec:test-ncdy}

\begin{figure}[htpb]
\centering
\includegraphics[scale=0.29]{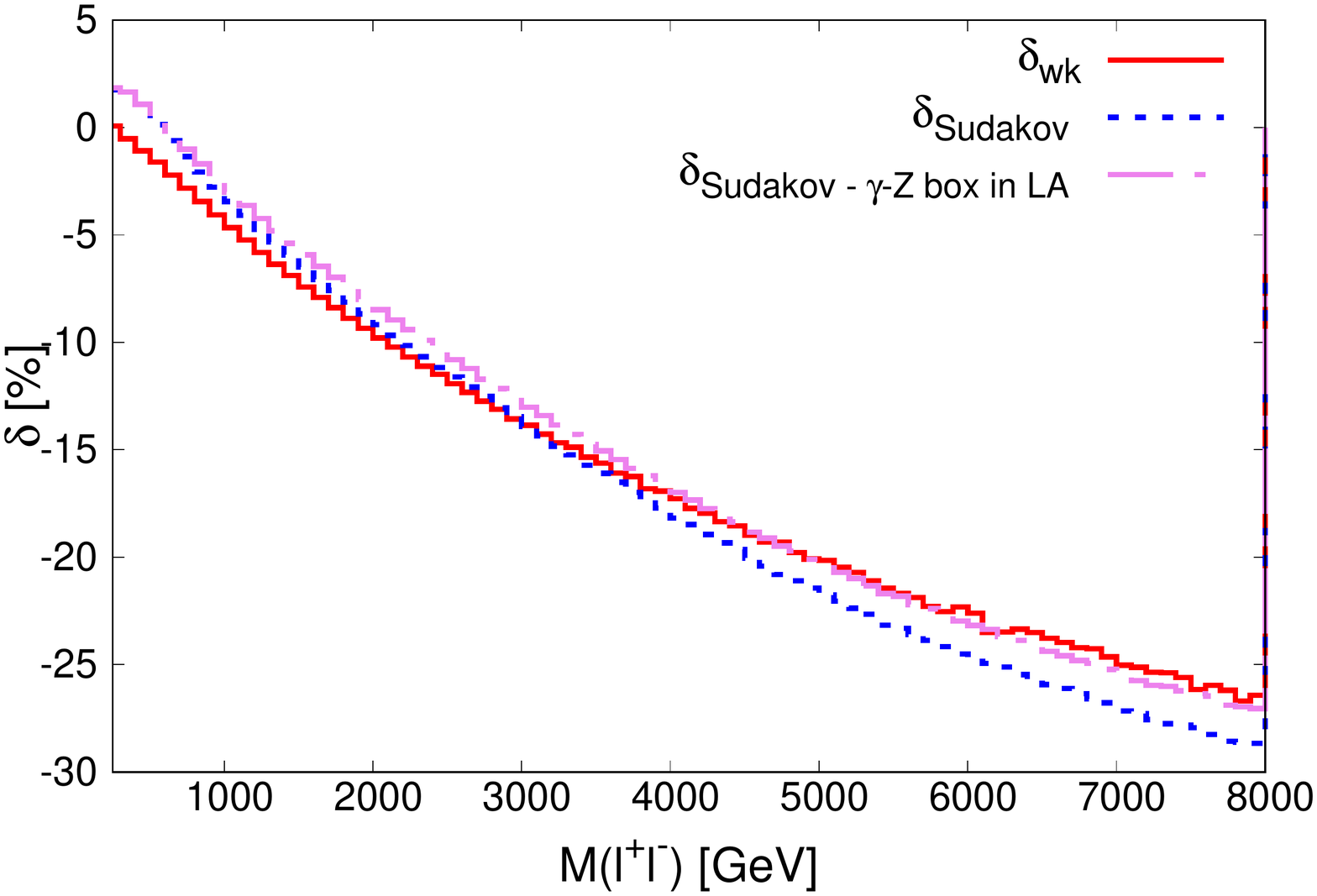}
\includegraphics[scale=0.29]{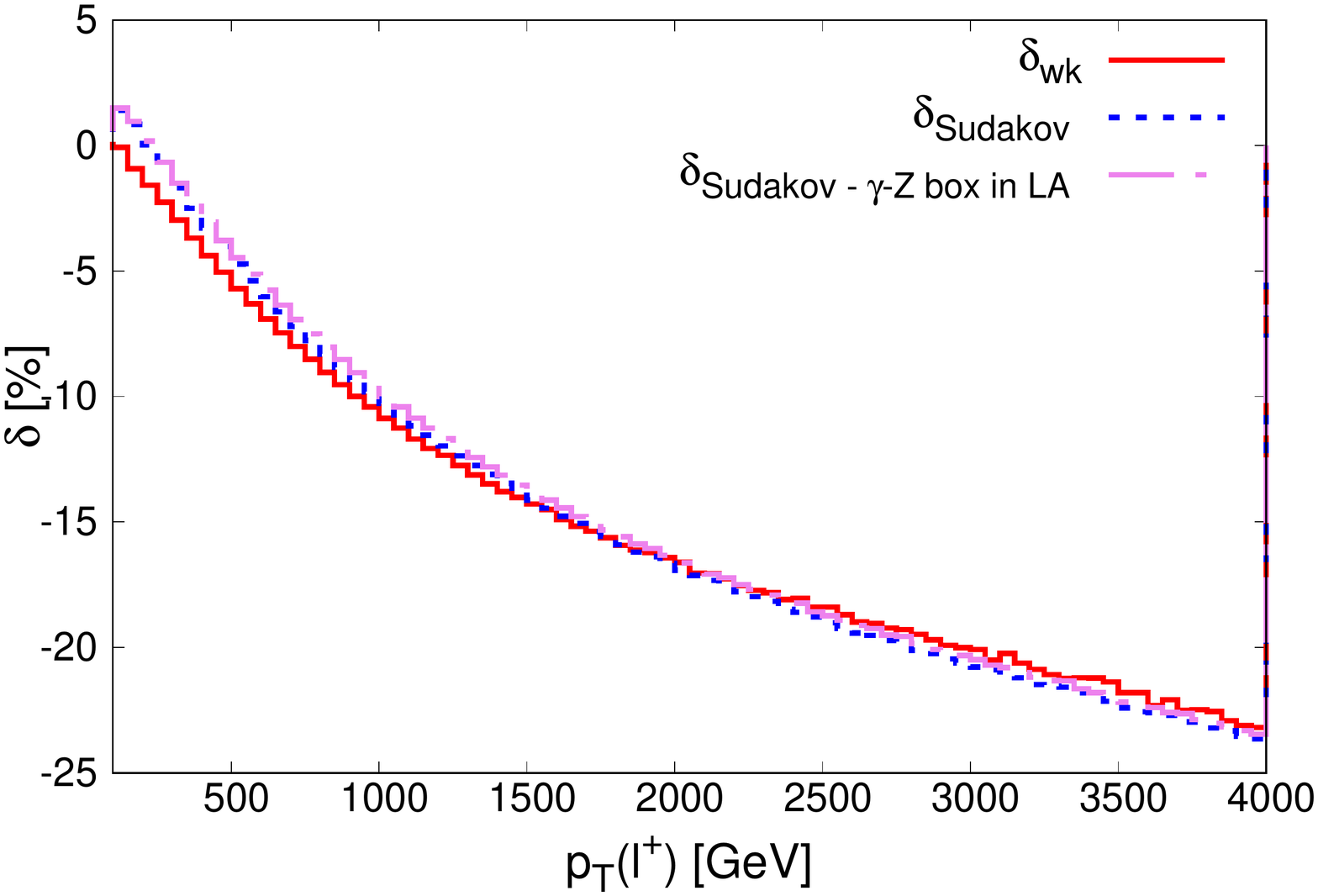} 
\caption{Relative weak one-loop corrections from {\tt MCFM} to the
  invariant lepton-pair mass ($M(l^+l^-$, $l=e,\mu)$) (left) and lepton transverse
  momentum ($p_T(l^+)$) (right) distributions in the NC DY process at the 13
  TeV LHC. The correction is expressed as a percentage of the LO result and results are shown for both the exact ($\delta_{\wk}$ of Eq.~(\ref{eq:relcorr})) (red,
  solid) and approximate Sudakov ($\delta_{\Sudakov}$ of Eq.~(\ref{eq:relcorr_suda})) calculation,
  the latter with (blue, dotted) and without (pink, long-dashed-dotted) the
  $\gamma-Z$ box contribution of Eq.~(\ref{eq:gamzbox}).
\label{fig:z:minv:pt}}
\end{figure}

As we have seen in Section~\ref{sec:ncdycomp} the effect of the weak
one-loop corrections on the total rate for the NC DY process is rather
small.  However the situation is quite different when investigating
the effect on kinematic distributions such as the invariant mass of
the lepton pair and the transverse momentum of the leptons.  These are
shown for both the exact weak corrections $\delta_{\wk}$ and the
Sudakov approximation $\delta_{\Sudakov}$, over ranges extending to
multi-TeV values, in Fig.~\ref{fig:z:minv:pt}. We have used the same
setup and cuts described in Section~\ref{sec:ncdycomp} apart from
increasing the cut on $M(l^+l^-)$ to 200~GeV.  The Sudakov
approximation shows good agreement with the exact NLO calculation in
the $p_T(l^+)$ distribution but there is a discrepancy in the
$M(l^+l^-)$ distribution at the level of about $3\%$ for $M(l^+ l^-)
\sim 8$~TeV.

We can trace this remaining difference to the contribution of the
$\gamma-Z$ box in the Sudakov approximation, which is not included in
the exact weak one-loop correction, since it is considered part of the
QED ${\cal O}(\alpha)$ correction to the NC DY process. To illustrate the impact of
the $\gamma-Z$ box we evaluate the contribution of this diagram to the 
matrix element squared at ${\cal O}(\alpha^3)$ in the leading
approximation (LA) at high energies.
We can then identify the part proportional to
$\log\brc{\hat{t}/ \hat{u}}\log\brc{\hat{s} / M_Z^2}$
as the contribution of the $\gamma-Z$ box to the Sudakov
approximation of Eq.~(\ref{eq:z:dl}), which reads:
\begin{align}\label{eq:gamzbox}
  \overline{\sum}\mathrm{Re}\brc{\delta^{\mathrm{SSC,Z}}\mm_0\times\mm_0^*}
  & = \frac{2}{3}\alpha^3\pi\cdot\frac{-16}{\hat{s}^2}\cdot\log\brc{\frac{\hat{t}}{\hat{u}}}\log\brc{\frac{\hat{s}}{M_Z^2}}\cdot\Bigg\{ \nonumber \\
  & + Q_q^2Q_l^2\left[g_v^qg_v^l\brc{\hat{t}^2+\hat{u}^2}-g_a^qg_a^l\brc{\hat{t}^2-\hat{u}^2}\right] \nonumber \\
  & + 2~Q_qQ_l\left[\brc{{g_v^q}^2+{g_a^q}^2}\brc{{g_v^l}^2+{g_a^l}^2}\brc{\hat{t}^2+\hat{u}^2}
    -4g_v^qg_v^qg_a^qg_a^l\brc{\hat{t}^2-\hat{u}^2}\right] \nonumber \\
  & +(g_v^qg_v^l+g_a^qg_a^l)\left[\brc{g_v^qg_v^l+g_a^qg_a^l}^2+3\brc{g_v^qg_a^l+g_a^qg_v^l}^2\right]\hat{u}^2 \nonumber \\
  & -(g_v^qg_v^l-g_a^qg_a^l)\left[\brc{g_v^qg_v^l-g_a^qg_a^l}^2+3\brc{g_v^qg_a^l-g_a^qg_v^l}^2\right]\hat{t}^2 \Bigg\}
\end{align}
In Fig.~\ref{fig:z:minv:pt} we also show the effect of subtracting this
contribution from the Sudakov approximation of Eq.~(\ref{eq:z:dl}). As
expected, this modified Sudakov approximation now represents an excellent
description of the full one-loop weak correction to the
lepton-pair invariant mass distribution.

\begin{figure}[htpb]
  \begin{center}
    \includegraphics[scale=0.29]{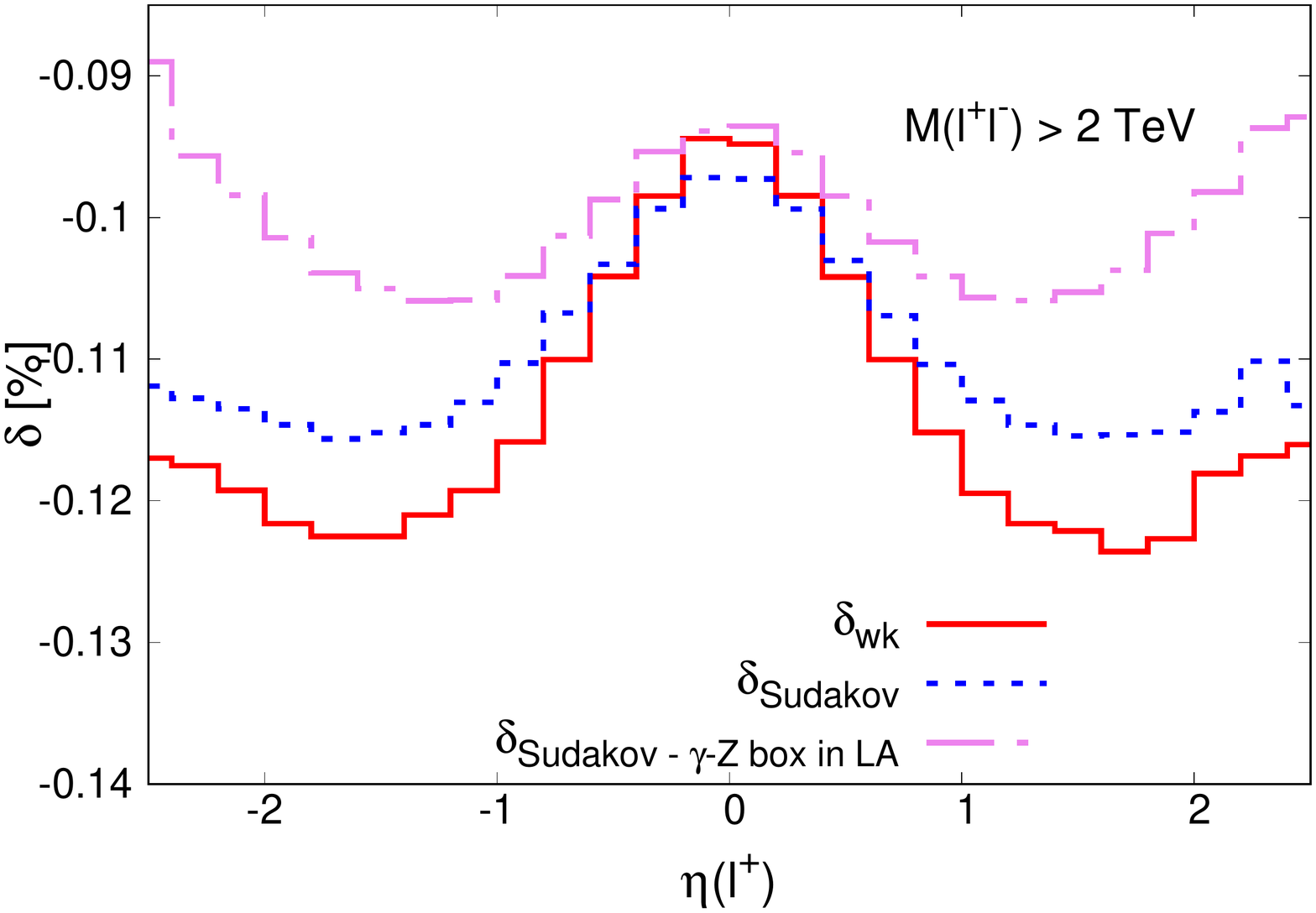}
    \includegraphics[scale=0.29]{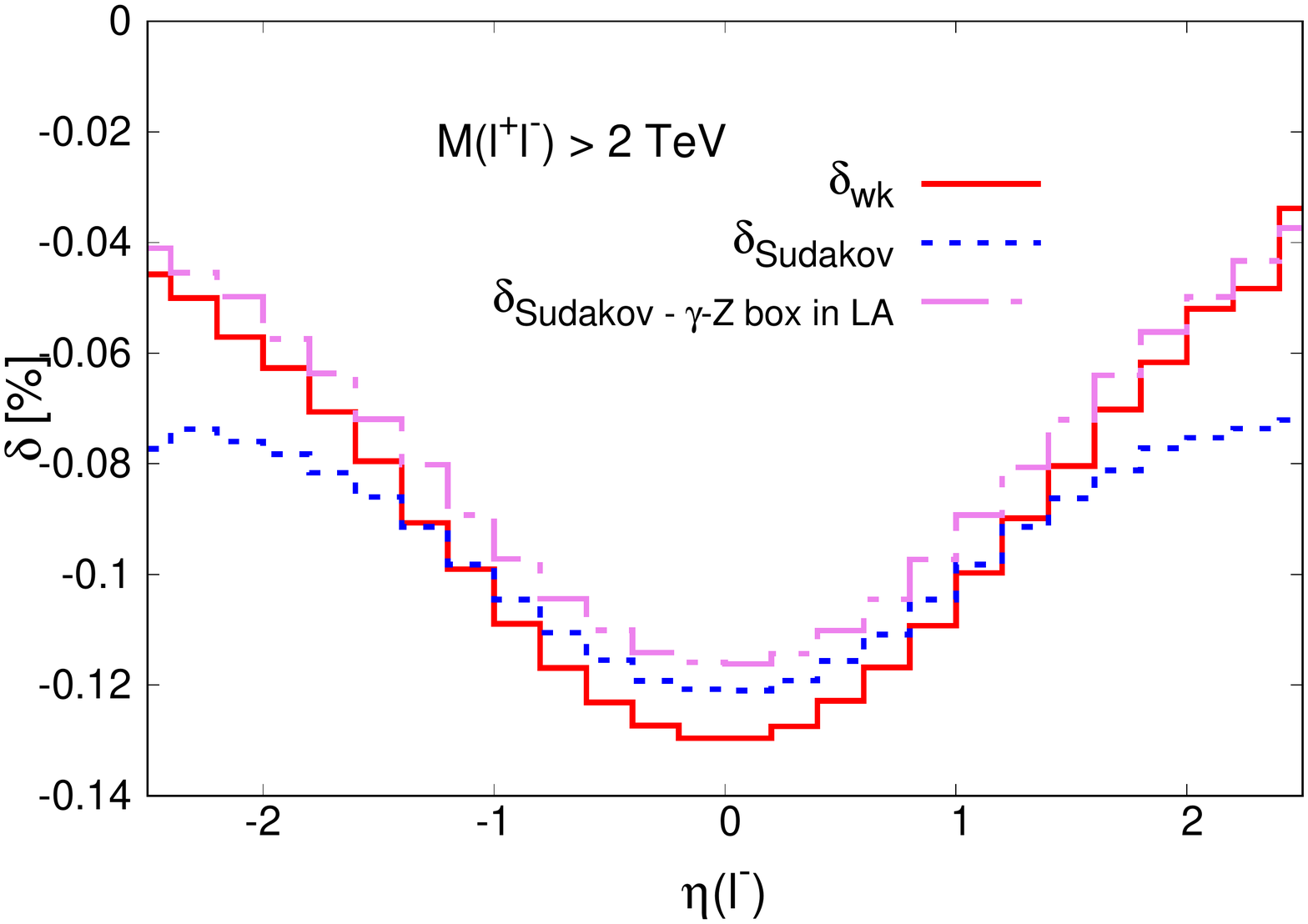} \\
    \includegraphics[scale=0.29]{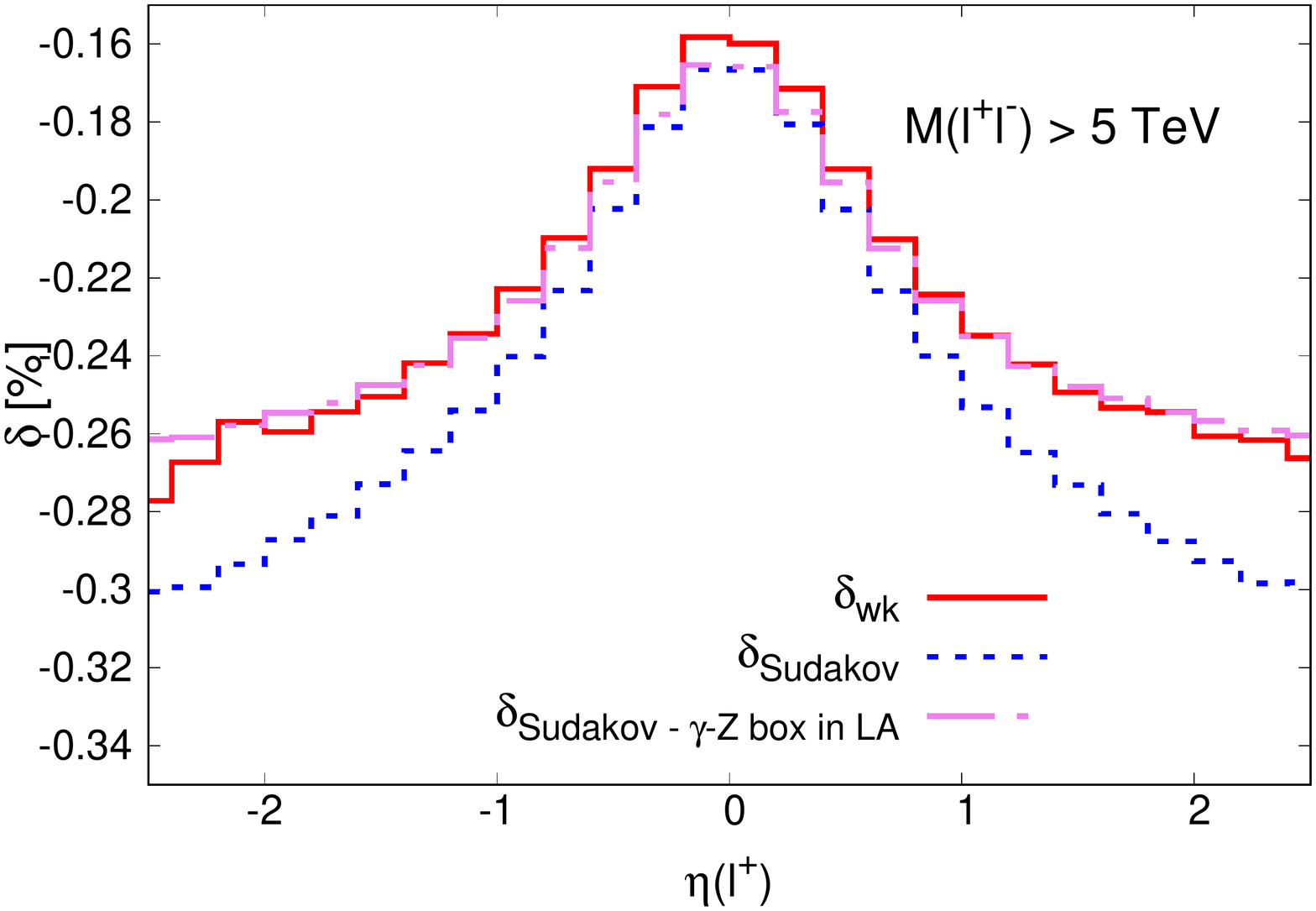}
    \includegraphics[scale=0.29]{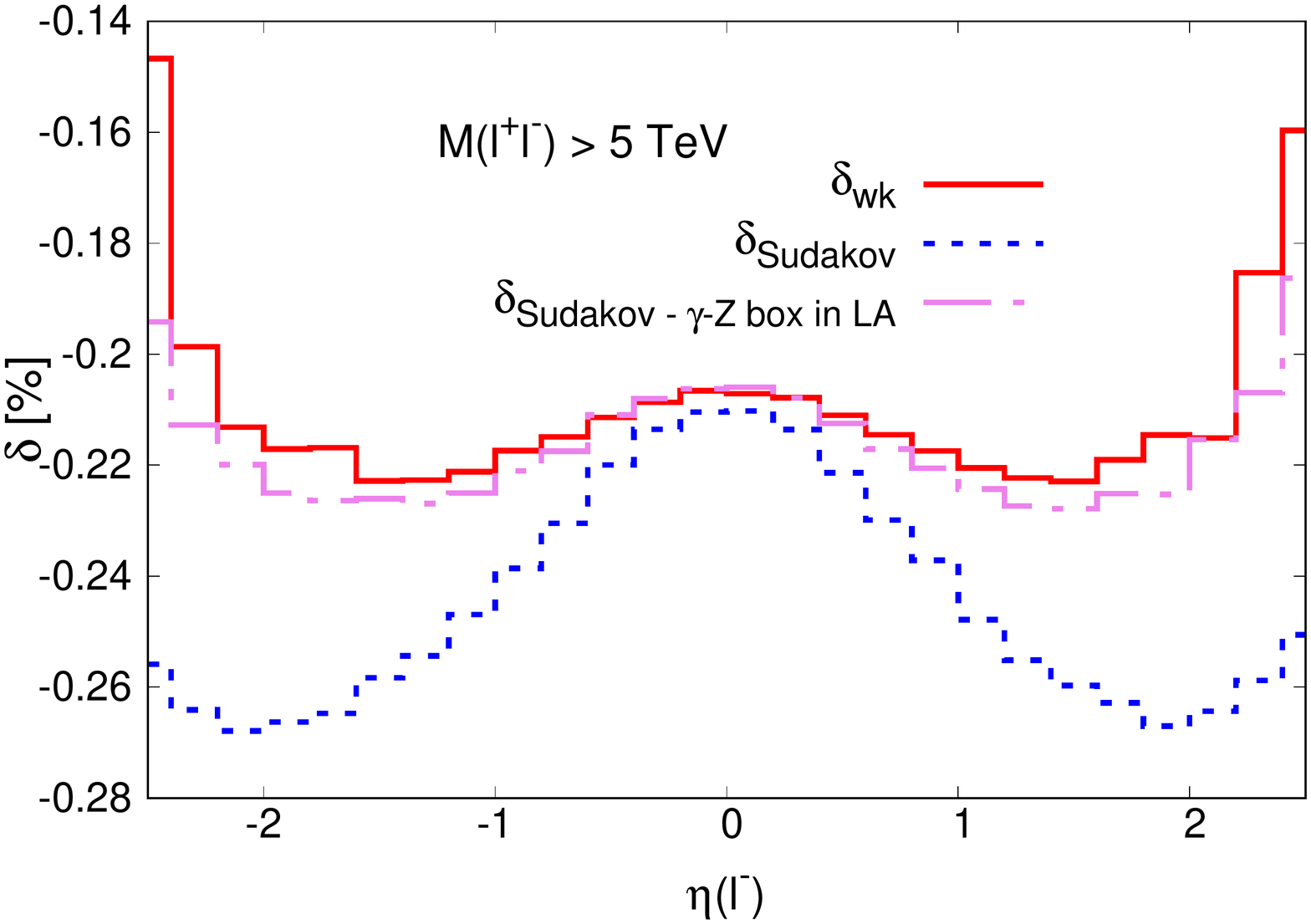} \\
    \caption{Relative weak one-loop corrections to the pseudorapidity
      distributions of the positively (left) and negatively (right)
      charged leptons in the NC DY process at the 13 TeV LHC.  
The correction is expressed as a percentage of the LO result and
results are shown for both the exact ($\delta_{\wk}$ of Eq.~(\ref{eq:relcorr})) (red,
  solid) and approximate Sudakov ($\delta_{\Sudakov}$ of Eq.~(\ref{eq:relcorr_suda})) calculation,
  the latter with (blue, dotted) and without (pink, long-dashed-dotted) the
  $\gamma-Z$ box contribution of Eq.~(\ref{eq:gamzbox}).
    \label{fig:z:rap}}
  \end{center}
\end{figure}

In Fig.~\ref{fig:z:rap}, we compare the relative weak one-loop corrections 
$\delta_{\wk}$ and $\delta_{\Sudakov}$ to the pseudo-rapidity distribution
of the charged leptons, where we apply successive cuts on the lepton-pair invariant mass 
at 2~TeV and 5~TeV in order to focus on the high-energy behavior.  
Despite this cut,
the exact and Sudakov calculations are not in good agreement outside the very 
central rapidity region unless the $\gamma-Z$ box contribution of Eq.~(\ref{eq:z:dl}) 
is subtracted from $\delta_{\Sudakov}$.
When this is the case, this modified Sudakov approximation agrees well with the exact 
calculation at $M(l^+l^-)>5$~TeV.
However, overall the effect of the weak one-loop corrections on
the lepton rapidity distributions is rather mild, since they are not
very sensitive to the presence of the weak Sudakov logarithms.

Another observable that is interesting to measure at the LHC is the
forward-backward asymmetry of the charged leptons as a function of the invariant
mass of the lepton pair, $A_{FB}(M(l^+l^-))$. It is 
defined by~\cite{Baur:1997wa},
\begin{equation}
A_{\rm FB}=\frac{F-B}{F+B}
\label{EQ:DEFAFB}
\end{equation}
where
\begin{equation}
F=\int_0^1 \frac{{\rm d}\sigma}{{\rm d}\cos\theta^*}\,{\rm d}\cos\theta^*, \qquad
B=\int_{-1}^0\frac{{\rm d}\sigma}{{\rm d}\cos\theta^*}\,{\rm d}\cos\theta^*.
\label{EQ:DEFFB}
\end{equation}
$\cos\theta^*$ is given in the Collins-Soper frame~\cite{Collins:1977iv} by,
\begin{equation}
\cos\theta^*  =  \frac{|p_z(l^+l^-)|}{p_z(l^+l^-)}~\frac{2}{M(l^+l^-)\sqrt{M^2(l^+l^-) +p_T^2(l^+l^-)}}
\left [p^+(l^-)p^-(l^+)-p^-(l^-) p^+(l^+)\right ] \;,
\label{EQ:CSTAR}
\end{equation}
where,
\begin{equation}
p^\pm=\frac{1}{\sqrt{2}}\left (E\pm p_z\right ),
\end{equation}
and $E$, $p_z$ are the energy and longitudinal component of the momentum respectively.

This observable is sensitive to the weak mixing angle and, in the
vicinity of the $Z$ resonance where the number of events is very high,
precision measurements of this quantity have been made both at the
Tevatron~\cite{Aaltonen:2016nuy,Abazov:2014jti} and the
LHC~\cite{Chatrchyan:2011ya,Aad:2015uau,Aaij:2015lka}.  However, it is
also interesting to study this observable far from the resonance
region.  For instance, in the high-invariant mass region $A_{FB}$ can
be used in the search for extra gauge bosons ($Z'$) (see, for example,
Ref.~\cite{Accomando:2015cfa}).

\begin{figure}[htpb]
  \begin{center}
    \includegraphics[scale=0.32]{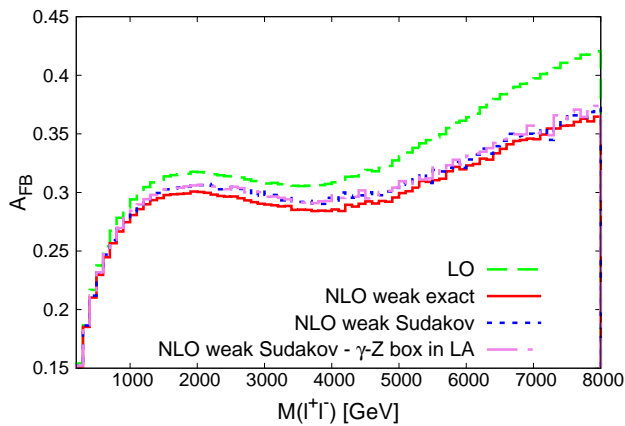}
    \caption{Differential lepton forward-backward asymmetry $A_{FB}$ with
      respect to the invariant mass of the charged lepton pair in the NC DY
      process at the 13 TeV LHC.  Results are shown for the LO (green,
      dashed) prediction and for both the exact (red, solid) and
      approximate Sudakov calculation, the latter with (blue, dotted)
      and without (pink, long-dashed-dotted) the $\gamma-Z$ box contribution of
      Eq.~(\ref{eq:gamzbox}).
\label{fig:z:afb}}
  \end{center}
\end{figure}

The impact of the exact weak one-loop corrections on $A_{FB}$,
compared to the Sudakov approximation with and without the
contribution from the $\gamma-Z$ box diagram of Eq.~(\ref{eq:gamzbox}), is shown in
Fig.~\ref{fig:z:afb}.  We note that the effect of the weak corrections
is well-described by the Sudakov approximation throughout the
distribution.  These effects are relatively mild for invariant masses
that have been probed with good precision so far (around 1 TeV), but
grow as large as $-12\%$ in the far tail.

\subsection{Top-quark pair production}\label{sec:test-ttb}

\begin{figure}[htpb]
  \includegraphics[scale=0.29]{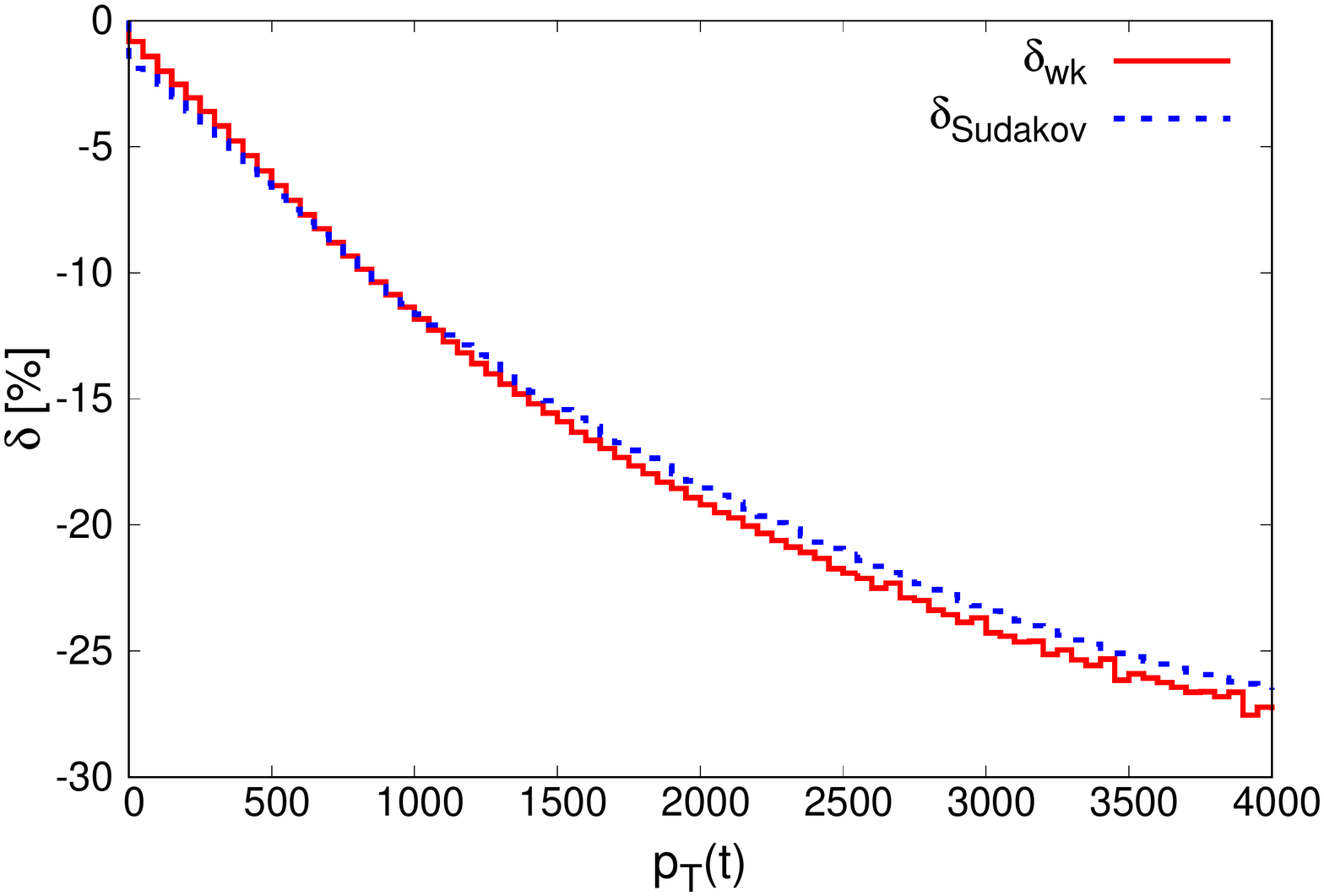}
  \includegraphics[scale=0.29]{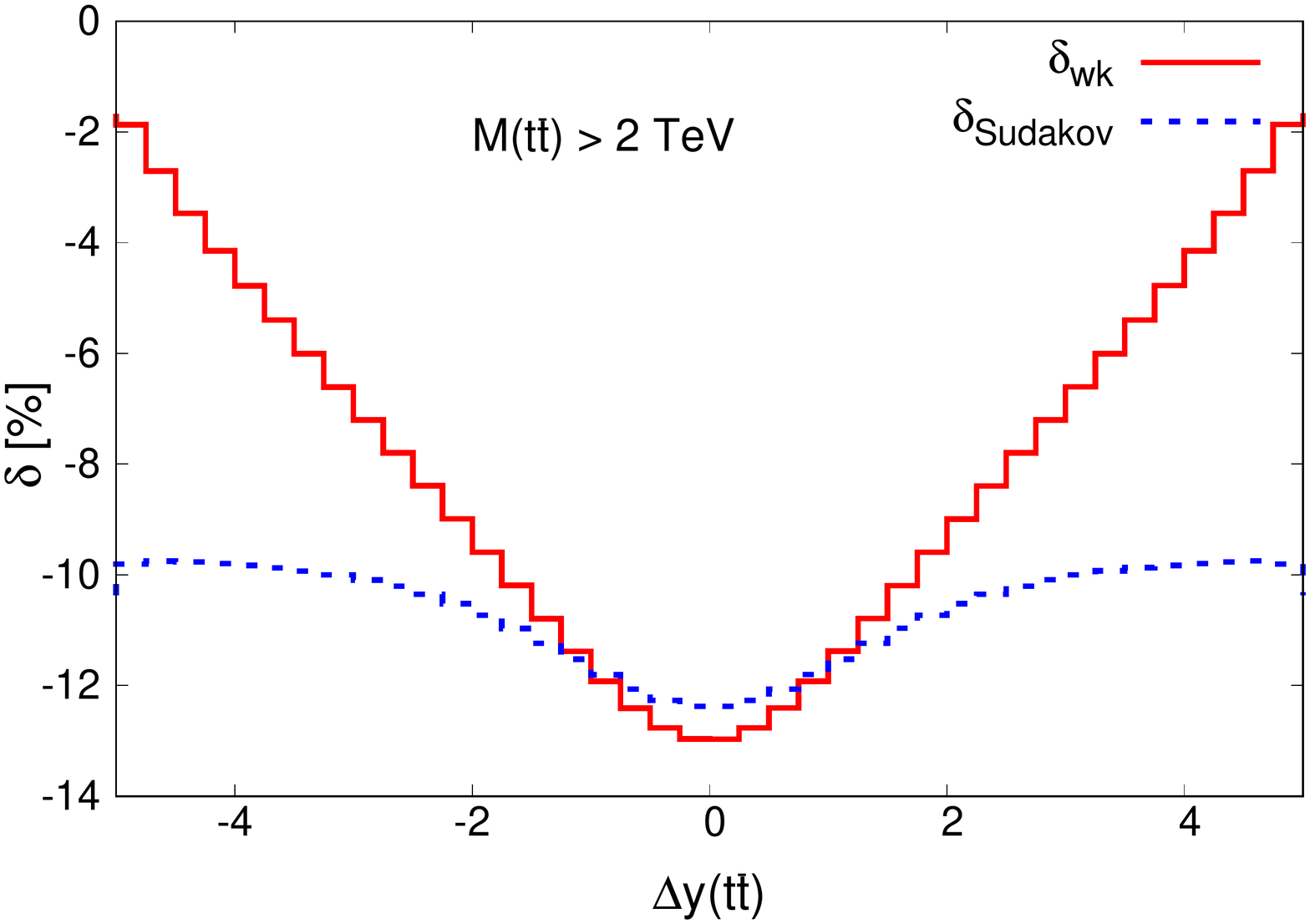}
    \caption{Relative weak corrections to the $p_T(t)$
      (left) and $\Delta y(t\bar t)$ (right) distributions in $\ttb$ production at the $13$~TeV LHC.
      The correction is expressed as a percentage of the LO result and is
    shown for the exact weak relative correction $\delta_{\wk}$ of
    Eq.~(\ref{eq:relcorr}) (red, solid) and the Sudakov approximation
    of $\delta_{\Sudakov}$ of Eq.~(\ref{eq:relcorr_suda}) (blue, dotted).}
  \label{fig:ttSudakov1}
\end{figure}

We now turn to the case of top-quark pair production, where we follow
the setup already used in Section~\ref{sec:ttbcomp}.
Figure~\ref{fig:ttSudakov1} shows the results of the comparison in the
cases of the $p_T(t)$ and $\Delta y(t\bar t)$
distributions. For the distribution of the
rapidity difference we have applied an additional $\ttb$ invariant
mass cut of $M(t\bar t) > 2$~TeV. 
Agreement between the exact and approximate
calculations is almost perfect for $p_T(t)$, but this is not the case
for $\Delta y(t\bar t)$.  There the Sudakov approximation is only
close to the exact result for small rapidity differences, $|\Delta
y(t\bar t)| < 2$, due to angular dependence in the corrections that
is not captured in the Sudakov approximation.

\begin{figure}[htpb]
  \includegraphics[scale=0.29]{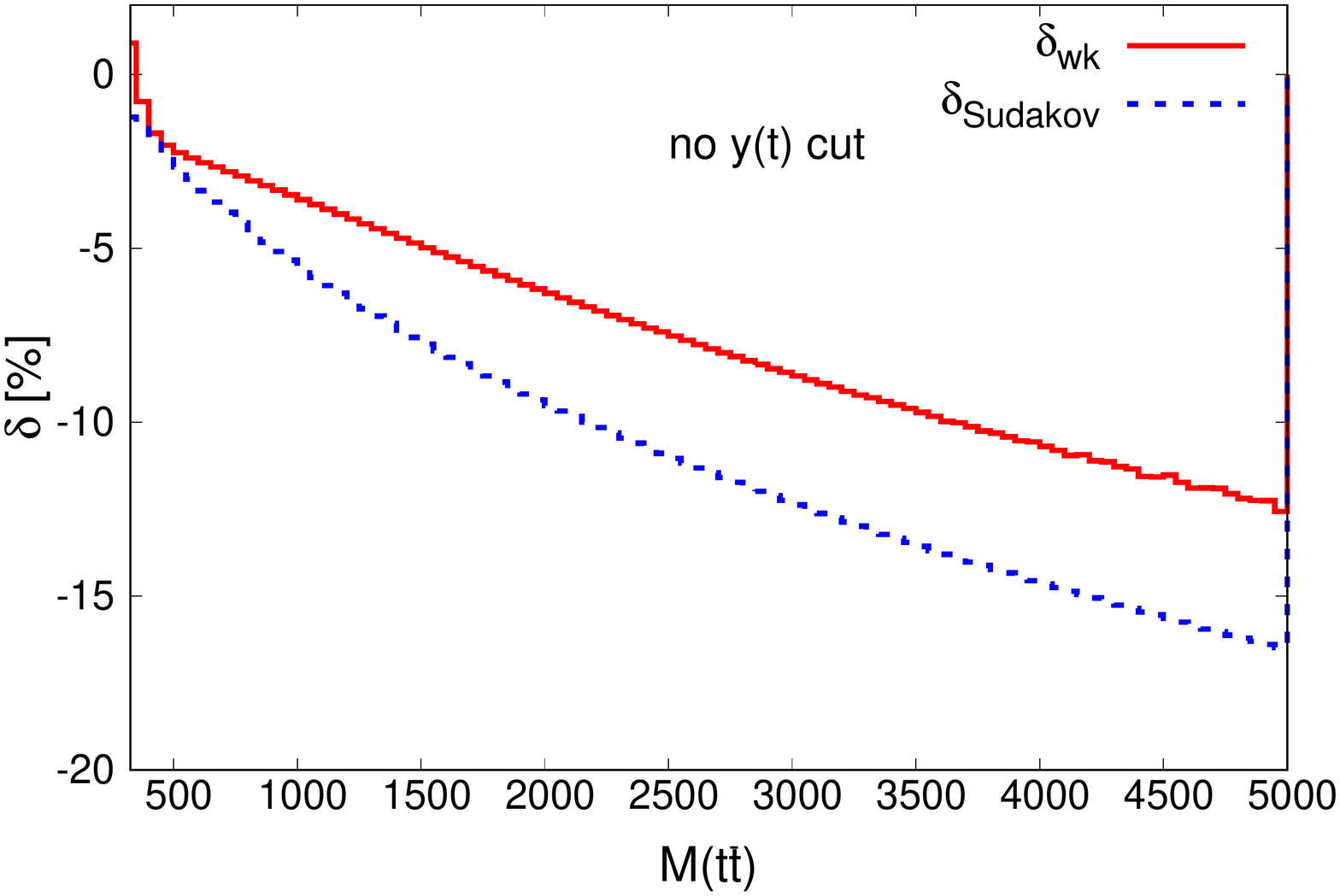}
  \includegraphics[scale=0.29]{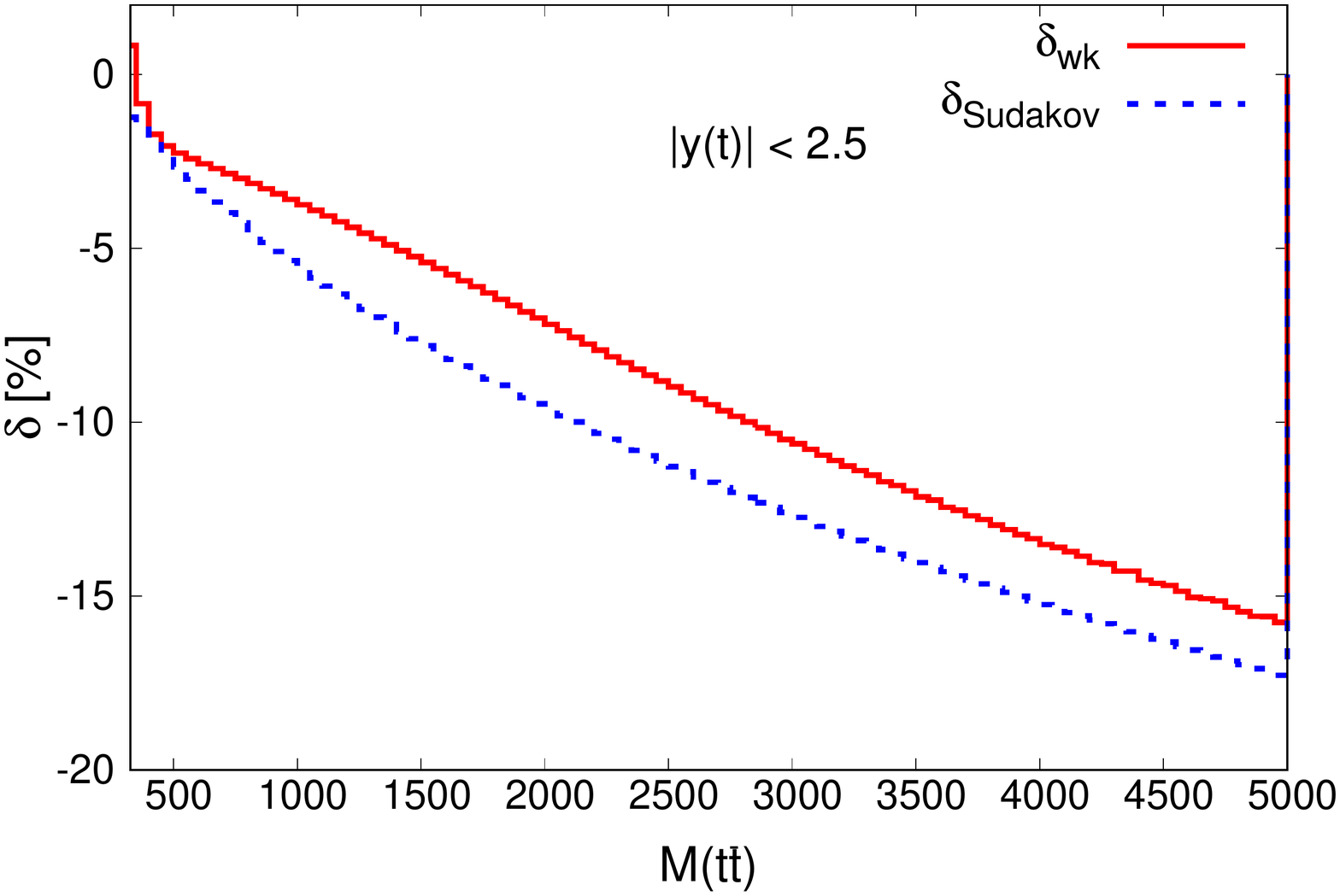}
  \includegraphics[scale=0.29]{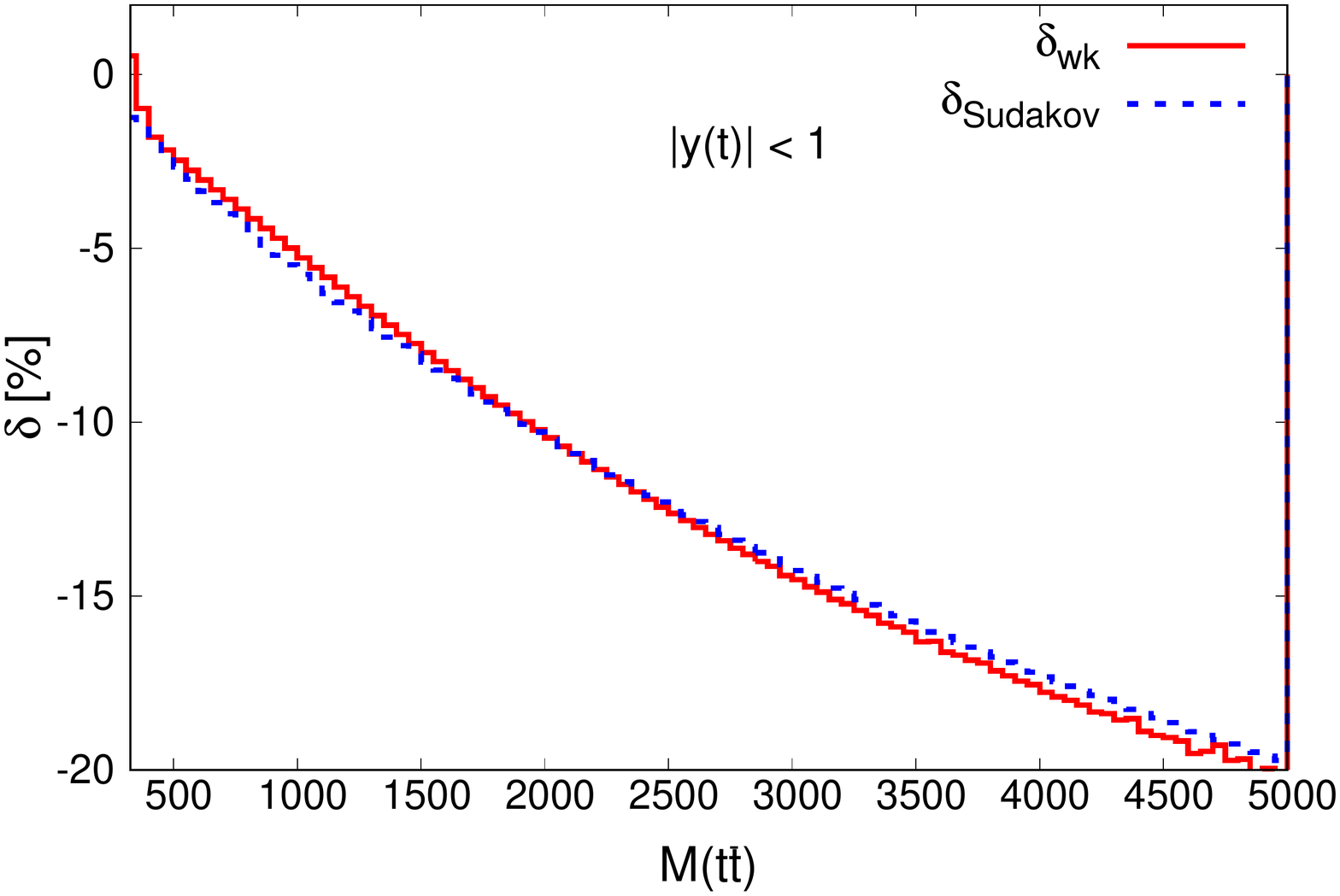} 
  \caption{Relative weak corrections to $M(t{\bar t})$ in $\ttb$ production at the $13$~TeV
    LHC with no cuts applied (top left), after application of a moderate
    rapidity cut (top right) and for central top quarks (center).  The
    correction is expressed as a percentage of the LO result and is
    shown for the exact weak relative correction $\delta_{\wk}$ of
    Eq.~(\ref{eq:relcorr}) (red, solid) and the Sudakov approximation $\delta_{\Sudakov}$ 
    of Eq.~(\ref{eq:relcorr_suda}) (blue, dotted).
  \label{fig:ttSudakov2}}
\end{figure}

The situation for the distribution of the invariant mass of the top pair is
shown in Fig.~\ref{fig:ttSudakov2}.  In this case the Sudakov
approximation also does not describe the effect of the weak
corrections on the $M(t{\bar t})$ distribution very well.  Since
Fig.~\ref{fig:ttSudakov1} demonstrates that the approximation works
best for more central rapidities, we repeat the comparison of the
$M(t{\bar t})$ distribution after application of rapidity cuts on the
top and anti-top quarks.  We consider both central production of top
quarks, $|y(t,\bar t)| < 1$ and an intermediate case, $|y(t,\bar t)|
< 2.5$.  Agreement is substantially improved after the application of
a moderate rapidity cut on the top quarks, while for highly-central
top quarks the approximation describes the exact result extremely well
over the entire range.

\subsection{Di-jet Production}\label{sec:test_dijet}

\begin{figure}[ht]
  \includegraphics[scale=0.29]{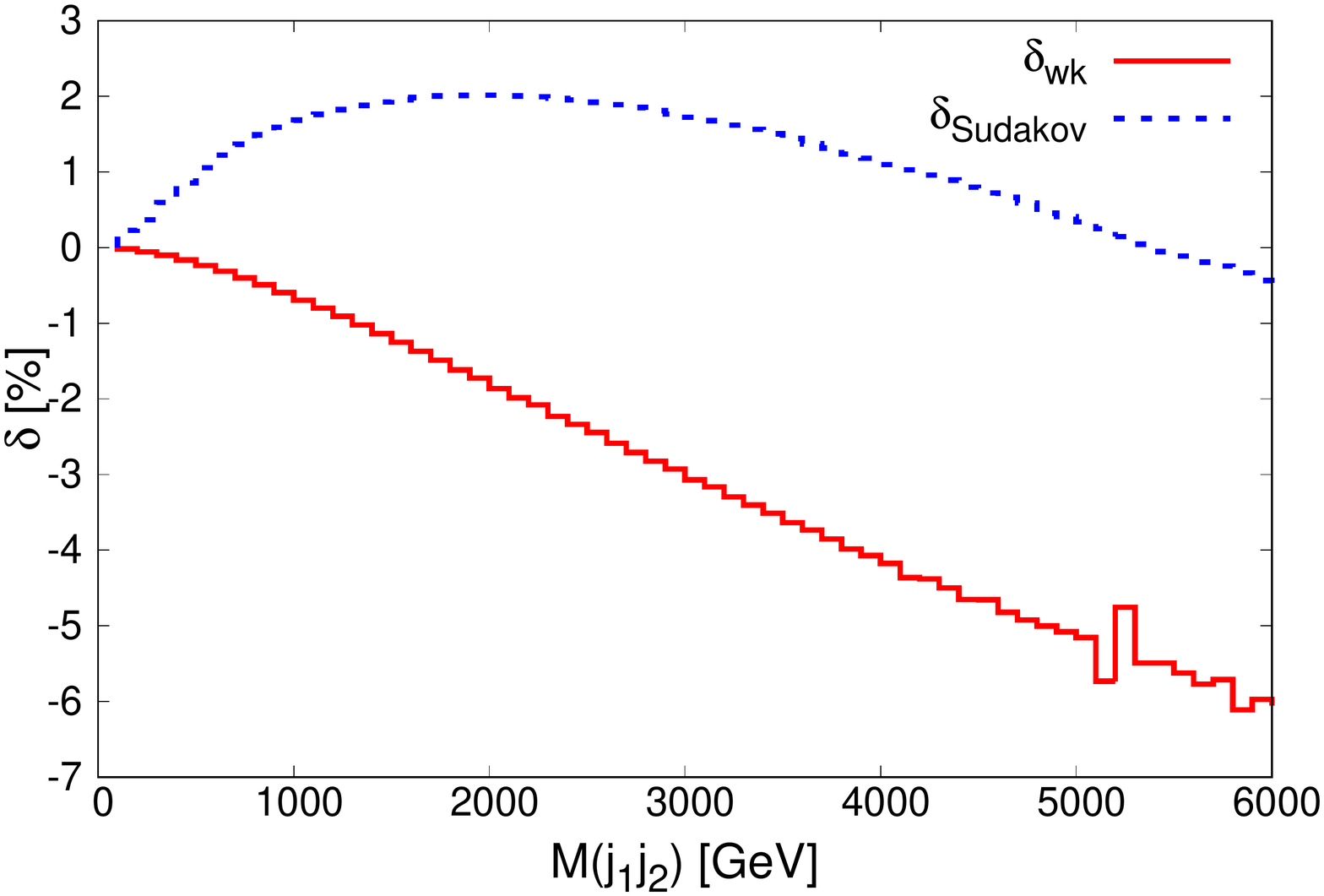}
  \includegraphics[scale=0.29]{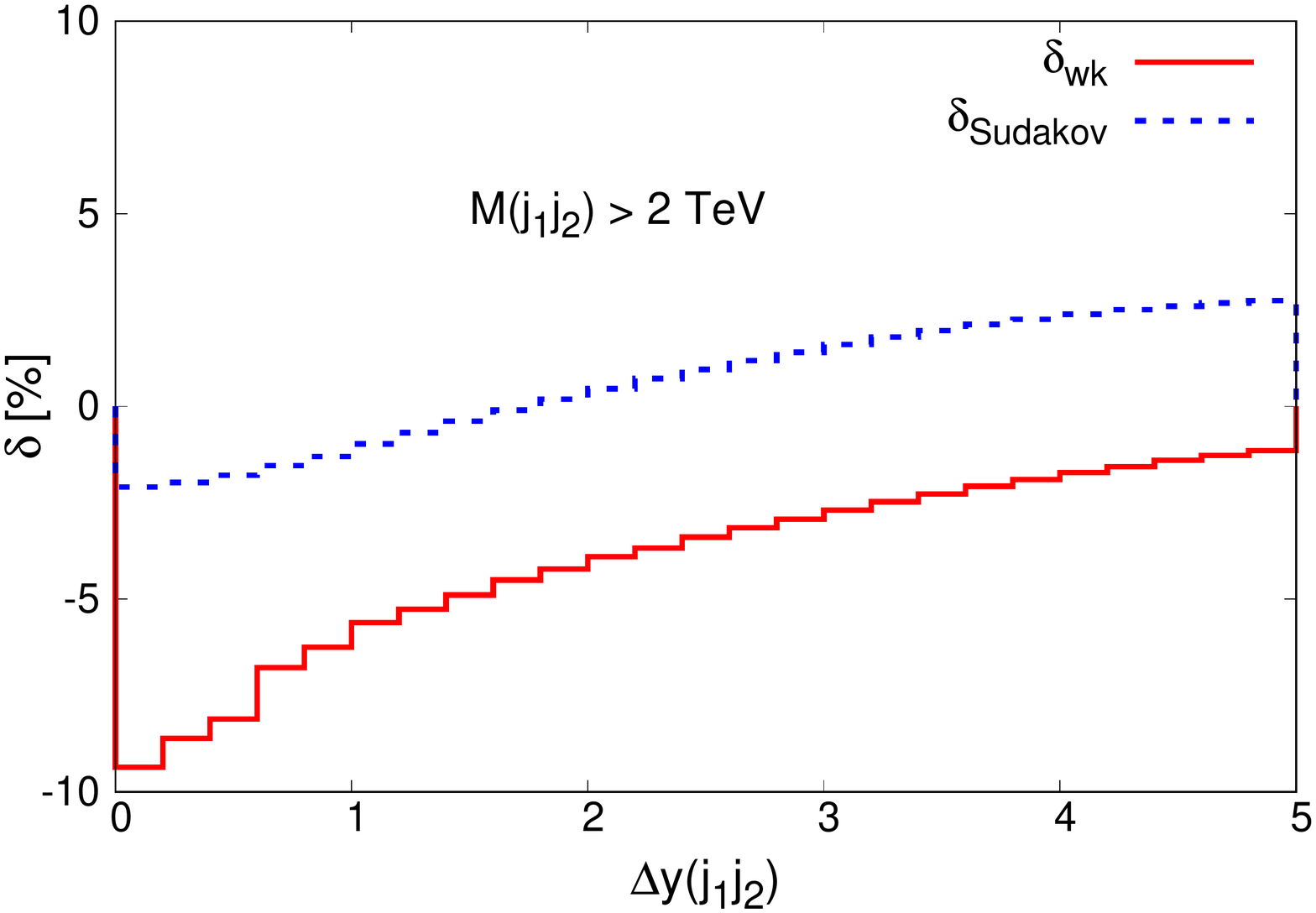} \\
  \includegraphics[scale=0.29]{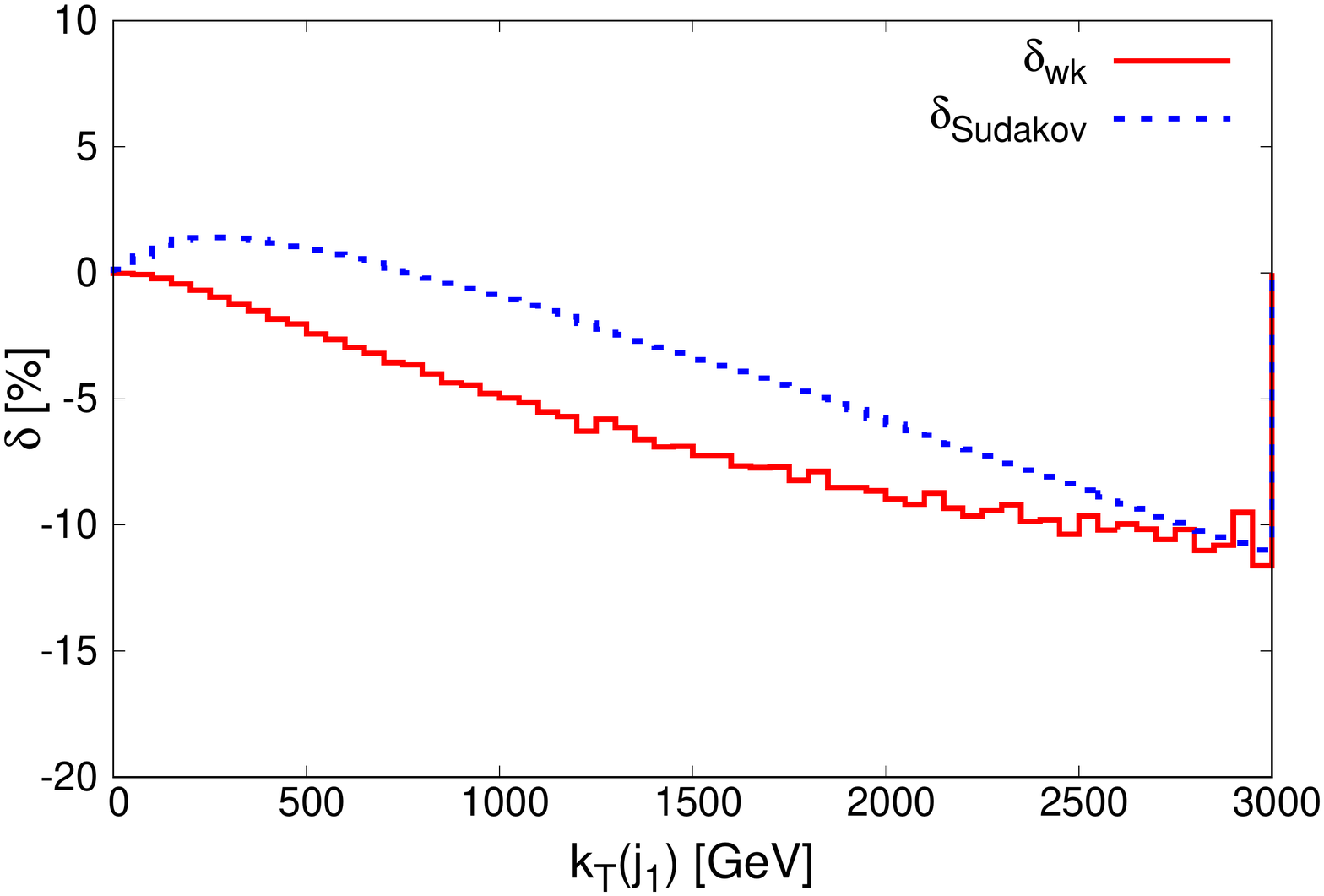}
  \includegraphics[scale=0.29]{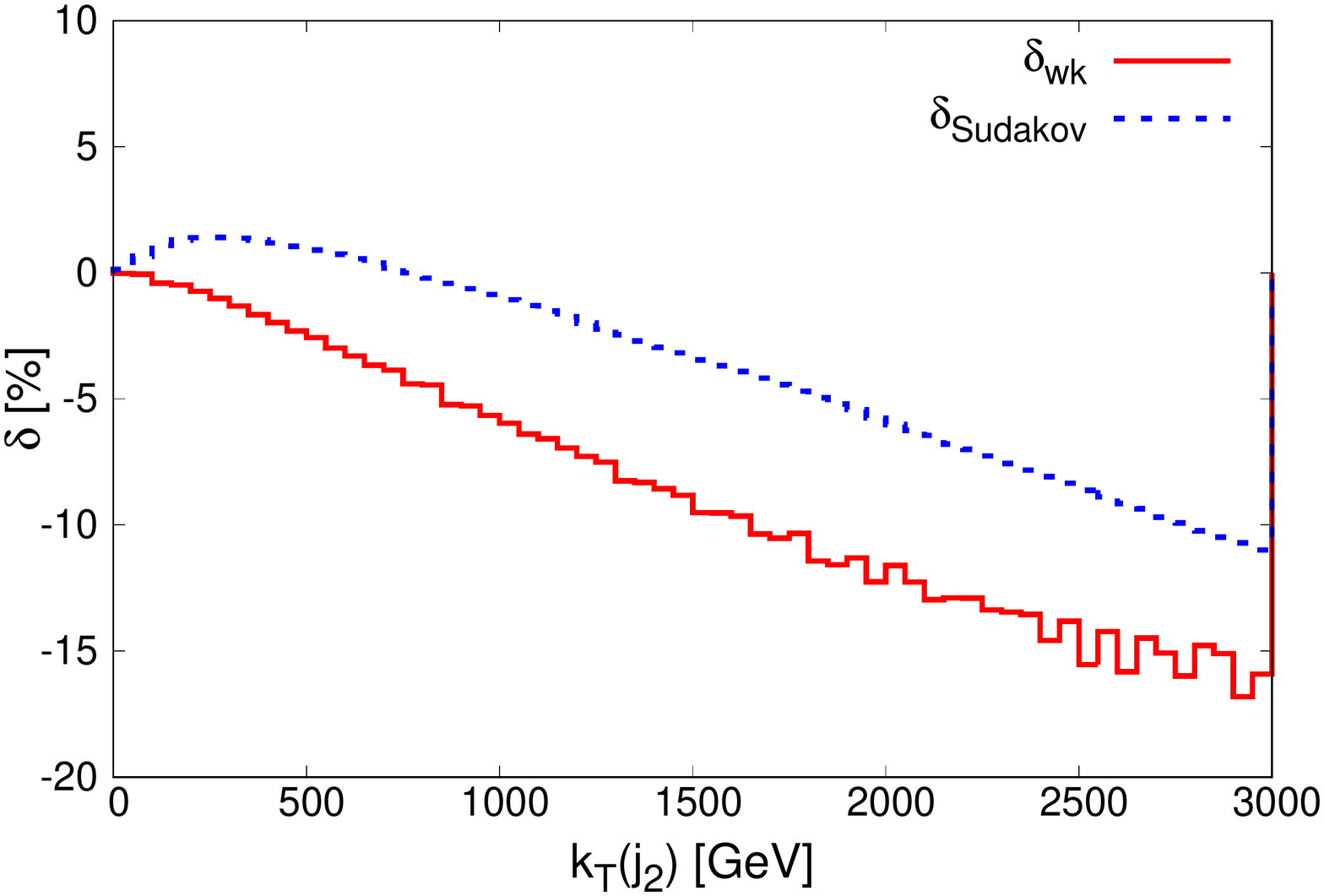}
  \caption{
Relative weak corrections to the invariant mass (upper left), the
    absolute rapidity difference between the two leading jets (upper right), and 
    the transverse momentum distributions of leading
    (lower left) and subleading (lower right) jets in di-jet production at the 13 TeV LHC. The correction is expressed as a percentage of the LO ${\cal O}(\alpha_s^2)$ 
    cross section and is shown for the exact weak relative corrections of Eq.~(\ref{eq:relcorr})
    (red, solid) and the Sudakov approximation of Eq.~(\ref{eq:relcorr_suda}) (blue, dotted). 
    \label{fig:jj:13teva}}
\end{figure}

Here we compare the exact calculation of di-jet production at 
${\cal O}(\alpha_s^2 \alpha)$ described in Section~\ref{sec:dijet}
with the Sudakov approximation described in
Section~\ref{sec:suda-dijet}.  In
Fig.~\ref{fig:jj:13teva} we show the comparison for the di-jet
invariant mass ($M(j_1j_2)$), the transverse momenta of the leading and
next-to-leading (in $k_T$) jets, and the absolute rapidity difference
between these two jets ($\Delta y(j_1j_2)$).  Results are shown for the relative
corrections for the 13~TeV LHC in the setup used in Section~\ref{sec:dijetcomp}.  
For the distribution of the rapidity
difference we have applied an additional di-jet invariant mass cut of
$M(j_1j_2) > 2$~TeV.

We observe that the Sudakov approximation has little utility in this
case, with substantial differences from the exact calculation in each
distribution.  This can be traced to the rich angular structure of the
weak corrections, especially in the four-quark subprocesses, whose
admixture is impossible to capture in an approximate form of
Sudakov-type. For example, the QCD virtual corrections to the
four-quark amplitude shown in Figs.~\ref{fig:2j:4q:v2},
\ref{fig:2j:4q:v3} (and the corresponding real corrections),  
which also contribute at ${\cal
  O}(\alpha_s^2 \alpha)$, are not captured by the Sudakov
approximation, but still have a sizeable impact on the distributions
shown here.  This is also highlighted by the fact that, in contrast to the
case of top-quark pair production, there is no region in $\Delta
y(t\bar t)$ in Fig.~\ref{fig:jj:13teva} where the results of the exact weak and
approximate calculations are close, so that the application of a
central rapidity cut does little to improve the effectiveness of the
Sudakov approximation.

\section{Combination of QCD and Weak Corrections}\label{sec:combination}

In this section we will consider the combination of the exact NLO weak
corrections that we have presented, with QCD corrections at NLO and
beyond.  The aim of this section is to compare the sizes of the two
effects and to demonstrate the inherent ambiguity in how the two
should be combined, particularly in cases where either correction, or
both, is large.

To illustrate this we will consider two procedures for combining the
corrections.  One straightforward method is to simply add the weak
corrections, $\sigma_{\wk}$ to the NLO or NNLO QCD cross section,
\begin{equation}
\sigma_{QCD+\wk} = \sigma_{(N)NLO \, QCD}+\sigma_{\wk} \; ,
\label{eq:QCDweakadd}
\end{equation}
An alternative is to combine them using a ``multiplicative'' procedure,
\begin{equation}
\sigma_{QCD\times \wk} = \sigma_{(N)NLO \, QCD} \left( 1 + \frac{\sigma_{\wk}}{\sigma_{LO}} \right)\; .
\label{eq:QCDweakmult}
\end{equation}
This procedure should better account for factorizable higher-order
mixed QCD-weak corrections. Compared to the additive procedure, this
approach enhances the impact of weak corrections in regions where the
QCD corrections are large. To illustrate the numerical impact of these
two approaches we discuss in the following the relative corrections
with respect to the (N)NLO QCD cross section defined as,
\begin{equation}
\delta_{\add} = \frac{\sigma_{QCD+\wk}-\sigma_{(N)NLO \, QCD}}{\sigma_{(N)NLO \, QCD}}=\frac{\sigma_{\wk}}{\sigma_{(N)NLO \, QCD}} \; 
\label{eq:dQCDweakadd}
\end{equation}
for the additive approach, and 
\begin{equation}
\delta_{\prod} = \frac{\sigma_{QCD \times \wk}-\sigma_{(N)NLO \, QCD}}{\sigma_{(N)NLO \, QCD}}=\frac{\sigma_{\wk}}{\sigma_{LO}} 
\label{eq:dQCDweakmult}
\end{equation}
for the ``multiplicative'' procedure. Similar studies of different combinations of QCD and EW corrections
can also be found for instance in Ref.~\cite{Kuhn:2013zoa} for $\ttb$ production and 
in Refs.~\cite{Li:2012wna,Dittmaier:2012kx,Mishra:2013una,Alioli:2016fum} for NC DY production. In case of DY processes,
the mixed QCD-electroweak corrections at ${\cal O}(\alpha_s \alpha)$ have been calculated in the pole approximation
in Refs.~\cite{Dittmaier:2014qza,Dittmaier:2015rxo}, which considerably improves predictions in the resonance region.

\subsection{NNLO QCD and Weak Corrections to the NC Drell-Yan process}
\label{sec:dycomb}

The NNLO QCD corrections to the Drell-Yan process can be computed in
{\tt MCFM}~\cite{Boughezal:2016wmq} and their combination with the
weak corrections is therefore particularly straightforward.  For the
results presented here we retain the parameters and setup of the
previous sections, with the exception that all computations are now
performed with the NNLO MSTW2008 set.

\begin{figure}[htpb]
  \includegraphics[scale=0.29]{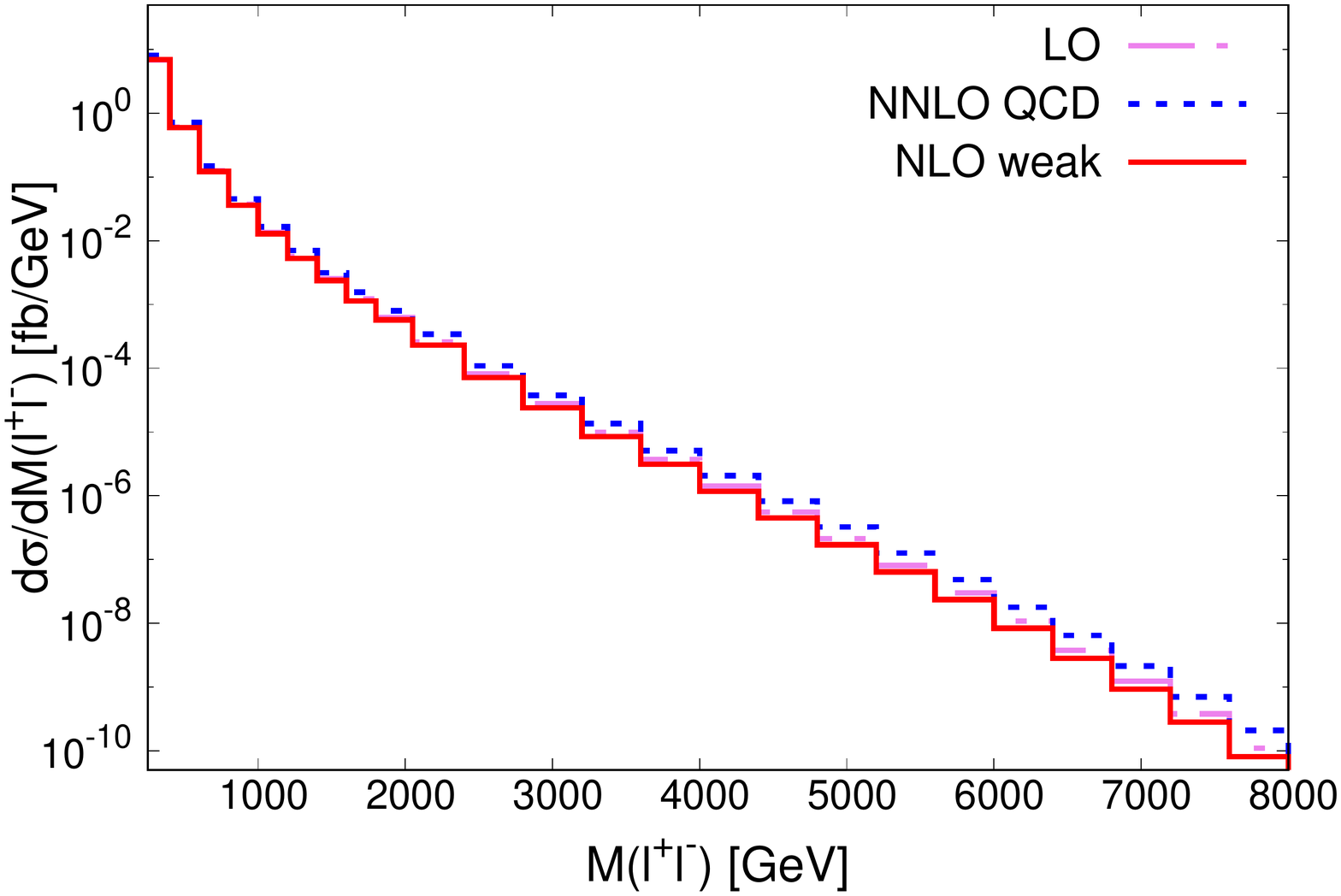}
  \includegraphics[scale=0.29]{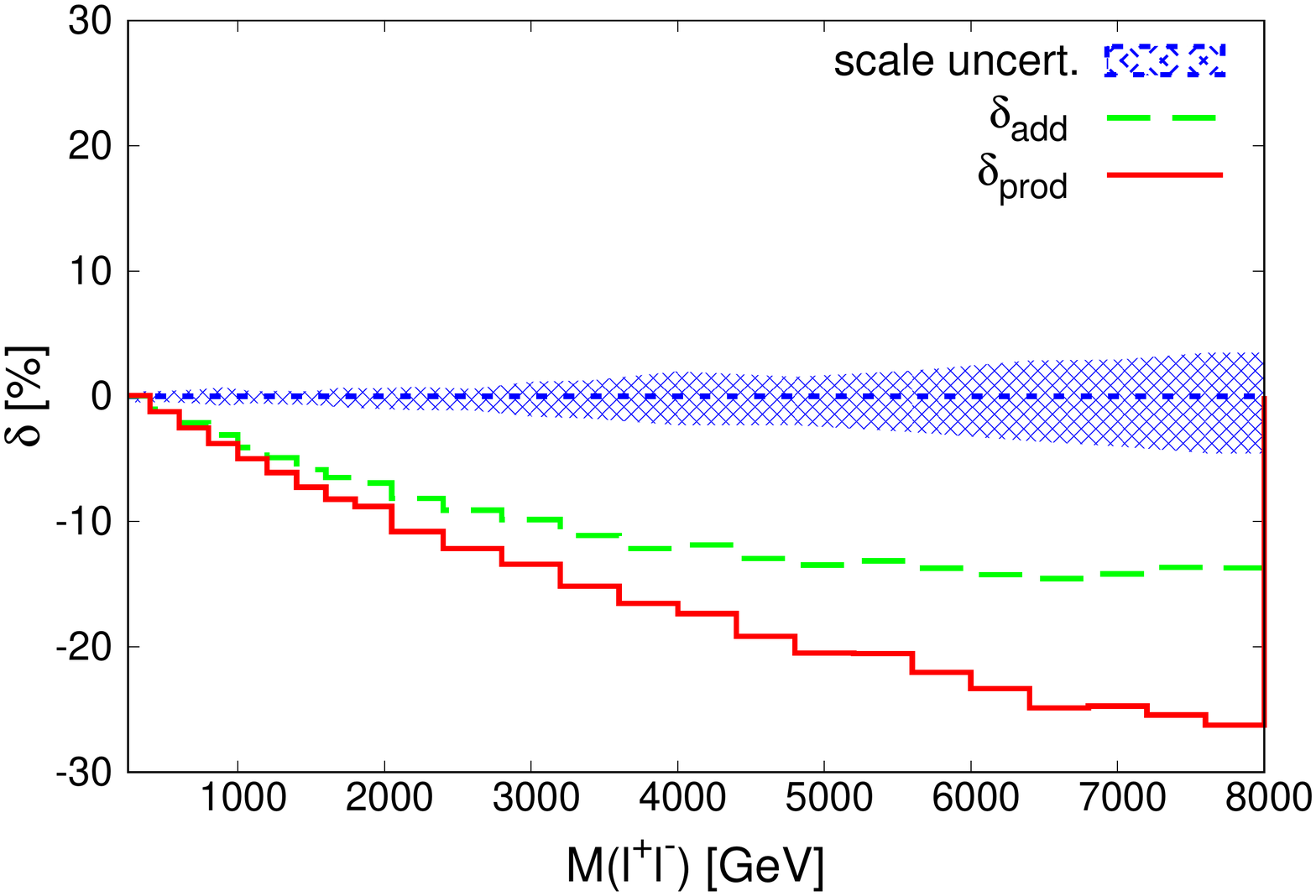}
  \caption{Left: comparison of the effect of NLO weak (red, solid) and NNLO QCD
    corrections (blue, dotted) on the invariant mass distribution of the di-lepton pair in the
    NC DY process at 13~TeV. The LO distribution is also shown (pink, long-dashed-dotted).  Right: a comparison
    of the two procedures ($\delta_{\add}$ of Eq.~(\ref{eq:dQCDweakadd}) (green, dashed) and $\delta_{\prod}$ of Eq.~(\ref{eq:dQCDweakmult}) (red, solid)) used to combine NNLO QCD and NLO weak
    effects, together with the scale uncertainty in the pure NNLO QCD
    calculation (blue band). The NNLO QCD results have been obtained with {\tt
      MCFM}~\cite{Boughezal:2016wmq}.
  \label{fig:DYcombined}}
\end{figure}

Our results are shown in Figure~\ref{fig:DYcombined}.  As is
well-known, the effect of NNLO QCD corrections, relative to LO, is
large and positive throughout the distribution.  This is apparent from
the left-hand plot.  The right-hand plot compares the two combination
procedures by plotting the relative corrections with respect to the NNLO QCD result,
$\delta_{\add}$ and $\delta_{\prod}$ as defined in
Eqs.~(\ref{eq:QCDweakadd}) and~(\ref{eq:QCDweakmult}), respectively.
Since these are normalized to the NNLO QCD prediction, the result for
$\delta_{\prod}$ could have been read-off directly from
Figure~\ref{fig:z:minv:pt} (c.f. Eq.~(\ref{eq:dQCDweakmult})).  The
fact that both corrections are substantial means that the two
procedures give noticeably different results for $M(l^+l^-)>2$~TeV.  As a
point of reference, in this plot we also show the theoretical
uncertainty resulting from the choice of scale in the NNLO QCD
calculation.  This is obtained by considering the envelope of
predictions obtained when using alternative scales given by,
\begin{equation}
(\mu_F/M_Z, \mu_R/M_Z) = \left\{ (0.5, 0.5),~(2, 2),~(0.5, 1),~(0.5, 2),~(1, 0.5),~(2, 0.5)\right\} \;.
\end{equation}
With this prescription the scale uncertainty is as large as $5$\%
in the tail of the distribution, but is still smaller than the effect
of combining with weak corrections using either procedure.  It
is therefore clear that both effects must be included, with an
accompanying uncertainty associated with the choice of combination
procedure, in order to provide the best theoretical prediction.  As a
conservative estimate of the combination uncertainty one might simply
take the difference between the two procedures, which is of the same
order as the QCD scale uncertainty.

\subsection{NNLO QCD and Weak One-loop Corrections to Top-Quark Pair Production}
\label{sec:ttbcomb}

We now consider the combination of corrections to the top-quark pair process, namely
exact NLO weak corrections with
QCD corrections at NNLO. Since these corrections are not yet
available in differential form in a public code, we will compare our
NLO weak results with the NNLO results that have been published so
far~\cite{Czakon:2015owf,Czakon:2016ckf}.

We first focus on results for the Tevatron collider, where the NNLO results
are easily read-off from tables presented in
Ref.~\cite{Czakon:2016ckf}.  In order to match the results of that
study we modify our input parameters from the previous sections
slightly, to those shown in
Table~\ref{table:parameters_top_nnlocomparison}.  As before, we use
$\mu_F = \mu_R = m_t$ and we employ the NNLO MSTW2008 PDF
set~\cite{Martin:2009iq}.  Note that, in order to validate our setup,
we have recomputed the predictions at LO and NLO QCD using {\tt MCFM} and found perfect
agreement.
\begin{table}[htpb]
  \begin{tabular}{|l|l|}
  \hline
  $M_W = 80.398\,\GeV$ &
  $M_Z = 91.1876\,\GeV$ \\  
  $M_H = 126\,\GeV$ &
  $m_t = 173.3\,\GeV$ \\ 
  $m_b = 4.82\,\GeV$ &
  $\alpha(m_t) = 1/132.3407$ \\
  $\sin^2\theta_W = 1 - M_W^2/M_Z^2$ &  
  $\alpha_s(m_t) = 0.125666$ \\ 
  \hline
\end{tabular}
\caption{Input parameters used for the calculation of the
  NLO weak corrections in the setup used to calculate the NNLO QCD
  corrections to top-quark pair production at the Tevatron in
  Ref.~\cite{Czakon:2016ckf}.
\label{table:parameters_top_nnlocomparison}}
\end{table}

Our results are presented in the form of per-bin corrections to a
selection of observables, following the original presentation of NNLO
QCD results in Ref.~\cite{Czakon:2016ckf}.  Results are shown for 
$M(t\bar t)$ (Table~\ref{tab:Mtt}) and $p_T(t)$
(Table~\ref{tab:PTt}). We first note that the effect of the NNLO QCD
corrections is typically very small, at the level of a few percent of the NLO QCD result,
but is as large as almost $20\%$ for the highest $M(t\bar t)$ bin.  
As expected, the effect of the NLO weak corrections is readily apparent in the
$M(t\bar t)$ and $p_T(t)$ distributions.  The onset of the Sudakov
logarithms is clear in the $M(t\bar t)$ results, although the weak
corrections are non-negligible (and positive) in the first bin.  The
NLO weak effects are of a similar size to the corrections from NNLO
QCD and the two clearly must be taken into account together.  This is
even more clear in the $p_T(t)$ distribution, where the effects of the
NLO weak corrections are larger than those due to NNLO QCD for $p_T(t)
> 200$~GeV.
%
\begin{table}[h]
\begin{center}
\begin{tabular}{| c | c | c | c |}
\hline
$\Mtt$ [GeV] & \multicolumn{3}{|c|}{$d\sigma/d\Mtt$ [pb/bin]} \\
\hline 
  & {\rm NLO QCD} & {\rm NNLO QCD corr} & {\rm NLO weak corr} \\ \hline
	[240	 ; 	412.5]	                 &  $	   2.96     ~ \times ~ 10^{	  0	  }$	  &	  $	  0.17     ~ \times ~ 10^{	 0	 }$	&  $	   0.05     ~ \times ~ 10^{	  0	  }$\\     \hline
	[412.5	 ; 	505]	                 &  $	   2.47     ~ \times ~ 10^{	  0	  }$	  &	  $	  0.12     ~ \times ~ 10^{	 0	 }$	&  $	  -0.01     ~ \times ~ 10^{	  0	  }$\\     \hline
	[	505	 ; 	615	]	 &  $	   9.20     ~ \times ~ 10^{	  -1	  }$	  &	  $	  0.30     ~ \times ~ 10^{	 -1	 }$	&  $	  -0.15     ~ \times ~ 10^{	  -1	  }$\\     \hline
	[	615	 ; 	750	]	 &  $	   2.66     ~ \times ~ 10^{	  -1	  }$	  &	  $	  0.07     ~ \times ~ 10^{	 -1	 }$	&  $	  -0.08     ~ \times ~ 10^{	  -1	  }$\\     \hline
	[750	 ; 	1200]	                 &  $	   6.20     ~ \times ~ 10^{	  -2	  }$	  &	  $	  0.16     ~ \times ~ 10^{	 -2	 }$	&  $	  -0.27     ~ \times ~ 10^{	  -2	  }$\\     \hline
	[	1200	 ; 	$\infty$]	 &  $	   1.07     ~ \times ~ 10^{	  -4	  }$	  &       $       0.20     ~ \times ~ 10^{	 -4	 }$	&  $	  -0.10     ~ \times ~ 10^{	  -4	  }$\\      \hline
\end{tabular}
\caption{\label{tab:Mtt} The $\Mtt$ differential distribution in NLO QCD and the corrections due to NNLO QCD and NLO weak effects in $\ttb$ production at the Tevatron. The NLO and NNLO QCD results are taken from Ref.~\cite{Czakon:2016ckf}.}
\end{center}
\end{table}
%
\begin{table}[h]
\begin{center}
\begin{tabular}{| c | c | c | c |}
\hline
$\PTt$ [GeV]& \multicolumn{3}{|c|}{$d\sigma/d\PTt$ [pb/bin]} \\
\hline 
  & {\rm NLO QCD} & {\rm NNLO QCD corr} & {\rm NLO weak corr} \\ \hline
	[	0	 ; 	45	]	&	$	1.15    ~ \times ~ 10^{       0       }$      &       $       0.08    ~ \times ~ 10^{	    0	    }$      & $       0.02    ~ \times ~ 10^{	    0	    }$  \\    \hline
	[	45	 ; 	90	]	&	$	2.27    ~ \times ~ 10^{       0       }$      &       $       0.12    ~ \times ~ 10^{	    0	    }$      & $       0.02    ~ \times ~ 10^{	    0	    }$  \\    \hline
	[	90	 ; 	140]	        &	$	1.88    ~ \times ~ 10^{       0       }$      &       $       0.09    ~ \times ~ 10^{	    0	    }$      & $       0.00    ~ \times ~ 10^{	    0	    }$  \\    \hline
	[140	 ; 	200]	                &	$	9.81    ~ \times ~ 10^{       -1      }$      &       $       0.29    ~ \times ~ 10^{	    0	    }$      & $      -0.01    ~ \times ~ 10^{	    0	    }$  \\    \hline
	[200	 ; 	300]	                &	$	3.67    ~ \times ~ 10^{       -1      }$      &       $      -0.02    ~ \times ~ 10^{	    -1      }$      & $      -0.11    ~ \times ~ 10^{	    -1      }$  \\    \hline
	[300	 ; 	500]	                &	$	4.20    ~ \times ~ 10^{       -2      }$      &       $      -0.13    ~ \times ~ 10^{	    -2      }$      & $      -0.24    ~ \times ~ 10^{	    -2      }$  \\    \hline
	[500	 ; 	$\infty$ ]	        &	$	2.21    ~ \times ~ 10^{       -4      }$      &       $       0.04    ~ \times ~ 10^{	    -4      }$      & $      -0.25    ~ \times ~ 10^{	    -4      }$  \\    \hline 
\end{tabular}
\caption{\label{tab:PTt} The $\PTt$ differential distribution in NLO QCD and the corrections due to NNLO QCD and NLO weak effects in $\ttb$ production at the Tevatron. The NLO and NNLO QCD results are taken from Ref.~\cite{Czakon:2016ckf}.}
\end{center}
\end{table}

In Ref.~\cite{Czakon:2016ckf} the NNLO QCD predictions have been
compared with data from the
D\O~collaboration~\cite{Abazov:2014vga}. We note that, although the
size of the NLO weak corrections is comparable to the NNLO QCD ones in
some of the bins, even the combined effects remain rather small.  As a
result, the inclusion of the NLO weak corrections does not
significantly alter the extent of the agreement of the Standard Model
prediction with the experimental data.

\begin{figure}[ht]
  \includegraphics[scale=0.31]{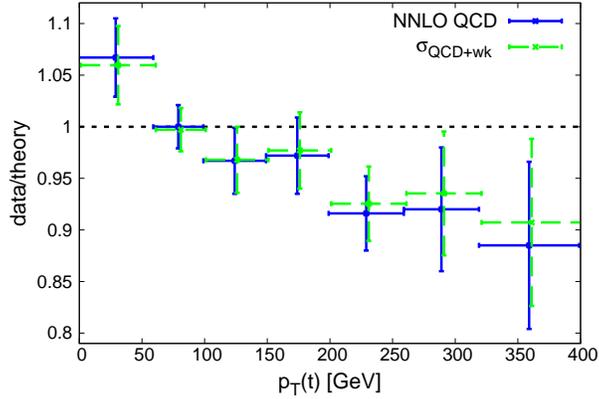}
    \caption{Comparison of NNLO QCD (blue, solid) and combined NNLO QCD+NLO weak (green, dashed)
    predictions for the $p_T(t)$ distribution in $\ttb$ production with 8~TeV CMS data~\cite{Khachatryan:2015oqa}.  The data 
    is divided by the theoretical prediction in each bin of the top quark $p_T$.
    The NNLO QCD predictions are taken from Ref.~\cite{Czakon:2015owf}. 
  \label{fig:pttcms}}
\end{figure}

As we have already observed, the effects of the weak corrections
should be larger at the LHC.  The amount of data collected means that
the ATLAS and CMS collaborations are sensitive to top quarks produced
further above threshold, and the data is subject to significantly
smaller experimental uncertainties.  In order to combine our
calculations with NNLO QCD corrections we use the predictions of
Ref.~\cite{Czakon:2015owf}, which were compared with results from the
CMS collaboration~\cite{Khachatryan:2015oqa}.  These results represent
an analysis of the full $19.6$~fb$^{-1}$ data set taken at $8$~TeV.
The distribution that is most sensitive to the weak corrections, and
for which we can readily extract the effect of NNLO QCD, is the
transverse momentum of the top quarks.  For this analysis we do not
distinguish between top and anti-top quarks, instead including both in
the distribution, and normalize to the total cross-section.  Our
results are shown in Fig.~\ref{fig:pttcms}.  The NNLO QCD comparison
was shown in Ref.~\cite{Czakon:2015owf}.  Here we ameliorate that
analysis by including also the NLO weak corrections, which are simply
added on top of the NNLO predictions according to Eq.~(\ref{eq:QCDweakadd}).
We see that, although the shape
of this distribution is slightly better described throughout, a
difference in shape between the data and NNLO QCD+NLO weak theory
remains.  Since the effect of the weak corrections is rather small the
alternative combination of Eq.~(\ref{eq:QCDweakmult}) would yield almost
identical results.

\subsection{NLO QCD and Weak Corrections to Di-jet Production}\label{sec:dijetcomb}

Almost-complete results for di-jet production at hadron colliders have
recently been presented through NNLO
QCD~\cite{Currie:2013dwa,Currie:2013dza,Currie:2014upa,Pires:2014rxa}.
However, here we restrict ourselves to the NLO QCD results that can be
easily computed using the publicly available Monte-Carlo program {\tt
  MEKS}~\cite{Gao:2012he} (higher-order QCD corrections to di-jet production are not
available in {\tt MCFM}).

\begin{figure}[htpb]
  \includegraphics[scale=0.29]{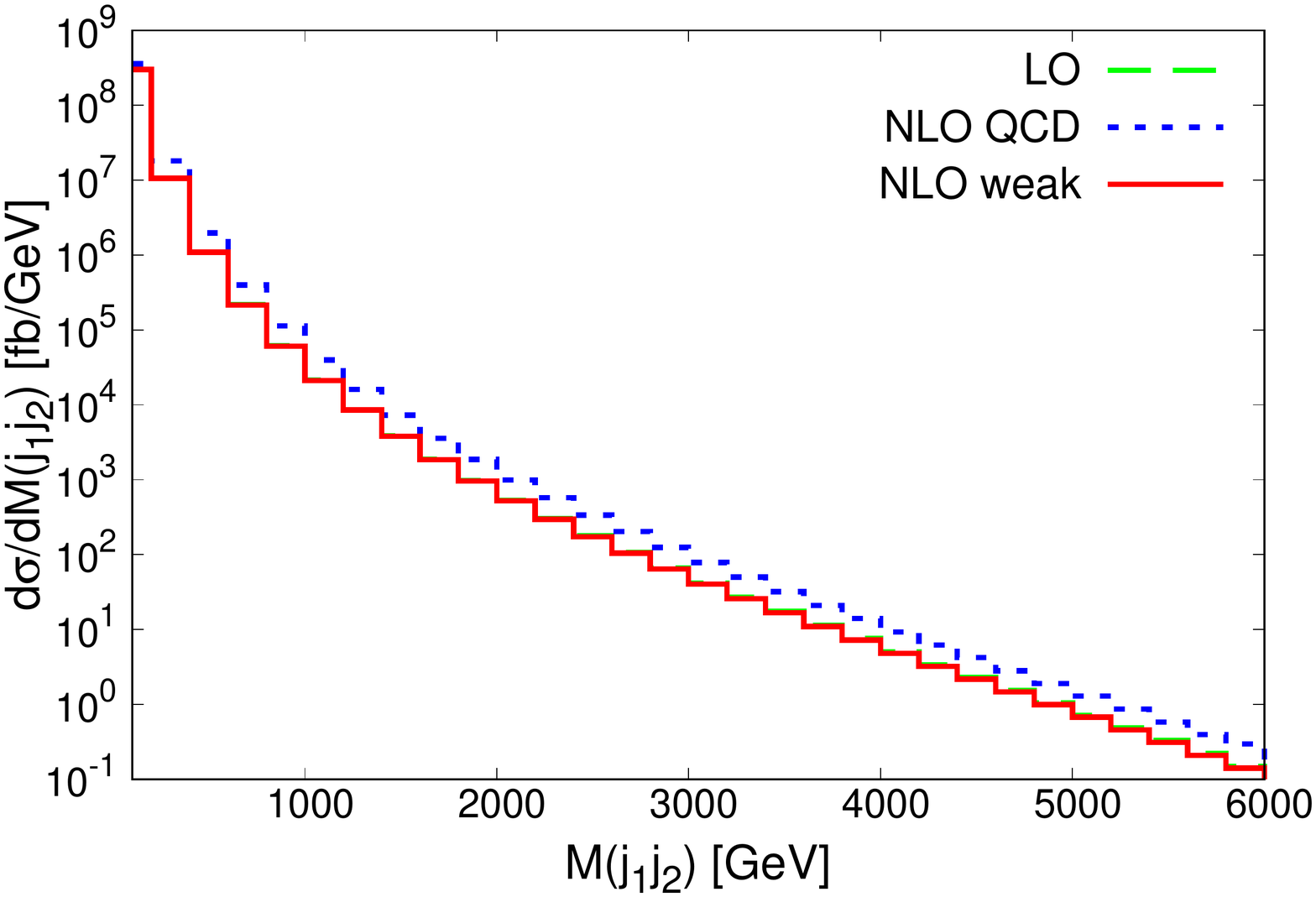}
  \includegraphics[scale=0.29]{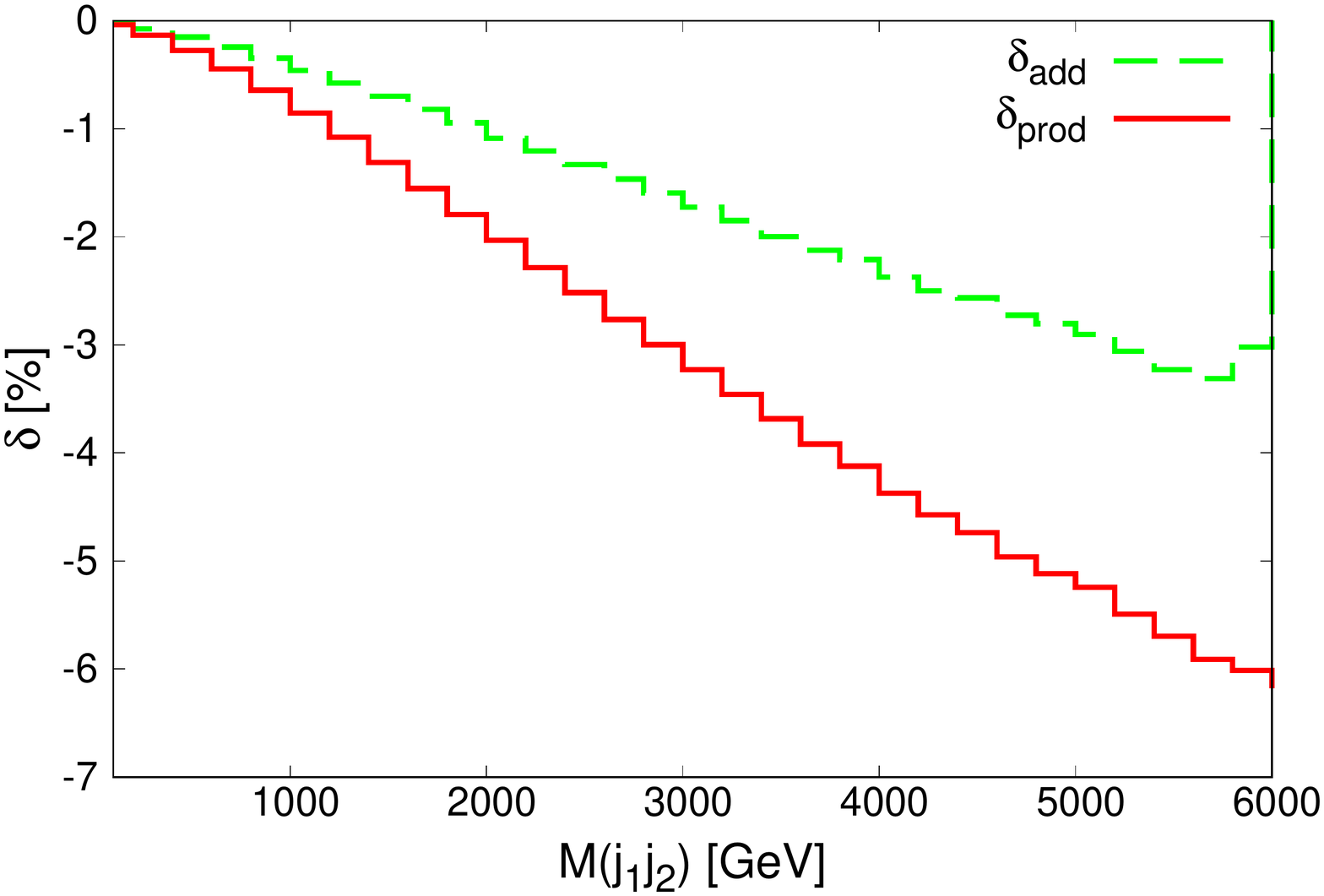}
  \caption{ Left: comparison of the effect of NLO weak (red, solid) and NLO QCD (blue, dotted)
    corrections on the invariant mass distribution of the di-jet pair in di-jet
    production at the 13~TeV LHC.  The LO QCD distribution at ${\cal O}(\alpha_s^2)$ is also shown (green, dashed). Right: a comparison of the two
    procedures 
($\delta_{\add}$ of Eq.~(\ref{eq:dQCDweakadd}) (green, dashed) and $\delta_{\prod}$ of
Eq.~(\ref{eq:dQCDweakmult}) (red, solid))
used to combine NLO QCD and NLO weak effects.  The NLO
    QCD results have been obtained with {\tt
      MEKS} (version 1.0)~\cite{Gao:2012he}.
  \label{fig:dijetcombined}}
\end{figure}

Figure~\ref{fig:dijetcombined}~(left) shows the distribution of the
invariant mass of the two leading jets at LO, at NLO QCD or after
inclusion of NLO weak corrections.  The NLO QCD corrections are rather
mild at small invariant masses, but increase the cross-section by a
factor of around $1.7$ in the multi-TeV range.  This leads to a
substantial difference between $\delta_{\add}$ and $\delta_{\prod}$ in
the tail of the distribution, as shown in
Fig.~\ref{fig:dijetcombined}~(right).  However the size of the
combined correction, in either approach, is relatively small, for
instance in comparison with the impact of the weak corrections in the
NC DY case (Fig.~\ref{fig:DYcombined}).

One of the interesting analyses of di-jet production at the LHC is the
search for new physics beyond the Standard Model through a study of
the scattering angle between the two jets.  The production of jets
through QCD is dominated by small-angle scattering, while additional
interactions, for instance through a contact
term~\cite{Chiappetta:1990jd}, lead to jet production at much wider
angles~\cite{Harris:2011bh}.  Both ATLAS~\cite{ATLAS:2015nsi} and
CMS~\cite{Khachatryan:2014cja,CMS:2015djr} have taken advantage of this observation in
order to place stringent constraints on various models of new
physics.

\begin{figure}[ht]
  \includegraphics[scale=0.29]{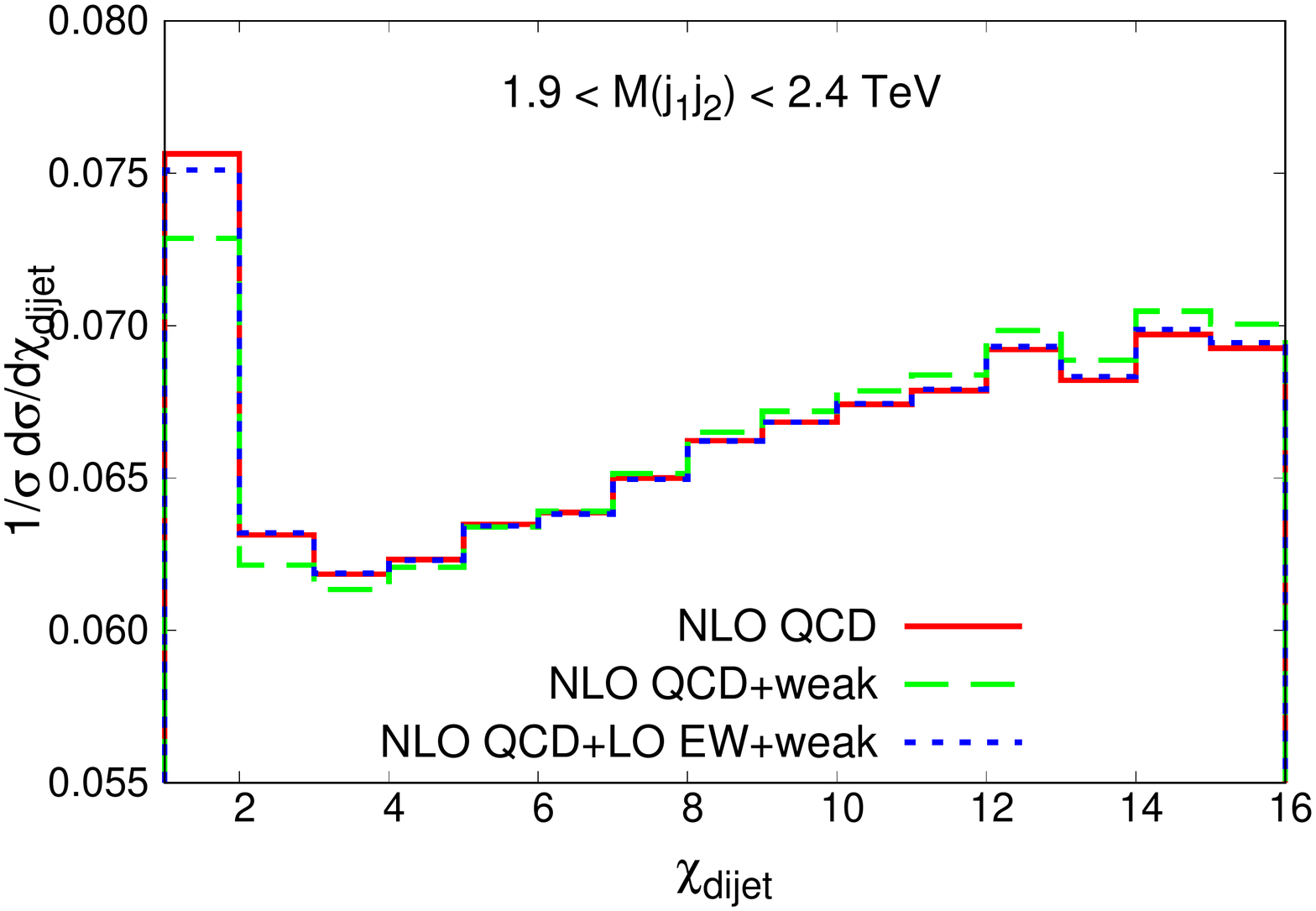}
  \includegraphics[scale=0.29]{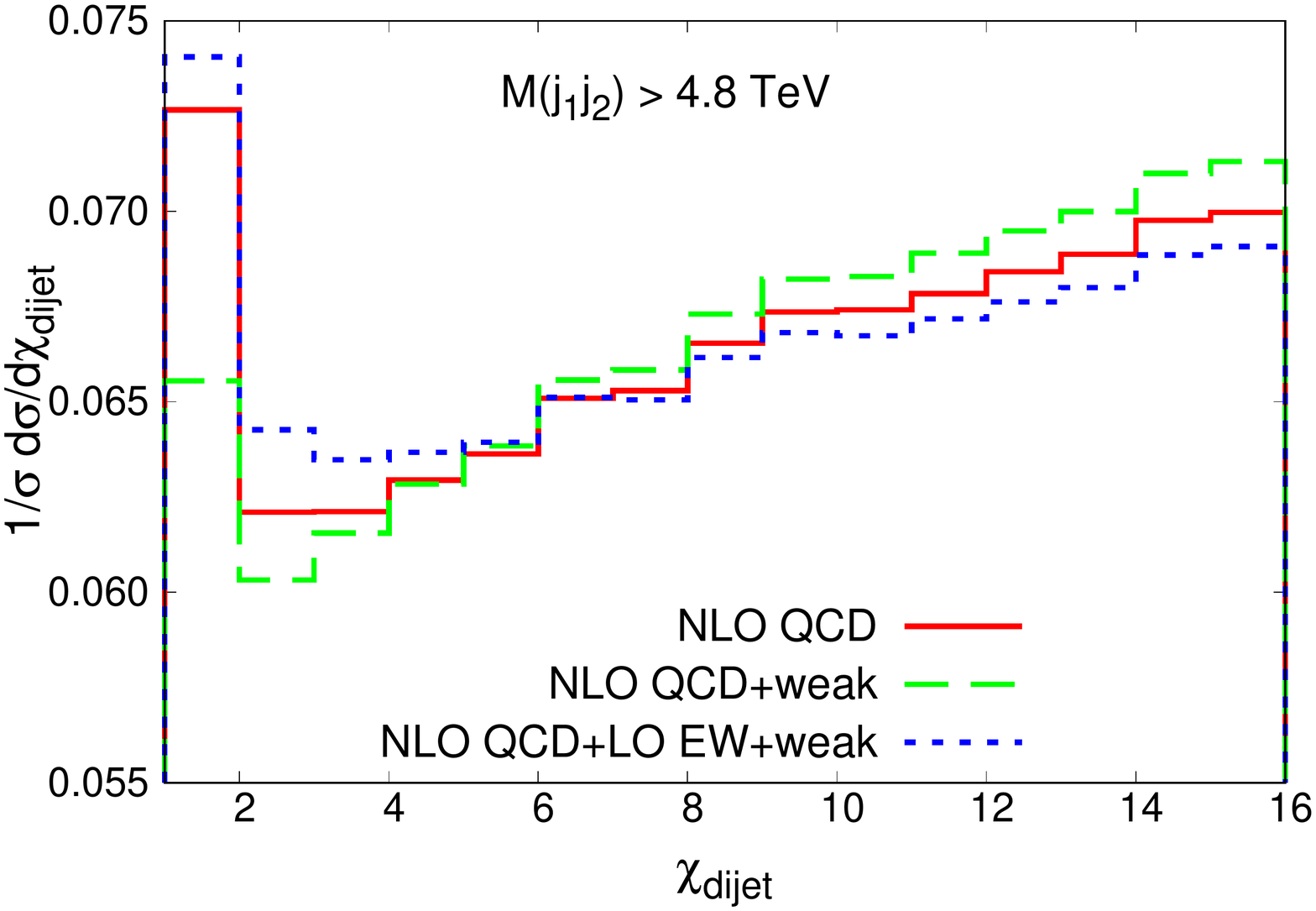}
  \caption{Normalized $\chi_\dijet$ distributions in di-jet production
    at the 13 TeV LHC in a low (left) and high (right) invariant
    di-jet mass bin at NLO QCD (red, solid) and when adding the ${\cal
      O}(\alpha_s^2 \alpha)$ (green, dashed) and LO EW (blue, dotted)
    contributions.  The NLO QCD results have been obtained with {\tt
      MEKS} (version 1.0)~\cite{Gao:2012he}.
    \label{fig:jj:chi1-6}}
\end{figure}

Here we will consider the effect of weak corrections under a set of
cuts used by a recent CMS analysis~\cite{CMS:2015djr}.  The key observable,
$\chi_\dijet$, is simply related both to the scattering angle and to the
rapidity between the two leading jets ($y(j_1,j_2)$),
\begin{equation}
\chi_\dijet = \exp\left(|y(j_1) - y(j_2)|\right) \;.
\end{equation}
We used the CT14 PDF set to produce the result in the same setup as
used by CMS in Ref.~\cite{CMS:2015djr}. We use the anti-$k_T$ jet
algorithm with $R = 0.4$ and apply a cut $y_{\mathrm{boost}} =
\frac{1}{2}|y(j_1)+y(j_2)| < 2.22$.  In Fig.~\ref{fig:jj:chi1-6} we
show the normalized $\chi_\dijet$ distribution for a low and high
invariant di-jet mass bin, also used in the CMS analysis, calculated
at NLO QCD with {\tt MEKS} (version 1.0)~\cite{Gao:2012he} and when
adding the {\tt MCFM} prediction for the LO EW and ${\cal
  O}(\alpha_s^2 \alpha)$ contribution. As expected, the weak one-loop
corrections are most significant in the highest mass bin where the
${\cal O}(\alpha_s^2 \alpha)$ contribution reduces the NLO QCD
distribution by 9.8\% at small values of $\chi_\dijet$. The LO EW
contribution largely cancels the weak corrections so that the overall
effect is an increase of the NLO QCD result by 2\% in the first
$\chi_\dijet$ bin.  Our results are consistent with the findings
presented in the CMS analysis~\cite{CMS:2015djr}, which is based on
the calculation of Ref.~\cite{Dittmaier:2012kx}. 
It is interesting to note that the new physics scenarios
under consideration in Ref.~\cite{CMS:2015djr} have their largest
impact in the high-mass bin for small values of $\chi_\dijet$,
i.~e. exactly in the same kinematic regime where weak corrections
become important.

\section{Conclusions}\label{sec:conclusions}

The role of electroweak corrections in the comparison of future LHC
data with theoretical predictions in the Standard Model is becoming
increasingly important.  As the availability of higher-order
perturbative QCD corrections extends past NLO, to NNLO and beyond, the
resulting cross-sections often suffer from a residual theoretical
uncertainty that is comparable in size to the expected size of
electroweak corrections.  Moreover, as the LHC collects more data it
will begin to probe, with reasonable precision, final states with
energies in the multi-TeV region.  Such configurations receive
one-loop electroweak corrections that are especially enhanced, by
Sudakov logarithms of the form $\alpha
\log^2\left(M_{\rm{final}}/M_W\right)$ with $M_{\rm final}$ being the
invariant mass of the leading pair of final-state particles, so that
including these effects is particularly important.

In this paper we have recomputed one-loop electroweak corrections at
the LHC, to three processes of considerable importance:
Neutral-Current Drell-Yan, top-quark pair and di-jet production.  As
well as performing exact calculations of these corrections, we have
also considered the approximation obtained by retaining only leading
and subleading Sudakov logarithms, following the approach of
Refs.~\cite{Denner:2000jv,Denner:2001gw}.  We have also performed a
detailed comparison of the efficacy of this approximation in order to
glean insight into situations in which it is less effective or fails
altogether.  For the processes at hand, the Sudakov approximation is
excellent for the case of NC DY, less accurate for top-quark pair
production and poor for the di-jet process.

Our calculations have been implemented in the framework of the parton-level Monte
Carlo code {\tt MCFM}, a general purpose program that had previously
been focussed on the calculation of higher-order corrections in QCD.
Although the electroweak calculations considered here have already
been presented in the literature, many of the results have not been
made available in a public code.  The inclusion of these results in a
portable code such as {\tt MCFM} will help to facilitate their use in
experimental analyses, particularly in combination with the NLO and
NNLO QCD corrections that are already available in the same framework.

\begin{figure}[htpb]
  \includegraphics[scale=0.32]{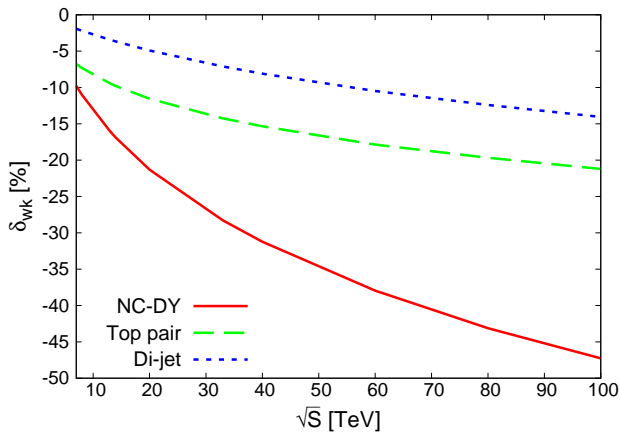}
  \caption{Relative weak corrections $\delta_{\wk}$ of
    Eq.~(\ref{eq:relcorr}) to the total cross sections 
in the high-energy region of the NC DY
    process (red, solid), $\ttb$ (green, dashed) and di-jet (blue,
    dotted) production, as a function of the proton-proton collision
    energy.  The high-energy region is defined by a cut on the
    invariant mass of the leading pair of final-state particles,
    $M_{\rm{final}} > \sqrt{S}/4$.
    \label{fig:sdep}}
\end{figure}

Finally, we note that the proper consideration of electroweak
corrections is even more important for any future hadron colliders
operating at higher energies.  This is illustrated in
Figure~\ref{fig:sdep}, which shows the relative EW correction in the
high-energy region (defined by $M_{\rm{final}} > \sqrt{S}/4$), at a
variety of machine center-of-mass energies ($\sqrt{S}$).  Particularly
in the case of the NC DY process, the inclusion of EW effects is
mandatory in order to have an accurate theoretical prediction for the
high-energy cross section at a $100$~TeV $pp$ machine (see also
Refs.~\cite{Mishra:2013una,Mangano:2016jyj} for recent reviews).

\appendix

\def\nlo{\mathrm{NLO}}
\def\rB{\mathrm{B}}
\def\rV{\mathrm{V}}
\def\rC{\mathrm{C}}
\def\rdip{\mathrm{dipoles}}
\def\qaiijj{q_i\bar{q}_i\ra q_j\bar{q}_j}
\def\ave#1{\langle#1\rangle}
\def\tp{\tilde{p}}

\section*{Appendix:~Real Corrections}\label{app:real}

{\tt MCFM} uses the Catani-Seymour dipole subtraction
method~\cite{Catani:1996vz} (and Ref.~\cite{Catani:2002hc} for massive
partons) to handle the cancellation of soft and collinear
singularities in NLO QCD calculations. For completeness, we present in
the following the explicit expressions we used for the {\tt MCFM}
implementation of the real ${\cal O}(\alpha_s)$ corrections to $\ttb$
and di-jet production at $\Oa$.  Symbolically, the corresponding cross
section can be written as
\begin{equation}\label{eq:master}
  \begin{aligned}
    \si_{real}\brc{p_a,p_b} = &\int_{3}\left[d\si^\rR\brc{p_a,p_b,p_1,p_2,p_3} - \sum_{\rdip}d\si^{\rB}\otimes dV_{\rdip}\brc{p_a,p_b,p_1,p_2,p_3}\right] \\
    & + \int_{2} d\si^{\rB}\brc{p_a,p_b,p_1,p_2}\otimes\bI
    + \int_{2}d\si^{\rC}\brc{p_a,p_b,p_1,p_2} \; ,
  \end{aligned}
\end{equation}
where $p_a$, $p_b$ are the momenta of the partons in the initial state
and $p_1$, $p_2$, $p_3$ are the momenta of the final state
partons. The terms involving a convolution (denoted by the symbol
$\otimes$) represent the dipole subtraction terms and their integrated
versions.  The two are related through the definition
\begin{equation}
  \bI = \sum_{\rdip}\int_{1} d\rV_{\rdip} \;.
\end{equation}
$d\si^R$ denotes the contribution of the real radiation
diagrams. $d\si^C$ contains the PDF counter-terms required to absorb
the remaining collinear singularity into the NLO PDFs, and reads
\begin{equation}
\begin{aligned}
  d\si^\rC_{ab}\brc{p_a,p_b,p_1,p_2} = & -\frac{\alpha_s}{2\pi}\frac{\brc{4\pi}^\epsilon}{\Gamma\brc{1-\epsilon}}
  \sum_{c}\int dx \bigg\{
  \left[
    -\frac{1}{\epsilon}\brc{\frac{\mu^2}{\mu_F^2}}^\epsilon P^{ac}\brc{x} \right.  \\
  & \left. + K^{ac}_{F.S.}\brc{x}
    \right]d\si^\rB_{cb}\brc{xp_a,p_b,p_1,p_2} +  \brc{a\lra b}
  \bigg\} \; , 
\end{aligned}
\end{equation}
where $\mu_F$ is the factorization scale and 
$K^{ac}_{F.S.}\brc{x}$ defines the factorization scheme. In the
\(\overline{\mathrm{MS}}\) scheme $K^{ac}_{F.S.}\brc{x}=0$.
$P^{ac}\brc{x}$ are Altarelli-Parisi probabilities which can be found, for instance,
in Ref.~\cite{Catani:1996vz}.

\subsection{Real corrections to $\ttb$ production}\label{app:realttb}

In $\ttb$ production $d\si^R$ describes the diagrams of
Fig.~\ref{fig:ttb:real} and reads
\begin{equation}\label{eq:ttb:real:head}
d\si^R \propto    \left|\mm^{\qqb\ra\ttb g}\right|^2 = \brc{4\pi}^3\ord\frac{\Nc^2-1}{4}~\frac{g_v^qg_v^t\brc{v_1+v_2}+g_a^qg_a^t\brc{a_1+a_2}}{{\hat{s}_{15}}~{\hat{s}_{25}}~{\hat{s}_{35}}~{\hat{s}_{45}}~{\hat{s}_{12}}~{\hat{s}_{34}}~\brc{\hat{s}_{12}-M_Z^2}~\brc{\hat{s}_{34}-M_Z^2}}
\end{equation}
with
\begin{equation}
  \begin{aligned}
    v_1 = & ~8 \brc{\hat{t}_{1}^2+\hat{t}_{2}^2+\hat{u}_{1}^2+\hat{u}_{2}^2} \big[\hat{s}_{12}^2 \brc{\hat{t}_{1}+\hat{t}_{2}-\hat{u}_{1}-\hat{u}_{2}}+\hat{s}_{12} \brc{\hat{t}_{1}+\hat{t}_{2}-\hat{u}_{1}-\hat{u}_{2}} \brc{\hat{t}_{1}+\hat{t}_{2}+\hat{u}_{1}+\hat{u}_{2}} \\
    & +\brc{\hat{t}_{1}+\hat{t}_{2}+\hat{u}_{1}+\hat{u}_{2}} \brc{\hat{t}_{1} \hat{t}_{2}-\hat{u}_{1} \hat{u}_{2}}\big]~\big[2 \hat{s}_{12} \brc{\hat{s}_{12}+\hat{t}_{1}+\hat{t}_{2}+\hat{u}_{1}+\hat{u}_{2}} -M_Z^2 \brc{\hat{t}_{1}+\hat{t}_{2}+\hat{u}_{1}+\hat{u}_{2}}\big] \\
     v_2 = & ~16~m_t^2 \bigg\{2 \hat{s}_{12}^3 \brc{\hat{t}_{1}+\hat{t}_{2}-\hat{u}_{1}-\hat{u}_{2}}+\hat{s}_{12}^2 \brc{\hat{t}_{1}^2+6 \hat{t}_{1} \hat{t}_{2}+\hat{t}_{2}^2-\hat{u}_{1}^2-6 \hat{u}_{1} \hat{u}_{2}-\hat{u}_{2}^2} \\
     & +\hat{s}_{12} \big[-\hat{t}_{1}^3+\hat{t}_{1}^2 \brc{3 \hat{t}_{2}-\hat{u}_{1}+\hat{u}_{2}} +\hat{t}_{1} \brc{3 \hat{t}_{2}^2+2 \hat{t}_{2} (\hat{u}_{1}+\hat{u}_{2})+\hat{u}_{1}^2-2 \hat{u}_{1} \hat{u}_{2}-\hat{u}_{2}^2} \\
     & -\hat{t}_{2}^3+\hat{t}_{2}^2 \brc{\hat{u}_{1}-\hat{u}_{2}}+\hat{t}_{2} \brc{-\hat{u}_{1}^2-2 \hat{u}_{1} \hat{u}_{2}+\hat{u}_{2}^2} + \brc{\hat{u}_{1}+\hat{u}_{2}} \brc{\hat{u}_{1}^2-4 \hat{u}_{1} \hat{u}_{2}+\hat{u}_{2}^2}\big] \\
     &-\brc{\hat{t}_{1}-\hat{t}_{2}+\hat{u}_{1}-\hat{u}_{2}}^2 \brc{\hat{t}_{1} \hat{t}_{2}-\hat{u}_{1} \hat{u}_{2}}\bigg\} \big[2 \hat{s}_{12} \brc{\hat{s}_{12}+\hat{t}_{1}+\hat{t}_{2}+\hat{u}_{1}+\hat{u}_{2}} -M_Z^2 \brc{\hat{t}_{1}+\hat{t}_{2}+\hat{u}_{1}+\hat{u}_{2}}\big] \\
     a_1 = & -8 \brc{\hat{t}_{1}^2+\hat{t}_{2}^2-\hat{u}_{1}^2-\hat{u}_{2}^2} \big[\hat{s}_{12}^2 \brc{\hat{t}_{1}+\hat{t}_{2}-\hat{u}_{1}-\hat{u}_{2}}+\hat{s}_{12} \brc{\hat{t}_{1}+\hat{t}_{2}-\hat{u}_{1}-\hat{u}_{2}} \brc{\hat{t}_{1}+\hat{t}_{2}+\hat{u}_{1}+\hat{u}_{2}} \\
     & +\brc{\hat{t}_{1}+\hat{t}_{2}+\hat{u}_{1}+\hat{u}_{2}} \brc{\hat{t}_{1} \hat{t}_{2}-\hat{u}_{1} \hat{u}_{2}}\big]~\big[2 \hat{s}_{12} \brc{\hat{s}_{12}+\hat{t}_{1}+\hat{t}_{2}+\hat{u}_{1}+\hat{u}_{2}}-M_Z^2 \brc{\hat{t}_{1}+\hat{t}_{2}+\hat{u}_{1}+\hat{u}_{2}}\big] \\
     a_2 = & -32~ m_t^2 \hat{s}_{12} \brc{\hat{s}_{12}+\hat{t}_{1}+\hat{u}_{1}} \brc{\hat{s}_{12}+\hat{t}_{2}+\hat{u}_{2}} \brc{\hat{t}_{1}+\hat{t}_{2}+\hat{u}_{1}+\hat{u}_{2}} \\
     &\times\brc{2 \hat{s}_{12}+\hat{t}_{1}+\hat{t}_{2}+\hat{u}_{1}+\hat{u}_{2}} \brc{M_Z^2+\hat{s}_{12}+\hat{t}_{1}+\hat{t}_{2}+\hat{u}_{1}+\hat{u}_{2}} \\
    \end{aligned}
\end{equation}
where the Lorentz invariants are defined as
\begin{equation}\label{eq:ttb:real:tail}
  \begin{aligned}
  &\hat{s}_{12} = \brc{p_q+p_{\bar{q}}}^2, \quad \hat{t}_1 = \brc{p_q+p_t}^2,
  \quad \hat{t}_2 = \brc{p_{\bar{q}}+p_{\bar{t}}}^2,
  \quad \hat{u}_1 = \brc{p_q+p_{\bar{t}}}^2,
  \quad \hat{u}_2 = \brc{p_{\bar{q}}+p_t}^2, \\
  & \begin{aligned}
      & \hat{s}_{15} = -\brc{\hat{s}_{12}+\hat{t}_1+\hat{u}_1}, \quad \hat{s}_{25} = -\brc{\hat{s}_{12}+\hat{t}_2+\hat{u}_2}, \quad \hat{s}_{35} = ~\brc{\hat{s}_{12}+\hat{t}_2+\hat{u}_1}, \\
      & \hat{s}_{45} = ~\brc{\hat{s}_{12}+\hat{t}_1+\hat{u}_2}, \quad \hat{s}_{34} = -\brc{\hat{s}_{12}+\hat{t}_1+\hat{t}_2+\hat{u}_1+\hat{u}_2+m_t^2}.
    \end{aligned}
  \end{aligned}
\end{equation}

In this case the unintegrated dipole contribution in Eq.~(\ref{eq:master}) consists
of two dipoles and is given by
\begin{equation}\label{eq:dipsttb}
  \begin{aligned}
    \sum_{\rdip}d\si^{\rB}\otimes dV_{\rdip}  =
    & ~d\Phi_3\brc{p_q,p_{\bar{q}};p_t,p_{\bar{t}},p_g}\frac{1}{S_3}\left\{
      \sum_{k=t,\bar{t}}\sum_{a=q,\bar{q}}\left[\m{D}_{gk}^a\Theta\brc{\alpha_{FI}-1+x_{gk,a}} \right. \right. \\
& \left. \left. +  \m{D}_k^{ag}\Theta\brc{\alpha_{IF}-u_g}\right]
      \right\}
  \end{aligned}
\end{equation}
with
\begin{equation}\label{eq:ttb:dip1}
  \begin{aligned}
    \m{D}_{gt}^q & =  -\frac{1}{\brc{p_g+p_t}^2-m_t^2}\frac{1}{x_{gt,q}}\frac{1}{C_F}
    \langle\tp_q,p_{\bar{q}},\tp_t,p_{\bar{t}}|\bT_q\bT_t\bV_{gt}^q|\tp_q,p_{\bar{q}},\tp_t,p_{\bar{t}}\rangle_{\alpha\alpha_s} \\
    & = \frac{1}{\brc{p_g+p_t}^2-m_t^2}\frac{1}{x_{gt,q}}\frac{1}{C_F}
    \ave{\bV_{gt}^q}\frac{\Nc^2-1}{4}\left|\mm^{\qqb\ra\ttb}\right|^2_{p_q\ra\tp_q,p_t\ra\tp_t}  
  \end{aligned}
\end{equation}
and
\begin{equation}\label{eq:ttb:dip2}
  \begin{aligned}
    \m{D}_t^{qg} & =  -\frac{1}{2p_qp_g}\frac{1}{x_{gt,q}}\frac{1}{C_F}
    \langle\tp_q,p_{\bar{q}},\tp_t,p_{\bar{t}}|\bT_t\bT_q\bV_t^{qg}|\tp_q,p_{\bar{q}},\tp_t,p_{\bar{t}}\rangle_{\alpha\alpha_s} \\
    & = \frac{1}{2p_qp_g}\frac{1}{x_{gt,q}}\frac{1}{C_F}
    \ave{\bV_t^{qg}}\frac{\Nc^2-1}{4}\left|\mm^{\qqb\ra\ttb}\right|^2_{p_q\ra\tp_q,p_t\ra\tp_t}
  \end{aligned}
\end{equation}
These are defined in terms of
\begin{equation}
  \begin{aligned}
  &\ave{\bV_{gt}^q} = 8\pi\alpha_sC_F\brc{\frac{2}{2-x_{gt,q}-\tilde{z}_t}-1-\tilde{z}_t-\frac{m_t^2}{p_gp_t}} \\
    &\ave{\bV_t^{qg}} = 8\pi\alpha_sC_F\brc{\frac{2}{1-x_{gt,q}-u_g}-1-x_{gt,q}}
  \end{aligned}
\end{equation}
where
\[
\begin{aligned}
  &\tilde{p}^{\mu}_q = x_{gt,q}p^{\mu}_q, \quad
  \tilde{p}^{\mu}_t = p^{\mu}_g+p^{\mu}_t-\brc{1-x_{gt,q}}p^{\mu}_q, \\
  &x_{gt,q} = \frac{p_qp_g+p_qp_t+p_gp_t}{p_qp_g+p_qp_t}, \quad
  \tilde{z}_t = \frac{p_qp_t}{p_qp_g+p_qp_t}, \quad
  u_g = \frac{p_gp_q}{p_gp_q+p_tp_q}.
\end{aligned}
\]

\begin{equation}
  \begin{aligned}
  &\m{D}_{gt}^{\bar{q}} = -\m{D}_{gt}^q\left|_{q\lra\bar{q}}\right., \quad
  \m{D}_{g\bar{t}}^q = -\m{D}_{gt}^q\left|_{t\lra\bar{t}}\right., \quad
  \m{D}_{g\bar{t}}^{\bar{q}} = \m{D}_{gt}^q\left|_{q\lra\bar{q}, t\lra\bar{t}}\right. \\
  &\m{D}_t^{g\bar{q}} = -\m{D}_t^{qg}\left|_{q\lra\bar{q}}\right., \quad
  \m{D}_{\bar{t}}^{gq} = -\m{D}_t^{qg}\left|_{t\lra\bar{t}}\right., \quad
  \m{D}_{\bar{t}}^{g\bar{q}} = \m{D}_t^{qg}\left|_{q\lra\bar{q}, t\lra\bar{t}}\right.,
  \end{aligned}
\end{equation}
The occurrence of the minus sign is due to the fact that
\[
\ave{~\cdots~|~\cdots~\bT_{\bar{q},\bar{t}}~\cdots~|~\cdots~} = - ~\ave{~\cdots~|~\cdots~\bT_{q,t}~\cdots~|~\cdots},
\]
and, for the same reason, it will appear again in the case of di-jet production.

The Born matrix elements squared, $\left|\mm^{\qqb\ra\ttb}\right|^2$ ,
used in the subtraction terms are stripped of their color
factors and read 
\begin{equation}
  \left|\mm^{\qqb\ra\ttb}\right|^2 = \brc{4\pi}^2\alpha\alpha_s
  \frac{8}{\hat{s}\brc{\hat{s}-M_Z^2}}\left[
    g_v^qg_v^t\brc{\hat{t}^2+\hat{u}^2+2m_t^2\hat{s}}
    -g_a^qg_a^t\brc{\hat{t}^2-\hat{u}^2}
    \right]
\end{equation}
The corresponding integrated dipoles are
\begin{equation}
  \begin{aligned}
    d\si^{\rB}\brc{p_q,p_{\bar{q}};p_t,p_{\bar{t}}}\otimes\bI = &-\frac{\alpha}{2\pi}\frac{\brc{4\pi}^\epsilon}{\Gamma\brc{1-\epsilon}}
    \int dxd\Phi_2\brc{xp_q,p_{\bar{q}};p_t,p_{\bar{t}}}\frac{1}{C_F} \times \\
    & \sum_{k=t,\bar{t}} \left[ \brc{-\frac{\mu^2}{s_{kq}}}^\epsilon \bI_{k,q}\brc{x,\epsilon,\mu_{kq};\alpha_{FI}}
      + \brc{-\frac{\mu^2}{s_{qk}}}^\epsilon \bI_{q,k}\brc{x,\epsilon,\mu_{qk};\alpha_{IF}} \right] \\ &
    \ave{xp_q,p_{\bar{q}},p_t,p_{\bar{t}}|\bT_a\bT_j|xp_q,p_{\bar{q}},p_t,p_{\bar{t}}}_{\alpha\alpha_s}
    + \brc{q\lra\bar{q}}
  \end{aligned}
\end{equation}
where
\begin{equation}\label{eq:ttb:intdip1}
  \begin{aligned}
    \bI_{t,q}\brc{x,\epsilon,\mu_{tq};\alpha_{FI}} =
    &~\delta\brc{1-x}~C_F~\Bigg[\frac{1}{\epsilon}\brc{1+\log\frac{\mu^2_{tq}}{1+\mu^2_{tq}}} + \log\mu^2_{tq}
      + \frac{1}{2}\log^2\mu^2_{tq} + \frac{1}{2}\log^2\brc{1+\mu^2_{tq}} \\
      &- 2\log\mu^2_{tq}\log\brc{1+\mu^2_{tq}} - 2\Li2\brc{-\mu^2_{tq}} + 2
      - \frac{\pi^2}{3} + 2\log\alpha_{FI}\brc{\log\frac{1+\mu^2_{tq}}{\mu^2_{tq}}-1}
      \Bigg]  \\
    &+\Theta\brc{x-1+\alpha_{FI}}~C_F~\Bigg\{
      \frac{1-x}{2\brc{1-x+x\mu^2_{tq}}^2} + \frac{2}{1-x}\log\frac{\mu^2_{tq}\brc{2-x+x\mu^2_{tq}}}{\brc{1+\mu^2_{tq}}\brc{1-x+x\mu^2_{tq}}} \\
      &+\left[\frac{2}{1-x}\brc{\log\frac{1+\mu^2_{tq}}{\mu^2_{tq}}-1}\right]_+
    \Bigg\}
  \end{aligned}
\end{equation}
and
\begin{equation}\label{eq:ttb:intdip2}
  \begin{aligned}
    \bI_{q,t}\brc{x,\epsilon,\mu_{qt};\alpha_{IF}} =
    &~\delta\brc{1-x}~C_F~\Bigg[
      \frac{1}{\epsilon^2} + \frac{1}{\epsilon}\log\brc{1+\mu^2_{qt}}
      -\frac{1}{2}\log^2\brc{1+\mu^2_{qt}} + 2\log\mu^2_{qt}\log\brc{1+\mu^2_{qt}} \\
      &+2\Li2\brc{-\mu^2_{qt}} + \frac{\pi^2}{6}
      \Bigg]
    -P^{qq}_{reg}\brc{x}\left[
      \frac{1}{\epsilon}-2\log\brc{1-x}+\log{x}+\log\brc{1-x+x\mu^2_{qt}}
      \right] \\
    &+ 1-x -\frac{2}{1-x}\brc{\log{x}+\log\frac{2-x+x\mu^2_{qt}}{1+\mu^2_{qt}}}
     \\
    &-\Theta\brc{z_+-\alpha_{IF}}~C_F~\bigg[\frac{2}{1-x}\log\frac{z_+\brc{1-x+\alpha_{IF}}}{\alpha_{IF}\brc{1-x+z_+}} \\
      &+ P^{qq}_{reg}\brc{x}\log\frac{z_+}{\alpha_{IF}}\bigg] - C_F~\Bigg\{
    \frac{2}{1-x}\left[\frac{1}{\epsilon}-2\log\brc{1-x}+\log\brc{1+\mu^2_{qt}}\right]
    \Bigg\}_+
  \end{aligned}
\end{equation}
with
\begin{equation}
  \begin{aligned}
    & s_{tq}=s_{qt} = 2\tp_tp_q, \quad \mu_{tq}=\mu_{qt}=\frac{m_t}{\sqrt{-2\tp_tp_q}}, \quad z_+ = \frac{1-x}{1-x+x\mu^2_{qt}}. \\
    &\bI_{t,\bar{q}} = \bI_{t,q}\left|_{q\lra\bar{q}}\right., \quad
    \bI_{\bar{t},q} = \bI_{t,q}\left|_{t\lra\bar{t}}\right., \quad
    \bI_{\bar{t},\bar{q}} = \bI_{t,q}\left|_{q\lra\bar{q},t\lra\bar{t}}\right. \\
    &\bI_{q,\bar{t}} = \bI_{q,t}\left|_{t\lra\bar{t}}\right., \quad
    \bI_{\bar{q},t} = \bI_{q,t}\left|_{q\lra\bar{q}}\right., \quad
    \bI_{\bar{q},\bar{t}} = \bI_{q,t}\left|_{q\lra\bar{q},t\lra\bar{t}}\right.
  \end{aligned}
\end{equation}

\subsection{Real corrections to di-jet production}\label{app:realdijet}

In di-jet production at $\Oa$ two sets of real corrections are
calculated directly, those that result from the interference between
$\hat{s}$- and $\hat{s}$-channel and $\hat{s}$- and $\hat{t}$-channel
matrix elements.  The complete real corrections for a given four-quark
or two-gluon-two-quark subprocess can be expressed in terms of
these two after using appropriate crossing relations.  We write the
two types of real correction to the $\qaiijj$ subprocess as
\begin{equation}\label{eq:2j:4q:real}
  \begin{aligned}
  \left|\mm^{\qaiijj g}\right|^2_{\sxs} & = \brc{4\pi}^3\ord\frac{\Nc^2-1}{4}
  \frac{R_{12}A^{\sxs_{12}} + R_{34}A^{\sxs_{34}}}{{\hat{s}_{15}}~{\hat{s}_{25}}~{\hat{s}_{35}}~{\hat{s}_{45}}~{\hat{s}_{12}}~{\hat{s}_{34}}~\brc{\hat{s}_{12}-M_{V^a}^2}~\brc{\hat{s}_{34}-M_{V^a}^2}} \\
  \left|\mm^{\qaiijj g}\right|^2_{\sxt} & = \brc{4\pi}^3\ord\frac{\Nc^2-1}{4}
  \frac{R_{12}A^{\sxt_{12}} + R_{34}A^{\sxt_{34}}}{{\hat{s}_{15}}~{\hat{s}_{25}}~{\hat{s}_{35}}~{\hat{s}_{45}}~\brc{\hat{s}_{12}-M_{V^a}^2}~\brc{\hat{s}_{34}-M_{V^a}^2}\hat{t}_1\hat{t}_2}
  \end{aligned}
\end{equation}
with
\begin{equation}
  R_{12} = \frac{\brc{\hat{s}_{12}-M_{V^a}^2}^2}{\brc{\hat{s}_{12}-M_{V^a}^2}^2 + \Gamma_{V^a}^2M_{V^a}^2}, \qquad
  R_{34} = \frac{\brc{\hat{s}_{34}-M_{V^a}^2}^2}{\brc{\hat{s}_{34}-M_{V^a}^2}^2 + \Gamma_{V^a}^2M_{V^a}^2}
\end{equation}
and 
\begin{equation}
  \begin{aligned}
    A^{\sxs_{12}} = &~16\hat{s}_{12} \brc{\hat{s}_{34}-M_{v^a}^2} \bigg[\hat{s}_{12}^2 \brc{\hat{t}_{1}+\hat{t}_{2}-\hat{u}_{1}-\hat{u}_{2}}+\hat{s}_{12} \brc{\hat{t}_{1}+\hat{t}_{2}-\hat{u}_{1}-\hat{u}_{2}} \brc{\hat{t}_{1}+\hat{t}_{2}+\hat{u}_{1}+\hat{u}_{2}} \\
    & +\brc{\hat{t}_{1}+\hat{t}_{2}+\hat{u}_{1}+\hat{u}_{2}} \brc{\hat{t}_{1} \hat{t}_{2}-\hat{u}_{1} \hat{u}_{2}}\bigg] \bigg[g_a^i g_a^f \brc{\hat{t}_{1}^2+\hat{t}_{2}^2-\hat{u}_{1}^2-\hat{u}_{2}^2}-g_v^i g_v^f \brc{\hat{t}_{1}^2+\hat{t}_{2}^2+\hat{u}_{1}^2+\hat{u}_{2}^2}\bigg] \\
    A^{\sxs_{34}} = &~16\brc{\hat{s}_{12} - M_{V^a}^2} \hat{s}_{34} \bigg[\hat{s}_{12}^2 \brc{\hat{t}_{1}+\hat{t}_{2}-\hat{u}_{1}-\hat{u}_{2}}+\hat{s}_{12} \brc{\hat{t}_{1}+\hat{t}_{2}-\hat{u}_{1}-\hat{u}_{2}} \brc{\hat{t}_{1}+\hat{t}_{2}+\hat{u}_{1}+\hat{u}_{2}} \\
    & +\brc{\hat{t}_{1}+\hat{t}_{2}+\hat{u}_{1}+\hat{u}_{2}} \brc{\hat{t}_{1} \hat{t}_{2}-\hat{u}_{1} \hat{u}_{2}}\bigg] \bigg[g_a^i g_a^f \brc{\hat{t}_{1}^2+\hat{t}_{2}^2-\hat{u}_{1}^2-\hat{u}_{2}^2}-g_v^i g_v^f \brc{\hat{t}_{1}^2+\hat{t}_{2}^2+\hat{u}_{1}^2+\hat{u}_{2}^2}\bigg] \\
    A^{\sxt_{12}} = &~\frac{16}{3} \brc{\hat{s}_{34} - M_{V^a}^2} \brc{\hat{u}_{1}^2+\hat{u}_{2}^2} \brc{g_a^i g_a^f+g_v^i g_v^f} \bigg\{9 \hat{s}_{12}^4+\hat{s}_{12}^3 \left[17 \hat{t}_{1}+17 \hat{t}_{2}+18 \brc{\hat{u}_{1}+\hat{u}_{2}}\right] \\
    & +\hat{s}_{12}^2 \left[8 \hat{t}_{1}^2+\hat{t}_{1} \brc{18 \hat{t}_{2}+17 \hat{u}_{1}+25 \hat{u}_{2}}+8 \hat{t}_{2}^2+\hat{t}_{2} \brc{25 \hat{u}_{1}+17 \hat{u}_{2}}+9 \brc{\hat{u}_{1}^2+4 \hat{u}_{1} \hat{u}_{2}+\hat{u}_{2}^2}\right] \\
    & +\hat{s}_{12} \big[\hat{t}_{2} \brc{\hat{t}_{1}^2+\hat{t}_{1} \brc{\hat{t}_{2}+8 \hat{u}_{1}}+8 \hat{u}_{1} \brc{\hat{t}_{2}+\hat{u}_{1}}}+\hat{u}_{2} \brc{25 \hat{u}_{1} \brc{\hat{t}_{1}+\hat{t}_{2}}+8 \hat{t}_{1} \brc{\hat{t}_{1}+\hat{t}_{2}}+18 \hat{u}_{1}^2} \\
    & +2 \hat{u}_{2}^2 \brc{4 \hat{t}_{1}+9 \hat{u}_{1}}\big]-\brc{\hat{t}_{1} \hat{t}_{2}-\hat{u}_{1} \hat{u}_{2}} \brc{7 \hat{t}_{1} \hat{t}_{2}+8 \hat{t}_{1} \hat{u}_{2}+8 \hat{t}_{2} \hat{u}_{1}+9 \hat{u}_{1} \hat{u}_{2}}\bigg\} \\
    A^{\sxt_{34}} = &~\frac{16}{3} \brc{\hat{s}_{12} - M_{V^a}^2} \brc{\hat{u}_{1}^2+\hat{u}_{2}^2} \brc{g_a^i g_a^f+g_v^i g_v^f} \bigg\{9 \hat{s}_{12}^4+\hat{s}_{12}^3 \left[19 \hat{t}_{1}+19 \hat{t}_{2}+18 \brc{\hat{u}_{1}+\hat{u}_{2}}\right] \\
    & +\hat{s}_{12}^2 \left[11 \hat{t}_{1}^2+4 \hat{t}_{1} \brc{6 \hat{t}_{2}+7 \hat{u}_{1}+5 \hat{u}_{2}}+11 \hat{t}_{2}^2+4 \hat{t}_{2} \brc{5 \hat{u}_{1}+7 \hat{u}_{2}}+9 \brc{\hat{u}_{1}^2+4 \hat{u}_{1} \hat{u}_{2}+\hat{u}_{2}^2}\right] \\
    & +\hat{s}_{12} \big[\hat{t}_{1}^3+2 \hat{t}_{1}^2 \brc{3 \hat{t}_{2}+5 \hat{u}_{1}+\hat{u}_{2}}+\hat{t}_{1} \brc{6 \hat{t}_{2}^2+16 \hat{t}_{2} \brc{\hat{u}_{1}+\hat{u}_{2}}+9 \hat{u}_{1}^2+29 \hat{u}_{1} \hat{u}_{2}+\hat{u}_{2}^2} \\
    & +9 \hat{u}_{2}^2 \brc{\hat{t}_{2}+2 \hat{u}_{1}}+\hat{u}_{2} \brc{\hat{t}_{2}+2 \hat{u}_{1}} \brc{10 \hat{t}_{2}+9 \hat{u}_{1}}+\hat{t}_{2} \brc{\hat{t}_{2}+\hat{u}_{1}}^2\big]+\hat{u}_{2} \big[\hat{t}_{1}^2 \brc{3 \hat{t}_{2}+\hat{u}_{1}} \\
    & -\hat{t}_{1} \brc{5 \hat{t}_{2}-9 \hat{u}_{1}} \brc{\hat{t}_{2}+\hat{u}_{1}}+\hat{t}_{2} \hat{u}_{1} \brc{\hat{t}_{2}+\hat{u}_{1}}\big]+\hat{t}_{1} \hat{t}_{2} \left[\hat{t}_{1}^2-5 \hat{t}_{1} \brc{\hat{t}_{2}+\hat{u}_{1}}+\brc{\hat{t}_{2}+\hat{u}_{1}} \brc{\hat{t}_{2}+2 \hat{u}_{1}}\right] \\
    & +\hat{u}_{2}^2 \left[\hat{t}_{1} (2 \hat{t}_{2}+\hat{u}_{1})+9 \hat{u}_{1} \brc{\hat{t}_{2}+\hat{u}_{1}}\right]\bigg\} \; ,
  \end{aligned}
\end{equation}
where $g_{v}^f(g_a^f)$ denote the vector(axial) vector coupling, $M_{V^a}$
the mass, and $\Gamma_{V^a}$ the total decay width of the weak gauge
boson.
The Lorentz invariants are defined as
\begin{equation}
    \begin{aligned}
  &\hat{s}_{12} = \brc{p_{q_i}+p_{\bar{q}_i}}^2, \quad \hat{t}_1 = \brc{p_{q_i}+p_{q_j}}^2,
  \quad \hat{t}_2 = \brc{p_{\bar{q}_i}+p_{\bar{q}_j}}^2,
  \quad \hat{u}_1 = \brc{p_{q_i}+p_{\bar{q}_j}}^2,
  \quad \hat{u}_2 = \brc{p_{\bar{q}_i}+p_{q_j}}^2, \\
  & \begin{aligned}
      & \hat{s}_{15} = -\brc{\hat{s}_{12}+\hat{t}_1+\hat{u}_1}, \quad \hat{s}_{25} = -\brc{\hat{s}_{12}+\hat{t}_2+\hat{u}_2}, \quad \hat{s}_{35} = ~\brc{\hat{s}_{12}+\hat{t}_2+\hat{u}_1}, \\
      & \hat{s}_{45} = ~\brc{\hat{s}_{12}+\hat{t}_1+\hat{u}_2}, \quad \hat{s}_{34} = -\brc{\hat{s}_{12}+\hat{t}_1+\hat{t}_2+\hat{u}_1+\hat{u}_2}.
    \end{aligned}
  \end{aligned}
\end{equation}

The real corrections to the subprocesses with quark-gluon initial
states can be obtained from Eq.~(\ref{eq:2j:4q:real}) by applying the
following crossing symmetries:
\begin{equation}\label{eq:2j:2g2q:real}
  \begin{aligned}
    &\left|\mm^{q_ig\ra q_iq_j\bar{q}_j}\right|^2 = -\left|\mm^{\qaiijj g}\right|^2_{s_{12}\ra t_1,~t_1\ra u_1,~t_2\ra s_{35},~u_1\ra s_{15},~u_2\ra s_{34}} \\
    &\left|\mm^{\bar{q}_ig\ra \bar{q}_iq_j\bar{q}_j}\right|^2 = -\left|\mm^{\qaiijj g}\right|^2_{s_{12}\ra t_1,~t_1\ra s_{35},~t_2\ra u_1,~u_1\ra s_{34},~u_2\ra s_{15}} \\
    &\left|\mm^{gq_i\ra q_jq_i\bar{q}_j}\right|^2 = -\left|\mm^{\qaiijj g}\right|^2_{s_{12}\ra s_{35},~t_1\ra s_{45},~t_2\ra u_2,~u_1\ra s_{25},~u_2\ra s_{34}} \\
    &\left|\mm^{g\bar{q}_i\ra \bar{q}_j\bar{q}_iq_j}\right|^2 = -\left|\mm^{\qaiijj g}\right|^2_{s_{12}\ra s_{35},~t_1\ra u_2,~t_2\ra s_{45},~u_1\ra s_{34},~u_2\ra s_{25}}
  \end{aligned}
\end{equation}

The unintegrated dipole contribution in Eq.~(\ref{eq:master}) for the
$\qaiijj$ subprocesses consist of four dipoles that are
written as follows:
\begin{equation}\label{eq:dipsdijet1}
  \begin{aligned}
    \sum_{\rdip}d\si^{\rB}\otimes dV_{\rdip} =
    & ~d\Phi_3\brc{p_{q_i},p_{\bar{q}_i};p_{q_j},p_{\bar{q}_j},p_g}\frac{1}{S_3}\bigg\{
    \sum_{k,l=q_j,\bar{q}_j}^{k\ne l}\m{D}_{gk,l}\Theta\brc{\alpha_{FF}-y_{gk,l}} \\
    & + \sum_{k=q_j,\bar{q}_j}\sum_{a=q_i,\bar{q}_i}\left[
      \m{D}_{gk}^a\Theta\brc{\alpha_{FI}-1+x_{gk,a}}
      + \m{D}_k^{ag}\Theta\brc{\alpha_{IF}-u_g}\right] \\
    & + \sum_{a=q_i,\bar{q}_i}\sum_{b=q_i,\bar{q}_i}^{a\ne b}\m{D}^{ag,b}\Theta\brc{\alpha_{II}-\tilde{v}_g}
      \bigg\}
  \end{aligned}
\end{equation}
with
\def\ave#1{\langle#1\rangle}
\def\tp{\tilde{p}}
\begin{eqnarray}\label{eq:dijet:dipole:head}
  \m{D}_{gq_j,\bar{q}_j} &=&
  -\frac{1}{2p_gp_{q_j}}\frac{1}{C_F}\ave{\bV_{gq_j,\bar{q}_j}}
    \langle p_{q_i},p_{\bar{q}_i},\tp_{q_j},\tp_{\bar{q}_j}|\bT_{\bar{q}_j}\bT_{q_j}|p_{q_i},p_{\bar{q}_i},\tp_{q_j},\tp_{\bar{q}_j}\rangle_{\alpha\alpha_s} \nonumber \\
  \m{D}_{gq_j}^{q_i} & =&
  -\frac{1}{2p_gp_{q_j}}\frac{1}{x_{gq_j,q_i}}\frac{1}{C_F}
    \ave{\bV_{gq_j}^{q_i}}\ave{\tp_{q_i},p_{\bar{q}_i},\tp_{q_j},p_{\bar{q}_j}|\bT_{q_i}\bT_{q_j}|\tp_{q_i},p_{\bar{q}_i},\tp_{q_j},p_{\bar{q}_j}}_{\alpha\alpha_s}
\nonumber \\
  \m{D}^{q_ig}_{q_j} & =&  
\frac{1}{2p_{q_i}p_g}\frac{1}{x_{gq_j,q_i}}\frac{1}{C_F}\ave{\bV^{q_ig}_{q_j}}
  \ave{\tp_{q_i},p_{\bar{q}_i},\tp_{q_j},p_{\bar{q}_j}|\bT_{q_j}\bT_{q_i}|\tp_{q_i},p_{\bar{q}_i},\tp_{q_j},p_{\bar{q}_j}}_{\alpha\alpha_s}
\nonumber \\
  \m{D}^{q_ig,\bar{q}_i}& =&
  \frac{1}{2p_{q_i}p_g}\frac{1}{x_{g,q_i\bar{q}_i}}\frac{1}{C_F}\ave{\bV^{q_ig,\bar{q}_i}}
  \ave{\tp_{q_i},p_{\bar{q}_{i}},\tp_{q_j},\tp_{\bar{q}_j}|\bT_{\bar{q}_i}\bT_{q_i}|\tp_{q_i},p_{\bar{q}_{i}},\tp_{q_j},\tp_{\bar{q}_j}}_{\alpha\alpha_s} \; .
\end{eqnarray}
These are defined in terms of
\begin{equation}
  \begin{aligned}
  &\ave{\bV_{gq_j,\bar{q}_j}} = 8\pi\alpha_sC_F
  \brc{
    \frac{2}{1-\tilde{z}_{q_j}\brc{1-y_{gq_j,\bar{q}_j}}} - 1-\tilde{z}_{q_j}
  }, \\
  &\tp_{q_j}^\mu = p_{q_j}^\mu + p_g^\mu - \frac{y_{gq_j,\bar{q}_j}}{1-y_{gq_j,\bar{q}_j}}p_{\bar{q}_j}, \quad
  \tp_{\bar{q}_j} = \frac{1}{1-y_{gq_j,\bar{q}_j}}p_{\bar{q}_j}, \\
  &y_{gq_j,\bar{q}_j} = \frac{p_gp_{q_j}}{p_gp_{q_j}+p_{q_j}p_{\bar{q}_j}+p_{\bar{q}_j}p_g}, \quad
  \tilde{z}_{q_j} = \frac{p_{q_j}p_{\bar{q}_j}}{p_gp_{\bar{q}_j}+p_{q_j}p_{\bar{q}_j}}. 
  \end{aligned}
\end{equation}

\begin{equation}
  \begin{aligned}
  &\ave{\bV_{gq_j}^{q_i}} = 8\pi\alpha_sC_F
  \brc{
    \frac{2}{2-x_{gq_j,q_i}-\tilde{z}_{q_j}} - 1 - \tilde{z}_{q_j}
  }, \\
  &\tp_{q_i}^\mu = x_{gq_j,q_i}p_{q_i}^\mu, \quad
  \tp_{q_j}^\mu = p_{q_j}^\mu + p_g^\mu - \brc{1-x_{gq_j,q_i}}p_{q_i}^\mu, \\
  &x_{gq_j,q_i} = \frac{p_{gq_j,q_i}p_{q_i}+p_gp_{q_i}+p_{q_j}p_g}{p_{q_j}p_{q_i}+p_gp_{q_i}}, \quad
  \tilde{z}_{q_j} = \frac{p_{q_j}p_{q_i}}{p_{q_j}p_{q_i}+p_gp_{q_i}}.
  \end{aligned}
\end{equation}

\begin{equation}
  \begin{aligned}
    &\ave{\bV_{q_j}^{q_ig}} = 8\pi\alpha_sC_F
    \brc{
      \frac{2}{1-x_{gq_j,q_g}-u_g}-1-x_{gq_j,q_i}
    }, \\
    &\tp_{q_i}^\mu = x_{gq_j,q_g}p_{q_i}^\mu, \quad
    \tp_{q_j}^\mu = p_{q_j}^\mu + p_g^\mu - \brc{1-x_{gq_j,q_g}}p_{q_i}^\mu, \\
    &x_{gq_j,q_g} = \frac{p_{q_j}p_{q_i} + p_gp_{q_i} + p_gp_{q_j}}{p_{q_j}p_{q_i}+p_gp_{q_i}}, \quad
    u_g = \frac{p_gp_{q_i}}{p_gp_{q_i} + p_{q_j}p_{q_i}}.
  \end{aligned}
\end{equation}

\begin{equation}
  \begin{aligned}
    &\ave{\bV^{q_ig,\bar{q}_i}} = 8\pi\alpha_sC_F
    \brc{
      \frac{2}{1-x_{g,q_i\bar{q}_i}} - 1 - x_{g,q_i\bar{q}_i}
    }, \\
    &\tp_{q_i}^\mu = x_{g,q_i\bar{q}_i}p_{q_i}^\mu, \quad
    \tp_{q_j(\bar{q}_j)}^\mu = p_{q_j(\bar{q}_j)}^\mu - \frac{2p_{q_j(\bar{q}_j)}\cdot\brc{K+\tilde{K}}}{\brc{K+\tilde{K}}^2}\brc{K+\tilde{K}}^\mu + \frac{2p_{q_j(\bar{q}_j)}\cdot K}{K^2}\tilde{K}^\mu, \\
    &x_{g,q_i\bar{q}_i} = \frac{p_{q_i}p_{\bar{q}_i}+p_gp_{q_i}+p_gp_{\bar{q}_i}}{p_{q_i}p_{\bar{q}_i}}, \quad
    K^\mu = p_{q_i}^\mu + p_{\bar{q}_i}^\mu + p_g^\mu, \quad
    \tilde{K}^\mu = \tp_{q_i}^\mu + p_{\bar{q}_i}^\mu, \quad \tilde{v}_g = -\frac{p_{q_i}p_g}{p_{q_i}p_{\bar{q}_i}}
  \end{aligned}
\end{equation}

The remaining dipole contributions can be obtained via the relations
\begin{equation}
  \begin{aligned}
    &\m{D}_{g\bar{q}_j,q_j} = \m{D}_{gq_j,\bar{q}_j}\left|_{q_j\lra\bar{q}_j}\right., \quad
     \m{D}^{\bar{q}_ig,q_i} = \m{D}^{q_ig,\bar{q}_i}\left|_{q_i\lra\bar{q}_i}\right.\\
    &\m{D}_{gq_j}^{\bar{q}_i} = -\m{D}_{gq_j}^{q_i}\left|_{q_i\lra\bar{q}_i}\right., \quad
    \m{D}_{g\bar{q}_j}^{q_i} = -\m{D}_{gq_j}^{q_i}\left|_{q_j\lra\bar{q}_j}\right., \quad
    \m{D}_{g\bar{q}_j}^{\bar{q}_i} = \m{D}_{gq_j}^{q_i}\left|_{q_i\lra\bar{q}_i, q_j\lra\bar{q}_j}\right. \\
    &\m{D}_{q_j}^{\bar{q}_ig} = -\m{D}_{q_j}^{q_ig}\left|_{q_i\lra\bar{q}_i}\right., \quad
    \m{D}_{\bar{q}_j}^{q_ig} = -\m{D}_{q_j}^{q_ig}\left|_{q_j\lra\bar{q}_j}\right., \quad
    \m{D}_{\bar{q}_j}^{\bar{q}_ig} = \m{D}_{q_j}^{q_ig}\left|_{q_i\lra\bar{q}_i, q_j\lra\bar{q}_j}\right. 
  \end{aligned}
\end{equation}

The following Born matrix elements squared stripped of their color
factors are to be used in these subtraction terms:
\begin{equation}
  \begin{aligned}
    \left|\mm^{\qaiijj}\right|^2_{\sxs} & = \brc{4\pi}^2\alpha\alpha_s
    \frac{\mathrm{prop}_{V^a}\brc{\hat{s}}}{\hat{s}^2}~8~\left[
      g_v^ig_v^f\brc{\hat{t}^2+\hat{u}^2} - g_a^ig_a^f\brc{\hat{t}^2-\hat{u}^2}
      \right] \\
    \left|\mm^{\qaiijj}\right|^2_{\sxt} & = \brc{4\pi}^2\alpha\alpha_s
    \frac{\mathrm{prop}_{V^a}\brc{\hat{s}}}{\hat{s}\hat{t}}~8~\brc{g_v^ig_v^f+g_a^ig_a^f}\hat{u}^2
  \end{aligned}
\end{equation}
with $\mathrm{prop}_{V^a}(\hat s)$ of Eq.~(\ref{eq:2j:propx}).
The integrated dipoles are combined according to
\begin{equation}
  \begin{aligned}
    d\si^{\rB}\brc{p_{q_i},p_{\bar{q}_i};p_{q_j},p_{\bar{q}_j}}\otimes\bI = &-\frac{\alpha}{2\pi}\frac{\brc{4\pi}^\epsilon}{\Gamma\brc{1-\epsilon}}\int dxd\Phi_2\brc{xp_{q_i},p_{\bar{q}_i};p_{q_j},p_{\bar{q}_j}}\frac{1}{C_F}\times\Bigg\{ \\
    & \brc{\frac{\mu^2}{s_{q_i\bar{q}_i}}}^\epsilon \bI_{q_i,\bar{q}_i}\brc{x,\epsilon;\alpha_{II}}
      + \sum_{k,l=q_j,\bar{q}_j}^{k\ne l}\brc{\frac{\mu^2}{s_{kl}}}^\epsilon \bI_{k,l}\brc{x,\epsilon;\alpha_{FF}} \\
    &+\sum_{k=q_j,\bar{q}_j}\bigg[
      \brc{-\frac{\mu^2}{s_{kq_i}}}^\epsilon \bI_{k,q_i}\brc{x,\epsilon;\alpha_{FI}} 
    + \brc{-\frac{\mu^2}{s_{q_ik}}}^\epsilon \bI_{q_i,k}\brc{x,\epsilon;\alpha_{IF}}
      \bigg] \Bigg\} \\
    & \times \ave{xp_{q_i},p_{\bar{q}_i},p_{q_j},p_{\bar{q}_j}|\bT_k\bT_l|xp_{q_i},p_{\bar{q}_i},p_{q_j},p_{\bar{q}_j}}_{\alpha\alpha_s}
    + \brc{q\lra\bar{q}}
  \end{aligned}
\end{equation}
where the four types of contribution can be written as
\begin{equation}\label{eq:dijet:dipole:tail}
  \begin{aligned}
  \bI_{q_i,\bar{q}_i}\brc{x,\epsilon;\alpha_{II}} =
  &~C_F\Bigg\{\delta\brc{1-x}\brc{\frac{1}{\epsilon^2} - \frac{\pi^2}{6}}
  + 1-x + \brc{1+x}\left[\frac{1}{\epsilon}-2\log\brc{1-x}\right] \\
  &-\frac{1+x^2}{1-x}\big[\log{x} - \log\alpha_{II}\brc{x}\big]
  -\bigg\{\frac{2}{1-x}\left[\frac{1}{\epsilon}-2\log\brc{1-x}\right]\bigg\}_+
  \Bigg\}
  \end{aligned}
\end{equation}
\begin{equation}
  \begin{aligned}
    \bI_{q_j,\bar{q}_j}\brc{x,\epsilon;\alpha_{FF}} =
    ~C_F\Bigg\{\delta\brc{1-x}\left[
      \frac{1}{\epsilon^2} + \frac{3}{2\epsilon} + \frac{7}{2} - \frac{\pi^2}{2}
      + \frac{3}{2}\brc{\alpha_{FF}-\log\alpha_{FF}} - \log^2\alpha_{FF}
      \right] \Bigg\}
  \end{aligned}
\end{equation}
\begin{equation}
  \begin{aligned}
    \bI_{q_j,q_i}\brc{x,\epsilon;\alpha_{FI}} =
   & ~C_F\Bigg\{\delta\brc{1-x}\left[\frac{1}{\epsilon^2} + \frac{3}{2\epsilon}
      + \frac{7}{2} - \frac{\pi^2}{2}
      - \log\alpha_{FI}\brc{\log\alpha_{FI} + \frac{3}{2}}\right] \\
   & + \Theta\brc{x-1+\alpha_{FI}}\bigg\{
    \frac{2}{1-x}\log\brc{2-x} -\frac{3}{2}\brc{\frac{1}{1-x}}_+
    - \left[\frac{2}{1-x}\log\brc{1-x}\right]_+
    \bigg\} \Bigg\}
  \end{aligned}
\end{equation}
\begin{equation}
  \begin{aligned}
    \bI_{q_i,q_j}\brc{x,\epsilon;\alpha_{IF}} =
    &~C_F\Bigg\{\delta\brc{1-x}\brc{\frac{1}{\epsilon^2} + \frac{\pi^2}{6}}
    + \brc{1+x}\left[\frac{1}{\epsilon} - \log\brc{1-x}\right]
    + 1-x - \frac{1+x^2}{1-x}\log{x} \\
    &- \frac{2}{1-x}\log\frac{1-x+\alpha_{IF}}{\alpha_{IF}}
    - \brc{1+x}\log\alpha_{IF}
    - \bigg\{\frac{2}{1-x}\left[\frac{1}{\epsilon}-2\log\brc{1-x}\right]
    \bigg\}_+ \Bigg\}
  \end{aligned}
\end{equation}
\begin{equation}
  \begin{aligned}
    &\bI_{\bar{q}_j,q_j} = \bI_{q_j,\bar{q_j}}, \quad \bI_{\bar{q}_i,q_i} = \bI_{q_i,\bar{q}_i} \\
    &\bI_{q_j,\bar{q}_i} = \bI_{\bar{q}_j,q_i} = \bI_{\bar{q}_j,\bar{q}_i} = \bI_{q_j,q_i} \\
    &\bI_{q_i,\bar{q}_j} = \bI_{\bar{q}_i,q_j} = \bI_{\bar{q}_i,\bar{q}_j} = \bI_{q_i,q_j}
  \end{aligned}
\end{equation}
The $\alpha_{II}(x)$ function is defined as follows, 
\begin{equation}\label{eq:al}
\alpha_{II}\brc{x} = \mathrm{min}\left\{\frac{\alpha_{II}}{1-x},1\right\} \, ,
\end{equation}
where $\alpha_{K}, K=II,IF,FI,FF$ is a variable which can be used to
limit the kinematic range for the subtraction of initial-initial ($K=II$),
initial-final ($K=IF,FI$), and final-final ($K=FF$) dipoles
as suggested in Ref.~\cite{Nagy:2003tz}. $\alpha_{K}=1$ corresponds
to standard Catani-Seymour subtraction.
The Altarelli-Parisi function for the $q\ra gq$ splitting reads
\begin{equation}
  P^{qq}\brc{x} = \frac{3}{2}C_F\delta\brc{1-x} + P^{qq}_{reg}\brc{x}
  + 2C_F\brc{\frac{1}{1-x}}_+,
\end{equation}
with
\[
P^{qq}_{reg} = -C_F\brc{1+x}.
\]

The real corrections in the quark-gluon-initiated subprocesses in the
two-gluon-two-quark category shown in Table~\ref{tab:2g2q} (processes B-E)
only exhibit initial-state collinear divergences that are eventually
absorbed into corresponding PDF counterterms.  The unintegrated dipole
contribution to these subprocesses at $\Oa$, given in Eq.~(\ref{eq:master}),
consists of two dipoles that can be written as:
\begin{equation}\label{eq:dipsdijet2}
  \begin{aligned}
    \sum_{\rdip}d\si^{\rB}\otimes dV_{\rdip} =
    & ~d\Phi_3\brc{p_g,p_{\bar{q}_i};p_{q_j},p_{\bar{q}_j},p_{\bar{q}_i}}\frac{1}{S_3}\Big\{
    \m{D}^{g\bar{q}_i,\bar{q}_i}\Theta\brc{\alpha_{II}-\tilde{v}_{\bar{q}_i}}
    \Big\} + \brc{q_i\lra\bar{q}_i}
  \end{aligned}
\end{equation}
with
\begin{eqnarray}\label{eq:2j:2g2q:dipole:head}
  \m{D}^{g\bar{q}_i,\bar{q}_i}& =& \frac{1}{2p_gp_{\bar{q}_i}}\frac{1}{x_{\bar{q}_i,gq_i}}\frac{1}{C_F}\ave{\bV^{g\bar{q}_i,\bar{q}_i}}\ave{\tp_{q_i},p_{\bar{q}_i},\tp_{q_j},\tp_{\bar{q}_j}|\bT_{q_i}\bT_{\bar{q}_i}|p_{q_i},\tp_{\bar{q}_i},\tp_{q_j},\tp_{\bar{q}_j}}_{\alpha\alpha_s} \; ,
\end{eqnarray}
where
\begin{equation}
  \begin{aligned}
    &\ave{\bV^{g\bar{q}_i,\bar{q}_i}} = 8\pi\alpha_sT_R
    \brc{
      1 - 2x_{\bar{q}_i,g\bar{q}_i} - 2x_{\bar{q}_i,g\bar{q}_i}^2
    }, \\
    &\tp_{q_i}^\mu = x_{\bar{q}_i,g\bar{q}_i}p_g^\mu, \quad
    \tp_{q_j(\bar{q}_j)}^\mu = p_{q_j(\bar{q}_j)}^\mu - \frac{2p_{q_j(\bar{q}_j)}\cdot\brc{K+\tilde{K}}}{\brc{K+\tilde{K}}^2}\brc{K+\tilde{K}}^\mu + \frac{2p_{q_j(\bar{q}_j)}\cdot K}{K^2}\tilde{K}^\mu, \\
    &x_{\bar{q}_i,g\bar{q}_i} = \frac{2p_gp_{\bar{q}_i}+p_{\bar{q}_i}^2}{p_gp_{\bar{q}_i}}, \quad
    K^\mu = p_g^\mu + 2p_{\bar{q}_i}^\mu, \quad
    \tilde{K}^\mu = \tp_{q_i}^\mu + p_{\bar{q}_i}^\mu, \quad \tilde{v}_{\bar{q}_i} = -1 \\
  & \m{D}^{gq_i,q_i} = \m{D}^{g\bar{q}_i,\bar{q}_i}\left|_{q_i\lra\bar{q}_i}\right.
  \end{aligned}
\end{equation}

The contribution of the corresponding integrated dipoles reads:
\begin{equation}
  \begin{aligned}
  & d\si^{\rB}\brc{p_g,p_{\bar{q}_i};p_{q_j},p_{\bar{q}_j}}\otimes\bI + \brc{q_i\lra\bar{q}_i} = -\frac{\alpha}{2\pi}\frac{\brc{4\pi}^\epsilon}{\Gamma\brc{1-\epsilon}}\int dxd\Phi_2\brc{xp_g,p_{\bar{q}_i};p_{q_j},p_{\bar{q}_j}}\frac{1}{C_F}\times\Bigg\{ \\
    &\brc{\frac{\mu^2}{s_{g\bar{q}_i}}}^\epsilon \bI_{g,\bar{q}_i}\brc{x,\epsilon;\alpha_{II}}\ave{xp_g(p_{q_i}),p_{\bar{q}_i},p_{q_j},p_{\bar{q}_j}|\bT_{q_i}\bT_{\bar{q}_i}|xp_g(p_{q_i}),p_{\bar{q}_i},p_{q_j},p_{\bar{q}_j}}_{\alpha\alpha_s}
    \Bigg\} + \brc{q_i\lra\bar{q}_i}
  \end{aligned}
\end{equation}
with
\begin{equation}\label{eq:2j:2g2q:dipole:tail}
  \begin{aligned}
    \bI_{g,\bar{q}_i}\brc{x,\epsilon;\alpha_{II}} = \bI_{g,q_i}\brc{x,\epsilon;\alpha_{II}} =
    & ~T_R~ \bigg\{
    \left[\brc{1-x}^2 + x^2\right]\left[2\log\brc{1-x} - \log{x} - \frac{1}{\epsilon}\right] + 2x - 2x^2 \\
    & + \left[\brc{1-x}^2 + x^2\right]\log\alpha_{II}\brc{x} \bigg\} \; ,
  \end{aligned}
\end{equation}
where the $\alpha_{II}(x)$ function is defined in Eq. (\ref{eq:al}) and
the Altarelli-Parisi function for the $g\ra q\bar{q}$ splitting reads
\begin{equation}
P^{gq}\brc{x} = P^{gq}_{reg}\brc{x} = T_R \left[\brc{1-x}^2 + x^2\right] \; .
\end{equation}

\begin{acknowledgments}
  This research is supported in part by the US DOE under contract
  DE-AC02-07CH11359 and the NSF under award no. PHY-1118138 and
  no. PHY-1417317. Part of this work was carried out at the KITP
  workshop {\em LHC Run II and the Precision Frontier} which is
  supported by NSF PHY11-25915. J.Z.'s work was supported in part by
  the Fermilab Graduate Student Research Program in Theoretical
  Physics.
\end{acknowledgments}

\bibliography{paper}

\end{document}